\documentclass[onecolumn,draftcls,10pt]{IEEEtran}
\usepackage{cite}
\ifCLASSINFOpdf
\else
   \usepackage[dvips]{graphicx}
\fi
\usepackage[cmex10]{amsmath}
\usepackage{amssymb,amsthm}
\newtheorem{assumption}{Assumption}
\newtheorem{definition}{Definition}
\newtheorem{proposition}{Proposition}
\newtheorem{lemma}{Lemma}

\newtheorem{remark}{Remark} 
\newcommand{\arginf}{\mathop{\mathrm{arginf}}\limits}

\begin{document}
%
% paper title
% can use linebreaks \\ within to get better formatting as desired
\title{Large-System Analysis of Joint Channel and Data Estimation for 
MIMO DS-CDMA Systems}
%
%
% author names and IEEE memberships
% note positions of commas and nonbreaking spaces ( ~ ) LaTeX will not break
% a structure at a ~ so this keeps an author's name from being broken across
% two lines.
% use \thanks{} to gain access to the first footnote area
% a separate \thanks must be used for each paragraph as LaTeX2e's \thanks
% was not built to handle multiple paragraphs
%

\author{Keigo~Takeuchi,~\IEEEmembership{Member,~IEEE,}
        Mikko~Vehkaper\"a,~\IEEEmembership{Member,~IEEE,}
        Toshiyuki~Tanaka,~\IEEEmembership{Member,~IEEE,}
        and~Ralf~R.~M\"uller,~\IEEEmembership{Senior Member,~IEEE}% <-this % stops a space
\thanks{Manuscript received February 24, 2010; revised March , 2011.
The work of K.~Takeuchi was in part supported by the Grant-in-Aid for 
Young Scientists (Start-up) (No. 21860035) from MEXT, Japan and by the 
Grant-in-Aid for Scientific Research on Priority Areas (No. 18079010) from 
MEXT, Japan. 
The work of M.~Vehkaper\"a was supported by the Norwegian Research Council 
under grant 171133/V30. 
The work of T.~Tanaka was in part supported by the Grant-in-Aid for 
Scientific Research on Priority Areas (No. 18079010) from MEXT, Japan. 
The material in this paper was presented in part at the 2008 IEEE 
International Symposium on Information Theory, Toronto, Canada, July 2008 
and at the 2nd Workshop on Physics-Inspired Paradigms in Wireless 
Communications and Networks, Seoul, Korea, June 2009.}
\thanks{K.~Takeuchi is with the Department of Communication Engineering and 
Informatics, the University of Electro-Communications, Tokyo 182-8585, Japan 
(e-mail: takeuchi@ice.uec.ac.jp).}
\thanks{M.~Vehkaper\"a was with the Department of 
Electronics and Telecommunications, the Norwegian University of Science and 
Technology (NTNU), NO--7491 Trondheim, Norway. He is currently with the School 
of Electrical Engineering, Royal Institute of Technology (KTH), SE--100 44 Stockholm, Sweden (e-mail: mikkok@kth.se).}% <-this % stops a space
\thanks{T.~Tanaka is with the Department of Systems Science, 
Graduate School of Informatics, Kyoto University, Kyoto, 
606-8501, Japan (e-mail: tt@i.kyoto-u.ac.jp).}
\thanks{R.~R.~M\"uller is with the Department of 
Electronics and Telecommunications, the Norwegian University of Science and 
Technology (NTNU), NO--7491 Trondheim, Norway (e-mail: ralf@iet.ntnu.no).}% <-this % stops a space
}

% note the % following the last \IEEEmembership and also \thanks - 
% these prevent an unwanted space from occurring between the last author name
% and the end of the author line. i.e., if you had this:
% 
% \author{....lastname \thanks{...} \thanks{...} }
%                     ^------------^------------^----Do not want these spaces!
%
% a space would be appended to the last name and could cause every name on that
% line to be shifted left slightly. This is one of those "LaTeX things". For
% instance, "\textbf{A} \textbf{B}" will typeset as "A B" not "AB". To get
% "AB" then you have to do: "\textbf{A}\textbf{B}"
% \thanks is no different in this regard, so shield the last } of each \thanks
% that ends a line with a % and do not let a space in before the next \thanks.
% Spaces after \IEEEmembership other than the last one are OK (and needed) as
% you are supposed to have spaces between the names. For what it is worth,
% this is a minor point as most people would not even notice if the said evil
% space somehow managed to creep in.

% The paper headers
\markboth{IEEE transactions on information theory,~Vol.~, No.~, 2010}%
{Takeuchi \MakeLowercase{\textit{et al.}}: Joint 
Channel and Data Estimation for MIMO DS-CDMA}
% The only time the second header will appear is for the odd numbered pages
% after the title page when using the twoside option.
% 
% *** Note that you probably will NOT want to include the author's ***
% *** name in the headers of peer review papers.                   ***
% You can use \ifCLASSOPTIONpeerreview for conditional compilation here if
% you desire.

% If you want to put a publisher's ID mark on the page you can do it like
% this:
\IEEEpubid{0000--0000/00\$00.00~\copyright~2010 IEEE}
% Remember, if you use this you must call \IEEEpubidadjcol in the second
% column for its text to clear the IEEEpubid mark.

% use for special paper notices
%\IEEEspecialpapernotice{(Invited Paper)}

% make the title area
\maketitle

\begin{abstract}
This paper presents a large-system analysis of the performance of 
joint channel estimation, multiuser detection, and per-user decoding (CE-MUDD) 
for randomly-spread multiple-input multiple-output (MIMO) direct-sequence 
code-division multiple-access (DS-CDMA) systems. 
A suboptimal receiver based on successive decoding in conjunction with 
linear minimum mean-squared error (LMMSE) channel estimation is investigated.  
The replica method, developed in statistical mechanics, is used to evaluate 
the performance in the large-system limit, where the number of users and 
the spreading factor tend to infinity while their ratio and the number of 
transmit and receive antennas are kept constant. The performance of the joint 
CE-MUDD based on LMMSE channel estimation is compared to the spectral 
efficiencies of several receivers based on one-shot LMMSE channel estimation, 
in which the decoded data symbols are not utilized to refine the initial 
channel estimates. The results imply that the use of joint CE-MUDD 
significantly reduces rate loss due to transmission of pilot signals, 
especially for multiple-antenna systems. As a result, joint CE-MUDD can 
provide significant performance gains, compared to the receivers based on 
one-shot channel estimation.  
\end{abstract}
% IEEEtran.cls defaults to using nonbold math in the Abstract.
% This preserves the distinction between vectors and scalars. However,
% if the journal you are submitting to favors bold math in the abstract,
% then you can use LaTeX's standard command \boldmath at the very start
% of the abstract to achieve this. Many IEEE journals frown on math
% in the abstract anyway.

% Note that keywords are not normally used for peerreview papers.

% For peer review papers, you can put extra information on the cover
% page as needed:
%\ifCLASSOPTIONpeerreview
% \begin{center} \bfseries EDICS Category: 3-BBND \end{center}
% \fi
%
% For peerreview papers, this IEEEtran command inserts a page break and
% creates the second title. It will be ignored for other modes.
\IEEEpeerreviewmaketitle

\section{Introduction}
\IEEEPARstart{D}{irect}-sequence code-division multiple-access (DS-CDMA) 
schemes are used in the air interface of third-generation (3G) mobile 
communication systems~\cite{Adachi98,Dahlman98,Ojanperae98}. 
In order to improve the spectral efficiency of DS-CDMA systems, 
the extension to multiple-input multiple-output (MIMO) DS-CDMA systems 
has been actively considered~\cite{Mantravadi03,Ni05,Juntti05,Li06,Nordio062,Buzzi06,Choi07,Takeuchi082,Hanly01,Cottatellucci07}. As a drawback of  
using multiple antennas at the transmitters, the receiver structure of MIMO 
DS-CDMA systems becomes more complex than that of conventional DS-CDMA 
systems. Therefore, it is important in MIMO DS-CDMA systems to construct 
receivers achieving an acceptable tradeoff between performance and complexity. 

A goal in this research area is to construct receivers which 
achieve near-optimal performance by using acceptable computational costs, 
since optimal joint decoding is infeasible in terms of complexity. 
Separation of detection and decoding significantly reduces the complexity of 
the receiver, although it is suboptimal in terms of performance. 
In multiuser detection (MUD)~\cite{Verdu98}, certain statistical properties 
of multiple-access interference (MAI) are used to detect data symbols.  
Large-system analysis is a powerful approach for evaluating the performance 
of MUD followed by per-user decoding for randomly-spread 
DS-CDMA systems. In this analysis an asymptotic limit is assumed, referred to 
as the large-system limit, in which the number of users and the spreading 
factor tend to infinity while their ratio is kept constant.   
A series of large-system analyses~\cite{Tse99,Verdu99,Shamai01,Tanaka02,Mueller04,Guo051} have revealed that the linear minimum mean-squared error (LMMSE) 
detection followed by per-user decoding achieves near-optimal 
performance for lightly-loaded systems with perfect channel state 
information (CSI) at the receiver. 
Furthermore, it has been shown 
numerically~\cite{Alexander98,Reed98,Moher98,Wang99} 
and analytically~\cite{Boutros02,Caire04} that 
iterative LMMSE-based multiuser detection and per-user decoding (MUDD) can 
achieve near-optimal performance even for highly-loaded systems with 
perfect CSI at the receiver.  
See \cite{Mantravadi03,Hanly01,Cottatellucci07} for the extension of these  
results to MIMO DS-CDMA systems with perfect CSI at the receiver. 
In the large-system analysis for MIMO DS-CDMA systems the numbers of transmit 
and receive antennas are commonly fixed, while the number of users and the 
spreading factor tend to infinity with their ratio fixed. Note that a 
many-antenna limit, in which the numbers of transmit and receive antennas 
tend to infinity, may be taken after this large-system 
limit~\cite{Takeuchi07}. In this paper, the numbers of transmit and receive 
antennas are kept finite. 

It is worth considering the no-CSI case since CSI is unknown in advance 
for practical MIMO DS-CDMA systems. Iterative MUDD was extended to iterative 
channel estimation (CE) and MUDD (CE-MUDD) in \cite{Alexander00}. 
In this scheme, the channel estimator utilizes soft feedback from the per-user 
decoders for refining the initial channel estimates. 
In terms of complexity, iterative CE-MUDD requires updating the filter 
coefficients of the channel estimator and the detector in every iteration. 
This implies that the computational complexity of iterative CE-MUDD is higher 
than that for receivers based on one-shot channel estimation, in which data 
estimation is performed without refining the initial channel estimates.  
In order to reduce the complexity of iterative CE-MUDD, linear channel 
estimators, such as the LMMSE channel estimator, have been commonly used. 
Linear channel estimators use known pilot symbols to obtain the initial 
channel estimates, while the optimal channel estimator can attain  
the initial channel estimates with no pilot symbols. 
Numerical simulations~\cite{Alexander00,Gamal00,Lampe02} 
demonstrated that iterative LMMSE-based CE-MUDD can provide significant 
performance gains for highly-loaded systems, compared to receivers based on 
one-shot channel estimation. 
However, these works did not discuss how to design the length of symbol 
periods assigned for transmission of pilot symbols, called training phase.   
Increasing the length of the training phase improves the accuracy of channel 
estimation, while transmission of pilot symbols reduces the transmission 
rate. Thus, how to design the length of the training phase should provide a 
great impact on achievable spectral efficiency for the no-CSI case. 
The goal of this paper is to optimize the length of the training phase on the 
basis of information-theoretical capacities. 

%the benefits obtained by using iterative LMMSE-based CE-MUDD, in 
%place of using receivers based on one-shot channel estimation, are not fully 
%understood. 

Simple modulation schemes, such as quadrature phase shift keying (QPSK) or 
quadrature amplitude modulation (QAM), are commonly used for coherent 
receivers that estimate CSI explicitly. Thus, the channel capacity for a fixed 
modulation scheme, called constrained capacity, corresponds to a performance 
bound\footnote{
Intuitively, one may expect that the optimal CE-MUDD can be implemented with 
iterative CE-MUDD. However, it is unclear whether the solution of the 
iterative CE-MUDD converges to the global optimal solution. Therefore, the 
constrained capacity might be a loose bound for the performance of iterative 
CE-MUDD.} for iterative receivers, while the true capacity might be achieved by 
using complicated modulation~\cite{Marzetta99}.  
It is a challenging issue to derive analytical formulas for the constrained 
capacities of wireless communication systems with no CSI, since the optimal 
channel estimator is nonlinear.   
Lower bounds for the constrained capacities have been derived 
instead~\cite{Medard00,Hassibi03}. The basic idea in these works for obtaining 
lower bounds is to replace the noise due to channel 
estimation errors by additive white Gaussian noise (AWGN). This replacement 
reduces optimal nonlinear channel estimation to LMMSE channel estimation. 
In this paper, this type of a lower bound is referred to as a lower 
bound based on LMMSE channel estimation.

We derive a lower bound for the constrained capacity of randomly-spread MIMO 
DS-CDMA systems with no CSI, on the basis of LMMSE channel estimation. 
The lower bound can be regarded as 
a performance index for iterative CE-MUDD based on LMMSE channel estimation. 
In an information-theoretical point of view, both LMMSE channel estimates and 
covariances of the estimation errors should be used in MUD~\cite{Evans00}, 
while the use of the covariances is uncommon in practice. In this paper, we 
consider {\em suboptimal} LMMSE channel estimation in which only a part of the 
covariances for the LMMSE estimation errors are used, while all covariances 
are used in the {\em true} LMMSE channel estimation. 
The use of suboptimal LMMSE channel estimation allows us to evaluate a lower 
bound for the constrained capacity. We refer to this lower bound as the 
performance of the joint CE-MUDD based on suboptimal LMMSE channel 
estimation, or simply as the performance of the joint CE-MUDD. 
In order to investigate the benefits obtained by using iterative CE-MUDD, 
we also analyze the performance of three receivers 
based on one-shot LMMSE channel estimation: one-shot CE-MUDD, an optimum 
separated receiver, and an LMMSE receiver. 
Table~\ref{table} lists the four receivers considered in this paper. 
The one-shot CE-MUDD performs 
joint MUDD based on one-shot LMMSE channel estimation. Intuitively, 
the performance of the one-shot CE-MUDD corresponds to a performance bound 
for an iterative receiver obtained by eliminating the feedback from the 
per-user decoders to the channel estimator in iterative CE-MUDD. 
The optimum separated receiver performs separated optimal detection and 
decoding on the basis of one-shot LMMSE channel estimation. The performance of 
the optimum separated receiver corresponds to a performance bound for a 
non-iterative receiver obtained by eliminating the feedback from the 
per-user decoders to the channel estimator and to the detector. 
The LMMSE receiver is obtained by replacing the optimal 
detector in the optimum separated receiver by the LMMSE detector. 
To the best of our knowledge, no analytical results for joint 
CE-MUDD are obtained, except for non-iterative linear receivers~\cite{Evans00} 
and iterative CE-MUDD based on hard decision feedback~\cite{Li07}. 
The methodology developed in this paper is applicable to the analysis of 
iterative LMMSE-based CE-MUDD. See \cite{Vehkaperae091} for details. 

\begin{table}[t]
\caption{
Four receivers considered in this paper. 
%$T_{\mathrm{ch}}$, $T_{\mathrm{det}}$, $T_{\mathrm{det}}^{(\mathrm{L})}$, 
%$T_{\mathrm{dec}}$, and $n$ denote the computational costs of the channel 
%estimator, the optimal detector, the LMMSE detector, a per-user decoder, 
%and the number of iterations, respectively. 
}
\label{table}
\begin{center}
\begin{tabular}{||c||c||c||c|c||}
\hline
&  & & \multicolumn{2}{|c||}{feedback} \\ 
\cline{4-5}
& channel estimation(CE) & MUD & to CE & to detector \\ 
\hline
\hline
joint CE-MUDD & suboptimal LMMSE & optimal & available & available \\ 
%& $O(n(T_{\mathrm{ch}}+T_{\mathrm{det}}+KT_{\mathrm{dec}}))$ \\
\hline
one-shot CE-MUDD & suboptimal LMMSE & optimal & & available \\ 
%& $O(T_{\mathrm{ch}} + n(T_{\mathrm{det}}+KT_{\mathrm{dec}}))$ \\
\hline
optimum separated receiver & suboptimal LMMSE & optimal & & \\ 
%& $O(T_{\mathrm{ch}} + T_{\mathrm{det}} + KT_{\mathrm{dec}})$ \\
\hline
LMMSE receiver & suboptimal LMMSE & LMMSE & & \\
%& $O(T_{\mathrm{ch}} + T_{\mathrm{det}}^{(\mathrm{L})} + KT_{\mathrm{dec}})$ \\
\hline
\end{tabular}
\end{center}
\end{table}

Our large-system analysis is based on the replica 
method~\cite{Nishimori01,Fischer91,Mezard87}, 
which is a powerful method for analyzing randomly-spread DS-CDMA 
systems~\cite{Guo051,Tanaka02,Mueller04,Takeda06,Takeuchi082} 
and MIMO systems~\cite{Moustakas03,Wen07,Mueller08,Zaidel10}.   
The replica method is based on several non-rigorous procedures at present 
time. In this paper, we assume that results obtained by using these procedures 
are correct since their proof is beyond the scope of this paper. 
See \cite{Guerra032,Talagrand06,Korada10} for recent progress with respect to 
the assumptions of the replica method. 

This paper is organized as follows: After summarizing the notation used in 
this paper, in Section~\ref{section_model} we 
introduce a discrete-time model of MIMO DS-CDMA systems. In 
Section~\ref{section_CE-MUDD}, the joint CE-MUDD based on suboptimal LMMSE 
channel estimation is defined. In Section~\ref{section_one_shot}, we define 
the three receivers based on one-shot LMMSE channel estimation. 
Section~\ref{section_main_result} presents the main results 
of this paper. In Section~\ref{section_numerical_result}, we compare the 
performance of the joint CE-MUDD with that of the three receivers based on 
one-shot LMMSE channel estimation. Numerical simulations for finite-sized 
systems are also performed to demonstrate the usefulness of our 
large-system analysis. In Section~\ref{section_conclusion}, 
we conclude this paper. The derivations of the main results are 
summarized in the appendices. 

\subsection{Notation} \label{section_notation} 
For a complex number $z\in\mathbb{C}$ and a real number $x\in\mathbb{R}$, 
$\mathrm{Re}(z)$, $\mathrm{Im}(z)$, $\mathrm{i}$, $z^{*}$, $\log x$, and 
$\ln x$ denote the real part, imaginary part, imaginary unit, complex 
conjugate, $\log_{2}x$, and $\log_{\mathrm{e}}x$, respectively. 
$|\mathcal{A}|$ stands for the number of elements of a set $\mathcal{A}$. 
For a matrix $\boldsymbol{A}$, $\boldsymbol{A}^{T}$, $\boldsymbol{A}^{H}$, 
$\mathrm{Tr}(\boldsymbol{A})$, and $\det(\boldsymbol{A})$ denote the 
transpose, conjugate transpose, trace, and the determinant, respectively. 
%$\mathrm{diag}(\boldsymbol{A}_{1},\ldots,\boldsymbol{A}_{N})$ represents the 
%block diagonal matrix with $\boldsymbol{A}_{n}$ as block~($n$, $n$). 
$\boldsymbol{I}_{N}$ stands for the $N\times N$ identity matrix.  
$\boldsymbol{1}_{N}$ denotes the $N$-dimensional vector whose elements are 
all one. $\boldsymbol{e}_{N}^{(n)}$ represents the $N$-dimensional vector in 
which the $n$th element is one and the other elements are all zero. 
$\mathcal{M}_{n}^{+}$ denotes the set of all positive definite $n\times n$ 
Hermitian matrices.  
%$\mathrm{vec}(\boldsymbol{A})$ stands for the vector 
%$\mathrm{vec}(\boldsymbol{A})
%=(\boldsymbol{a}_{1}^{T},\ldots,\boldsymbol{a}_{n}^{T})^{T}$ for a 
%matrix $\boldsymbol{A}=(\boldsymbol{a}_{1},\ldots,\boldsymbol{a}_{n})$.   
$\otimes$ denotes the Kronecker product operator between two matrices.  
$\delta(\cdot)$ represents the Dirac delta function, while $\delta_{i,j}$ 
denotes the Kronecker delta.  
$p(x)$ and $p(y|x)$ stand for the probability density function 
(pdf) of a continuous random variable $x$ and the conditional pdf of a 
continuous random variable $y$ given $x$, respectively. 
We use the same symbol $p(x)$ for the probability mass function 
(pmf) of a discrete random variable $x$.  
$a\sim P(a)$ indicates that the distribution of a random variable $a$ equals 
a distribution $P(a)$. If the pdf $p(a)$ of $a$ exists, we use the notation 
$a\sim p(a)$ to represent that $a$ follows the distribution whose 
pdf is given by $p(a)$. 
$\mathcal{CN}(\boldsymbol{m},\boldsymbol{\Sigma})$ denotes 
a proper $n$-dimensional complex Gaussian distribution with mean 
$\boldsymbol{m}\in\mathbb{C}^{N}$ and a covariance matrix 
$\boldsymbol{\Sigma}\in\mathcal{M}_{n}^{+}$~\cite{Neeser93}. 
The pdf of the $n$-dimensional complex Gaussian random vector 
$\boldsymbol{x}\sim\mathcal{CN}(\boldsymbol{m},\boldsymbol{\Sigma})$ is 
defined as $p(\boldsymbol{x})
=g_{n}(\boldsymbol{x}-\boldsymbol{m};\boldsymbol{\Sigma})$, given by 
\begin{equation} \label{Gauss} 
g_{n}(\boldsymbol{y};\boldsymbol{\Sigma}) = 
\frac{1}{\pi^{n}\det\boldsymbol{\Sigma}}\mathrm{e}^{
 - \boldsymbol{y}^{H}\boldsymbol{\Sigma}^{-1}\boldsymbol{y}
}. 
\end{equation}
$D(\boldsymbol{A} \|\boldsymbol{B})$ stands for the 
Kullback-Leibler divergence with the logarithm to base $2$ between  
$\mathcal{CN}(\boldsymbol{0},\boldsymbol{A})$ and 
$\mathcal{CN}(\boldsymbol{0},\boldsymbol{B})$. 
%In addition, $\mathrm{cov}[\boldsymbol{a}| b]$ stands for 
%the covariance matrix of a random vector $\boldsymbol{a}$ given a random 
%variable $b$. 
$I(x;y | z)$ denotes the conditional mutual information with the logarithm to 
base $2$ between a random variable $x$ and a random variable $y$ conditioned 
on a random variable $z$. 

The indices of chips, symbol periods, users, transmit antennas, and replicas 
are denoted by $l$, $t$, $k$, $m$, and $a$, respectively. In this paper, 
indices themselves have meanings, like the argument of distributions.  
Symbols with several superscripts or subscripts are used in this paper. 
We write sets of the symbols as follows: For a symbol $a_{i,j}^{k}$ and 
a subset $\mathcal{J}$ of indices $\{j\}$, the 
set $\mathcal{A}_{i,\mathcal{J}}^{k}$ denotes a subset 
$\{a_{i,j}^{k}:\hbox{for $j\in\mathcal{J}$}\}$ for fixed $i$ and $k$. 
When $\mathcal{J}$ equals the set of all indices $\{j\}$, 
$\mathcal{A}_{i,\mathcal{J}}^{k}$ is also written as $\mathcal{A}_{i}^{k}$.  
The sets $\mathcal{A}_{i}$, $\mathcal{A}$, and so on are defined in the same 
manner. The two sets $\mathcal{A}_{i}^{k}$ and $\mathcal{A}_{j}^{k}$ 
should not be confused with each other. 
The set $\mathcal{J}\backslash\{j\}=\{j'\in\mathcal{J}: j'\neq j\}$ denotes 
the set obtained by eliminating the element $j$ from $\mathcal{J}$. 
When $\mathcal{J}$ equals the set of all indices $\{j\}$, 
$\mathcal{J}\backslash\{j\}$ is simply written as $\backslash\{j\}$.  
As notational convenience for subsets of the natural numbers $\mathbb{N}$, 
we use $[a,b) = \{i\in\mathbb{N}: a\leq i < b\}$ for integers $a$ and 
$b(>a)$, which is always used as subscripts or superscripts for discrete sets. 
The other sets $[a,b]$, $(a,b)$, and so on are defined in the same manner. 
As exceptions, the two sets $\{t'\in\mathbb{N}: 1\leq t'\leq t\}$ and 
$\{t'\in\mathbb{N}:t\leq t'\leq T_{\mathrm{c}}\}$ for coherence time 
$T_{\mathrm{c}}$ are denoted by $\mathcal{T}_{t}$ and $\mathcal{C}_{t}$, 
respectively, instead of $[1,t]$ and $[t,T_{\mathrm{c}}]$. 

We use symbols with tildes and hats to represent random variables for 
postulated channels and estimates of random variables, respectively. 
Underlined symbols are used to represent random variables for decoupled 
single-user channels. Note that there are several exceptions in the replica 
analyses presented in Appendix~\ref{derivation_lemma_channel_estimation} and 
Appendix~\ref{appendix_derivation_proposition1}.  

\section{MIMO DS-CDMA Channel} \label{section_model}
We consider the uplink of a synchronous $K$-user frequency-flat fading MIMO 
DS-CDMA system with spreading factor~$L$, 
in which each user and the receiver have $M$ transmit antennas and $N$ 
receive antennas, respectively. A per-antenna spreading scheme is 
investigated in this paper: Different spreading sequences are used for 
different transmit antennas of each user. 
See \cite{Takeuchi081} for a generalization of spreading schemes. 
We assume block-fading channels with coherence time $T_{\mathrm{c}}$,  
i.e., fading coefficients do not change during $T_{\mathrm{c}}$ symbol 
periods, and they are independently sampled from a distribution at the 
beginning of the next coherent interval. 

The input symbol $u_{t,k,m}\in\mathbb{C}$ for the $m$th transmit antenna of 
user~$k$ in symbol period~$t$ is spread with a spreading sequence 
$\{s_{l,t,k,m}:l=1,\ldots,L\}$. 
%User~$k$ transmits the product of an input vector 
%$\boldsymbol{u}_{k,t}\in\mathbb{C}^{M}$ and the sequence 
%$\mathcal{S}_{k,t}=\{\boldsymbol{S}_{l,k,t}\in\mathbb{C}^{M\times M}
%: l=1,\ldots,L\}$ of spatial spreading matrices in symbol period~$t$. 
The chip-sampled received vectors  
$\{\boldsymbol{y}_{l,t}\in\mathbb{C}^{N}:l=1,\ldots,L\}$ 
in symbol periods~$t=1,\ldots,T_{\mathrm{c}}$ are given by 
\begin{equation} \label{MIMO_DS_CDMA}
\boldsymbol{y}_{l,t} = \frac{1}{\sqrt{L}}\sum_{k=1}^{K}\sum_{m=1}^{M} 
\boldsymbol{h}_{k,m}s_{l,t,k,m}u_{t,k,m} + \boldsymbol{n}_{l,t}. 
\end{equation}
In (\ref{MIMO_DS_CDMA}), $\boldsymbol{n}_{l,t}\sim\mathcal{CN}(\boldsymbol{0},
N_{0}\boldsymbol{I}_{N})$ represents the AWGN vector with variance 
$N_{0}$. Furthermore, 
$\boldsymbol{h}_{k,m}\in\mathbb{C}^{N}$ denotes the channel vector between 
transmit antenna~$m$ of the $k$th user and the receiver. Note that 
the channel vectors are fixed during $T_{\mathrm{c}}$ symbol periods. 

%Furthermore, the spatial spreading matrix 
%$\boldsymbol{S}_{l,k,t}\in\mathbb{C}^{M\times M}$ is a diagonal 
%matrix having the chip $s_{l,k,t,m}$ for the $m$th transmit antenna of 
%user~$k$ as the $m$th diagonal element, i.e., 
%$\boldsymbol{S}_{l,k,t}=\mathrm{diag}(s_{l,k,t,1},\ldots,s_{l,k,t,M})$. 

The assumption of frequency-flat fading channels might be an unrealistic 
assumption since practical MIMO DS-CDMA systems commonly operate over 
frequency-selective fading channels. For the sake of simplicity, however, 
we consider frequency-flat fading channels. An extension to 
frequency-selective fading channels is possible by considering the assumption 
of independent spreading sequences across different resolvable paths 
\cite{Evans00,Mantravadi03}. For details, see \cite{Takeuchi084}. 

The receiver does not have CSI in advance, which is information about all 
realizations of the channel vectors $\{\boldsymbol{h}_{k,m}: 
\hbox{for all $k$, $m$}\}$, while it knows all spreading 
sequences, the variance $N_{0}$ of the AWGN, and the statistical properties of 
all channel vectors and input symbols. 
In order for the receiver to estimate the channel vectors,  
consider that the first $\tau$ symbol periods $\mathcal{T}_{\tau}
=\{1,\ldots,\tau\}$ 
in each coherent interval are assigned to a training phase, and that the 
remaining $\tau'=T_{\mathrm{c}}-\tau$ symbol periods 
$\mathcal{C}_{\tau+1}=\{\tau+1,\ldots,T_{\mathrm{c}}\}$ 
are assigned to a communication phase. 
The length of the training phase $\tau$ is a design parameter, which will be 
optimized on the basis of large-system results. User~$k$ transmits pilot 
symbols  $\{x_{t,k,m}\in\mathbb{C}\}$ known to the 
receiver from the $m$th transmit antenna in the training phase 
$t\in\mathcal{T}_{\tau}$, and 
subsequently sends data symbols $\{b_{t,k,m}\in\mathbb{C}\}$ in the 
communication phase $t\in\mathcal{C}_{\tau+1}$. Therefore, 
the input symbol $u_{t,k,m}$ is given by 
\begin{equation} \label{input_symbol} 
u_{t,k,m} = \left\{
\begin{array}{cl}
x_{t,k,m} & \hbox{for $t\in\mathcal{T}_{\tau}$,} \\
b_{t,k,m} & \hbox{for $t\in\mathcal{C}_{\tau+1}$.}  
\end{array}
\right. 
\end{equation}

Throughout this paper, we assume that the input symbols $\{u_{t,k,m}\}$ are 
mutually independent for all $t$, $k$, $m$, and that each $u_{t,k,m}$ is 
a zero-mean random variable satisfying $|u_{t,k,m}|^{2}=P/M$. In numerical 
results, unbiased QPSK input symbols with $|u_{t,k,m}|^{2}=P/M$ are used. 
Furthermore, it is straightforward to extend the 
results to the unequal power case.  
%$P$ corresponds to the receive power 
%$of the input vector $\boldsymbol{u}_{k,t}$ in each symbol period. 
%$\mathcal{P}=\{P:\hbox{for all $k$}\}$ are assumed to be 
%independently sampled from a distribution. For convenience of notation, 
%we omit conditioning with respect to $\mathcal{P}$. For example, 
%$p(b_{t,k,m})$ stands for $p(b_{t,k,m}|P)$. 
Next, we assume that the channel vectors $\{\boldsymbol{h}_{k,m}\}$ are 
mutually independent for all $k$, $m$, and that each $\boldsymbol{h}_{k,m}$ 
has independent and identically distributed (i.i.d.) circularly symmetric 
complex Gaussian (CSCG) elements with unit variance. Finally, we assume that 
the spreading sequences $\{s_{l,t,k,m}: \hbox{for all $l$, $t$, $k$, $m$}\}$ 
are i.i.d. for all $l$, $t$, $k$, $m$, 
and that each $s_{l,t,k,m}$ is a CSCG random variable with unit variance. 
We have made the CSCG assumption of each chip for the sake of simplicity 
in analysis. We believe that the main results presented in this paper hold 
for a general distribution of $s_{l,t,k,m}$ with zero mean and 
finite moments, as shown numerically in Section~\ref{section_numerical_result}. 
%Without loss of generality,  
%$\boldsymbol{S}_{l,k,t}$ is represented as~\cite{Takeuchi081} 
%\begin{equation} \label{random_spreading} 
%\boldsymbol{S}_{l,k,t} = \sum_{j=1}^{J_{k}}s_{l,k,t,j}
%\boldsymbol{\Gamma}_{k,j}, 
%\end{equation}
%where $J_{k}$ denotes the rank of 
%$\mathrm{cov}[\mathrm{vec}(\boldsymbol{S}_{l,k,t})]$; $\{s_{l,k,t,j}\}$ 
%are mutually independent CSCG random variables with unit variance; and where 
%a deterministic matrix 
%$\boldsymbol{\Gamma}_{k,j}\in\mathbb{C}^{M\times M}$ is given as 
%$\sum_{j=1}^{J_{k}}\mathrm{vec}(\boldsymbol{\Gamma}_{k,j})
%\mathrm{vec}(\boldsymbol{\Gamma}_{k,j})^{H}=
%\mathrm{cov}[\mathrm{vec}(\boldsymbol{S}_{l,k,t})]$. 
%Equation~(\ref{random_spreading}) implies that it is possible to 
%generate a sequence of spatial spreading matrices by spatially dispersing 
%$J_{k}$ different spreading sequences with dispersion matrices 
%$\{\boldsymbol{\Gamma}_{k,j}:\hbox{for all $j$}\}$. 

We shall present several sets used in this paper: The set 
$\mathcal{Y}_{t}=\{\boldsymbol{y}_{l,t}\in\mathbb{C}^{N}:l=1,\ldots,L\}$ 
denotes the received vectors in symbol period~$t$. 
The set $\mathcal{S}_{t}=\{s_{l,t,k,m}:\hbox{for all $l$, $k$, $m$}\}$ 
represents the spreading sequences in symbol period~$t$.  
The sets $\mathcal{B}_{t,k}=\{b_{t,k,m}:\hbox{for all $m$}\}$ and 
$\mathcal{X}_{t,k}=\{x_{t,k,m}:\hbox{for all $m$}\}$ represent  
the data and pilot symbols for user~$k$ in symbol period~$t$, respectively.  
All data and pilot symbols in symbol period~$t$ are denoted by 
$\mathcal{B}_{t}=\{\mathcal{B}_{t,k}:\hbox{for all $k$}\}$ and 
$\mathcal{X}_{t}=\{\mathcal{X}_{t,k}:\hbox{for all $k$}\}$.   
The set $\mathcal{U}_{t}=\{u_{t,k,m}:\hbox{for all $k$, $m$}\}$ represents 
the input symbols in symbol period~$t$. 
The channel vectors for user~$k$ are denoted by 
$\mathcal{H}_{k}=\{\boldsymbol{h}_{k,m}:\hbox{for all $m$}\}$. 
The set $\mathcal{I}_{t}=\{\mathcal{Y}_{t},\ \mathcal{S}_{t},\  
\mathcal{U}_{t}\}$ denotes the information about 
the received vectors $\mathcal{Y}_{t}$, 
the spreading sequences $\mathcal{S}_{t}$, and the input 
symbols $\mathcal{U}_{t}$ in symbol period~$t$. 
The set $\overline{\mathcal{I}}_{t}=\{\mathcal{Y}_{t},\ \mathcal{S}_{t}\}$ 
is obtained by eliminating the input symbols $\mathcal{U}_{t}$ from 
$\mathcal{I}_{t}$.  
The training phase in stage~$t$ of successive decoding, introduced in the 
next section, is denoted by $\mathcal{T}_{t-1}=\{1,\ldots,t-1\}$, while the 
following stages are denoted by 
$\mathcal{C}_{t+1}=\{t+1,\ldots,T_{\mathrm{c}}\}$. The set 
$\mathcal{C}_{\tau+1}$ is also used to represent the communication phase. 
Note that the same index~$t$ as for symbol periods is used for the indices of 
stages in successive decoding. A list for several sets used in this paper is 
summarized in Appendix~\ref{appen_set}.

\section{Joint CE-MUDD} \label{section_CE-MUDD} 
\subsection{Joint CE-MUDD Based on LMMSE Channel Estimation} 
In order to define joint CE-MUDD based on LMMSE channel estimation, 
we shall derive a lower bound of the constrained capacity based on LMMSE 
channel estimation. The definition of joint CE-MUDD considered in this paper 
will be presented in the next subsection. 

We start by the constrained capacity of the MIMO DS-CDMA 
channel~(\ref{MIMO_DS_CDMA}) with no CSI. 
Let $\mathcal{Y}_{\mathcal{T}_{\tau}}
=\{\mathcal{Y}_{t}:t\in\mathcal{T}_{\tau}\}$ and 
$\mathcal{Y}_{\mathcal{C}_{\tau+1}}
=\{\mathcal{Y}_{t}: t\in\mathcal{C}_{\tau+1}\}$ denote the 
received vectors in the training and communication phases, respectively. 
Furthermore, we write all data symbols $\mathcal{B}$ and all pilot symbols 
$\mathcal{X}$ as 
$\mathcal{B}=\{\mathcal{B}_{t}:\hbox{for all $t\in\mathcal{C}_{\tau+1}$}\}$  
and $\mathcal{X}=\{\mathcal{X}_{t}:\hbox{for all $t\in\mathcal{T}_{\tau}$}\}$. 
The constrained capacity for the no-CSI case is given by the mutual 
information per chip between all data symbols $\mathcal{B}$ and 
$\{\mathcal{Y}_{\mathcal{C}_{\tau+1}}, \mathcal{Y}_{\mathcal{T}_{\tau}}, 
\mathcal{S}, \mathcal{X}\}$ known to the receiver~\cite{Cover06}, 
with $\mathcal{S}=\{\mathcal{S}_{t}:\hbox{for all $t$}\}$ denoting all 
spreading sequences, 
\begin{equation}
C_{\mathrm{opt}} = 
\frac{1}{LT_{\mathrm{c}}}I(\mathcal{B};\mathcal{Y}_{\mathcal{C}_{\tau+1}}, 
\mathcal{Y}_{\mathcal{T}_{\tau}}, \mathcal{S}, \mathcal{X}). 
\end{equation}  
Using the chain rule for mutual information, we obtain 
\begin{IEEEeqnarray}{rl} 
C_{\mathrm{opt}} 
=& \frac{1}{LT_{\mathrm{c}}}I(\mathcal{B};\mathcal{Y}_{\mathcal{C}_{\tau+1}}|  
\mathcal{Y}_{\mathcal{T}_{\tau}}, \mathcal{S}, \mathcal{X}) 
+ \frac{1}{LT_{\mathrm{c}}}I(\mathcal{B};\mathcal{Y}_{\mathcal{T}_{\tau}}, 
\mathcal{S}, \mathcal{X}) \nonumber \\  
=& \frac{1}{LT_{\mathrm{c}}}I(\mathcal{B};\mathcal{Y}_{\mathcal{C}_{\tau+1}}|  
\mathcal{Y}_{\mathcal{T}_{\tau}}, \mathcal{S}, \mathcal{X}), 
\label{spectral_efficiency}  
\end{IEEEeqnarray}
where the last equality holds since the data symbols $\mathcal{B}$ are 
independent of $\{\mathcal{Y}_{\mathcal{T}_{\tau}}, \mathcal{S}, 
\mathcal{X}\}$. 
For notational convenience, we hereafter omit conditioning with respect to 
$\mathcal{Y}_{\mathcal{T}_{\tau}}$, $\mathcal{S}$, and $\mathcal{X}$. 
%where we have omitted conditioning with respect to fixed 
%$\mathcal{Y}_{\mathcal{T}_{\tau}}$, $\mathcal{S}$, and $\mathcal{X}$ in order 
%to prevent the confusion of (\ref{spectral_efficiency}) with the conditional 
%mutual information $I(\mathcal{B};\mathcal{Y}_{\mathcal{C}_{\tau+1}}|
%\mathcal{Y}_{\mathcal{T}_{\tau}}, \mathcal{S},\mathcal{X})$. 
%For notational convenience, we hereafter omit conditioning with respect to 
%$\mathcal{Y}_{\mathcal{T}_{\tau}}$, $\mathcal{S}$, and $\mathcal{X}$. 

It is well known that the constrained capacity can be achieved 
by successive decoding. In order to obtain a lower bound for the constrained 
capacity, a successive decoding strategy has been considered in 
\cite{Teng_Li07,Padmanabhan08}. 
In this strategy, channel estimation and MUD in successive decoding are 
replaced by suboptimal ones that allow us to evaluate the performance 
analytically. We present joint CE-MUDD based on LMMSE channel estimation, 
following the successive decoding strategy. In successive decoding with 
$\tau'=T_{\mathrm{c}}-\tau$ stages, the data symbols $\{\mathcal{B}_{t}\}$ are 
decoded in the order $t=\tau+1,\ldots,T_{\mathrm{c}}$. 
Stage~$t\in\mathcal{C}_{\tau+1}$ consists of 
$K$ substages, in which the data symbols $\{\mathcal{B}_{t,k}\}$ 
in symbol period~$t$ are decoded in the order $k=1,\ldots,K$. 
We focus on substage~$k$ within stage~$t$. Let $\mathcal{B}_{(\tau,t)} = 
\{\mathcal{B}_{t'}:t'=\tau+1,\ldots,t-1\}$ and  
$\mathcal{Y}_{\mathcal{C}_{\tau+1}\backslash\{t\}}=\{\mathcal{Y}_{t'}:
t'\in\mathcal{C}_{\tau+1}, t'\neq t\}$ denote the data symbols decoded 
successfully in the preceding stages and the received vectors in the 
communication phase except for $\mathcal{Y}_{t}$, respectively. 
The independency of $\{\mathcal{B}_{t}\}$ for all $t$ implies that 
$\mathcal{B}_{t}$ is independent of the received vectors 
$\{\mathcal{Y}_{t'}\}$ in different symbol periods $t'\neq t$. 
By using the chain rule for mutual information repeatedly, 
(\ref{spectral_efficiency}) yields 
\begin{IEEEeqnarray}{rl} 
I(\mathcal{B};\mathcal{Y}_{\mathcal{C}_{\tau+1}}) 
=& \sum_{t=\tau+1}^{T_{\mathrm{c}}}
I(\mathcal{B}_{t};\mathcal{Y}_{\mathcal{C}_{\tau+1}} | \mathcal{B}_{(\tau,t)}) 
\nonumber \\ 
=& \sum_{t=\tau+1}^{T_{\mathrm{c}}}\left[
 I(\mathcal{B}_{t};\mathcal{Y}_{\mathcal{C}_{\tau+1}\backslash\{t\}} | 
 \mathcal{B}_{(\tau,t)})
 + I(\mathcal{B}_{t};\mathcal{Y}_{t} | \mathcal{B}_{(\tau,t)}, 
 \mathcal{Y}_{\mathcal{C}_{\tau+1}\backslash\{t\}}) 
\right] \nonumber \\ 
=& \sum_{t=\tau+1}^{T_{\mathrm{c}}}I(\mathcal{B}_{t};\mathcal{Y}_{t} | 
\mathcal{B}_{(\tau,t)}, \mathcal{Y}_{\mathcal{C}_{\tau+1}\backslash\{t\}}), 
\label{mutual_inf_tmp} 
\end{IEEEeqnarray}
where the last equality holds since $\mathcal{B}_{t}$ is independent of 
$\mathcal{Y}_{\mathcal{C}_{\tau+1}\backslash\{t\}}$. If there were 
dependencies between the data symbols in different symbol periods, 
the equality would not hold. 
Applying the chain rule for mutual information to (\ref{mutual_inf_tmp}) gives  
\begin{equation} \label{mutual_inf_end} 
C_{\mathrm{opt}} = \frac{1}{LT_{\mathrm{c}}}
\sum_{t=\tau+1}^{T_{\mathrm{c}}}\sum_{k=1}^{K}C_{t,k}^{\mathrm{opt}}, 
\end{equation}
with 
\begin{equation} \label{each_mutual_information} 
C_{t,k}^{\mathrm{opt}} 
= I(\mathcal{B}_{t,k};\mathcal{Y}_{t} | \mathcal{B}_{t,[1,k)}, 
\mathcal{B}_{(\tau,t)}, \mathcal{Y}_{\mathcal{C}_{\tau+1}\backslash\{t\}}), 
\end{equation}
where $\mathcal{B}_{t,[1,k)}=\{\mathcal{B}_{t,k'}: k'=1,\ldots,k-1\}$ denotes 
the data symbols decoded in the preceding substages. 

We focus on each mutual information~(\ref{each_mutual_information}).  
In estimating $\mathcal{B}_{t,k}$, the data symbols $\mathcal{B}_{(\tau,t)}$ 
decoded in the preceding stages are available for channel estimation, while 
the data symbols $\mathcal{B}_{t,[1,k)}$ decoded in the preceding substages 
are used in MUD. The optimal receiver achieving the mutual 
information~(\ref{each_mutual_information}) consists of the optimal channel 
estimator, the optimal detector, and per-user decoders. 
The optimal channel estimator uses the information 
$\mathcal{I}_{\mathcal{T}_{t-1}}=\{\mathcal{I}_{t'}:t'\in\mathcal{T}_{t-1}\}$  
in the preceding stages and the information 
$\overline{\mathcal{I}}_{\mathcal{C}_{t+1}}=\{\overline{\mathcal{I}}_{t'}:t'\in
\mathcal{C}_{t+1}\}$ in the following stages to construct the posterior 
pdf $p(\mathcal{H}|\mathcal{I}_{\mathcal{T}_{t-1}}, 
\overline{\mathcal{I}}_{\mathcal{C}_{t+1}})$ of all channel 
vectors $\mathcal{H}=\{\mathcal{H}_{k}:\hbox{for all $k$}\}$, 
which is sent to the optimal detector. Note that 
$p(\mathcal{H}|\mathcal{I}_{\mathcal{T}_{t-1}}, 
\overline{\mathcal{I}}_{\mathcal{C}_{t+1}})$ 
is non-Gaussian, since $\overline{\mathcal{I}}_{\mathcal{C}_{t+1}}=
\{\mathcal{Y}_{t'},\mathcal{S}_{t'}:t'\in\mathcal{C}_{t+1}\}$ is an 
incomplete data set, i.e, it does not contain 
the data symbols $\{\mathcal{B}_{t'}\}$. Consequently, 
the optimal channel estimator is nonlinear. 
In the optimal detector, the posterior pdf 
$p(\mathcal{B}_{t,k}|\mathcal{Y}_{t}, \mathcal{S}_{t}, 
\mathcal{B}_{t,[1,k)}, \mathcal{I}_{\mathcal{T}_{t-1}}, 
\overline{\mathcal{I}}_{\mathcal{C}_{t+1}})$ is constructed and fed to the 
corresponding decoder, by utilizing the information about the 
received vector $\mathcal{Y}_{t}$, the spreading sequences 
$\mathcal{S}_{t}$, the data symbols $\mathcal{B}_{t,[1,k)}$ in the preceding 
substages, and the posterior pdf 
$p(\mathcal{H}|\mathcal{I}_{\mathcal{T}_{t-1}}, 
\overline{\mathcal{I}}_{\mathcal{C}_{t+1}})$ provided by the channel 
estimator.   

It is difficult to obtain an analytical expression for the constrained 
capacity~(\ref{mutual_inf_end}), since the optimal channel estimator is 
nonlinear. In order to obtain a lower bound based on LMMSE channel 
estimation, we consider a lower bound for (\ref{each_mutual_information}),  
\begin{equation} \label{lower_bound} 
C_{t,k}^{\mathrm{opt}} 
\geq I(\mathcal{B}_{t,k};\mathcal{Y}_{t} | \mathcal{B}_{t,[1,k)}, 
\mathcal{B}_{(\tau,t)}, \mathcal{Y}_{(\tau,t)}), 
\end{equation}
with $\mathcal{Y}_{(\tau,t)}=\{\mathcal{Y}_{t'}:t'=\tau+1,\ldots,
t-1\}$ denoting the received vectors in the preceding stages. 
In the derivation of (\ref{lower_bound}), we have used the assumption of 
independent data symbols and the fact that conditioning does not increase 
(differential) entropy. 
Recall that conditioning with respect to  
$\mathcal{Y}_{\mathcal{T}_{\tau}}$, $\mathcal{S}$, and $\mathcal{X}$ in the 
mutual information is omitted. 
Substituting (\ref{lower_bound}) to (\ref{mutual_inf_end}) 
yields the lower bound $C_{\mathrm{opt}}>C$, with 
\begin{equation} \label{target_spectral_efficiency}
C = \frac{1}{LT_{\mathrm{c}}}\sum_{t=\tau+1}^{T_{\mathrm{c}}}
\sum_{k=1}^{K}I(\mathcal{B}_{t,k};\mathcal{Y}_{t} | \mathcal{B}_{t,[1,k)}, 
\mathcal{I}_{\mathcal{T}_{t-1}}, S_{t}), 
\end{equation} 
where we have re-written the conditioning random variables as 
$\{\mathcal{B}_{t,[1,k)},\ \mathcal{I}_{\mathcal{T}_{t-1}},\ S_{t}\}$, 
using the fact that the lower bound (\ref{lower_bound}) is independent 
of the spreading sequences $\mathcal{S}_{\mathcal{C}_{t+1}}
=\{\mathcal{S}_{t'}:t'\in\mathcal{C}_{t+1}\}$ in the following 
stages anymore.  

The receiver achieving the lower bound~(\ref{target_spectral_efficiency}) 
is obtained by 
replacing the optimal channel estimator by a channel estimator that sends 
the posterior pdf $p(\mathcal{H}|\mathcal{I}_{\mathcal{T}_{t-1}})$, instead of 
$p(\mathcal{H}|\mathcal{I}_{\mathcal{T}_{t-1}}, 
\overline{\mathcal{I}}_{\mathcal{C}_{t+1}})$. 
It is straightforward to find that this channel estimator is equivalent 
to the LMMSE channel estimator, as shown in the following remark. Thus, 
we refer to the lower bound~(\ref{target_spectral_efficiency}) as 
the spectral efficiency of the joint CE-MUDD based on LMMSE channel estimation. 

\begin{remark} \label{remark1} 
We shall show that the channel estimator corresponding to the lower 
bound~(\ref{lower_bound}) is the LMMSE channel estimator. 
We first confirm that the channel estimator is linear. 
The information $\mathcal{I}_{\mathcal{T}_{t-1}}$ in the training phase is a 
complete data set for estimating the channel vectors $\mathcal{H}$, i.e., 
$\mathcal{I}_{\mathcal{T}_{t-1}}$ contains the received vectors, the spreading 
sequences, and the input symbols in each symbol period. 
The Gaussian assumption of the channel vectors 
$\boldsymbol{h}_{k}\sim\mathcal{CN}(\boldsymbol{0},\boldsymbol{I}_{N})$ implies 
that the posterior pdf $p(\mathcal{H}|\mathcal{I}_{\mathcal{T}_{t-1}})$ 
is a proper complex Gaussian pdf whose mean is given by a linear transform 
of the received vectors. Thus, the channel estimator based on the posterior 
mean estimator is linear. 

We next show that the channel estimator is equal to the LMMSE channel 
estimator. See Appendix~\ref{derivation_LMMSE} for derivations of the LMMSE 
estimator. Following \cite{Medard00,Hassibi03}, we replace the term 
$\boldsymbol{h}_{k,m}s_{l,t',k,m}b_{t',k,m}$ in (\ref{MIMO_DS_CDMA}) 
for the following stages $t'=t+1,\ldots,T_{\mathrm{c}}$ by a proper 
complex Gaussian random vector whose mean and covariance are 
given by $\boldsymbol{h}_{k,m}s_{l,t',k,m}\mathbb{E}[b_{t',k,m}]$ and 
$|s_{l,t',k,m}|^{2}\mathbb{E}[\boldsymbol{h}_{k,m}b_{t',k,m}
(\boldsymbol{h}_{k,m}b_{t',k,m})^{H}]$, 
respectively. However, the mean is equal to zero, because of 
$\mathbb{E}[b_{t',k,m}]=0$. As a result, the received vectors in the 
following stages is independent of the channel vectors anymore. In other 
words, the LMMSE channel estimator does not utilize the 
information $\mathcal{Y}_{\mathcal{C}_{t+1}}
=\{\mathcal{Y}_{t'}:t'\in\mathcal{C}_{t+1}\}$ 
in the following stages, since the LMMSE channel estimator postulates that 
the received vector $\boldsymbol{y}_{l,t'}$ in (\ref{MIMO_DS_CDMA}) is 
independent of the channel vectors for $t'\in\mathcal{C}_{t+1}$. 
Therefore, the lower bound~(\ref{lower_bound}) corresponds to the spectral 
efficiency of the receiver based on the LMMSE channel estimation. 
\end{remark}

\subsection{Joint CE-MUDD Based on Suboptimal LMMSE Channel Estimation} 
It is still hard to evaluate the spectral 
efficiency~(\ref{target_spectral_efficiency}) of the joint CE-MUDD based on 
LMMSE channel estimation, since 
the posterior pdf $p(\mathcal{H}|\mathcal{I}_{\mathcal{T}_{t-1}})$ sent by 
the LMMSE channel estimator is not factorized into the product of the 
marginal posterior pdfs 
$\prod_{k=1}^{K}p(\mathcal{H}_{k}|\mathcal{I}_{\mathcal{T}_{t-1}})$. 
Instead, we consider a suboptimal receiver in which the LMMSE channel 
estimator is replaced by a suboptimal LMMSE channel estimator that sends  
the product 
$\prod_{k=1}^{K}p(\mathcal{H}_{k}|\mathcal{I}_{\mathcal{T}_{t-1}})$. 
The spectral efficiency for the suboptimal receiver provides a 
lower bound for the spectral efficiency~(\ref{target_spectral_efficiency}) 
based on LMMSE channel estimation. Since the LMMSE detector uses no 
covariances of the channel estimation errors between different 
users~\cite{Evans00}, as noted in Section~\ref{section_LMMSE}, 
no performance loss due to this replacement occurs for the LMMSE receiver. 
On the other hand, the optimal detector utilizes all information about the 
channel vectors, i.e., 
the joint posterior pdf $p(\mathcal{H}|\mathcal{I}_{\mathcal{T}_{t-1}})$.    
It is unclear whether the performance loss caused by using the suboptimal 
LMMSE channel estimator is negligible in the large-system 
limit for the optimal detector. 

\begin{figure}[t]
\begin{center}
\includegraphics[width=0.5\hsize]{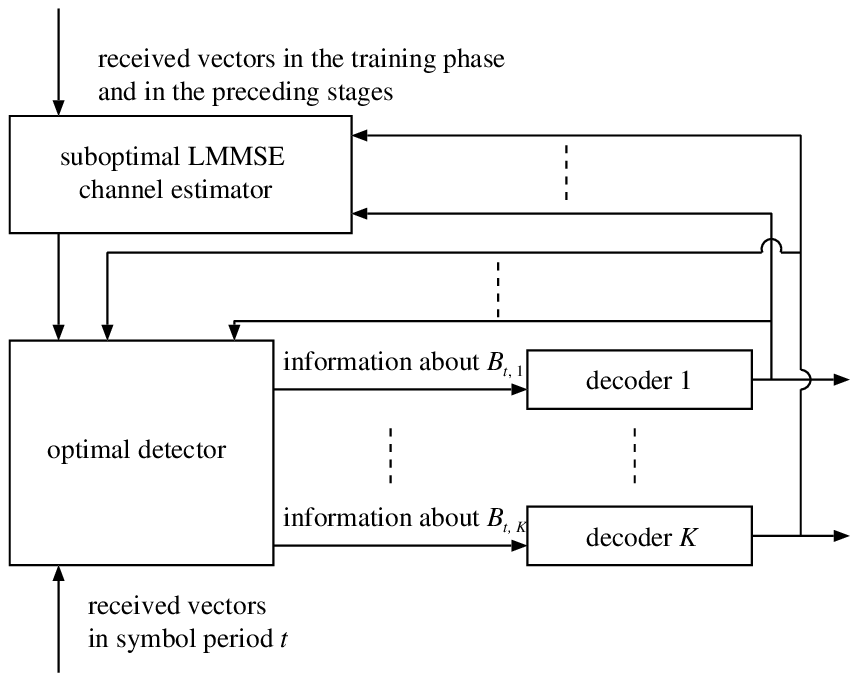}
\end{center}
\caption{
Joint CE-MUDD based on suboptimal LMMSE channel estimation. 
}
\label{fig1}
\end{figure}

Let us focus on substage~$k$ within stage~$t$.  
The suboptimal receiver consists of a suboptimal LMMSE channel estimator, 
the optimal detector, and the per-user decoders (See Fig.~\ref{fig1}). 
The channel estimator uses the pilot symbols 
and the data symbols decoded in the preceding stages to estimate the 
channel vectors. More precisely, the channel estimator constructs the 
posterior pdf $p(\mathcal{H}_{k}|\mathcal{I}_{\mathcal{T}_{t-1}})$ of 
the channel vectors $\mathcal{H}_{k}$ for each user 
by utilizing the known information 
$\mathcal{I}_{\mathcal{T}_{t-1}}=\{\mathcal{Y}_{t'},\ \mathcal{S}_{t'},\  
\mathcal{U}_{t'}:t'\in\mathcal{T}_{t-1}\}$, 
\begin{equation} \label{posterior_pdf_Hk} 
p(\mathcal{H}_{k} |\mathcal{I}_{\mathcal{T}_{t-1}}) = \frac{
 \int \prod_{t'=1}^{t-1}p(\mathcal{Y}_{t'} | \mathcal{H}, 
 \mathcal{S}_{t'}, \mathcal{U}_{t'})p(\mathcal{H})
 d\mathcal{H}_{\backslash\{k\}}
}
{
 \int \prod_{t'=1}^{t-1}p(\mathcal{Y}_{t'} | \mathcal{H}, 
 \mathcal{S}_{t'}, \mathcal{U}_{t'})p(\mathcal{H})d\mathcal{H}
},
\end{equation}
%$\mathcal{S}_{\mathrm{tr}}=\{\mathcal{S}_{t}:t\in\mathcal{T}_{\tau}\}$ 
%the spreading sequences in the training phase 
with $\mathcal{H}_{\backslash\{k\}}=\{\mathcal{H}_{k'}:\hbox{for all 
$k'\neq k$}\}$ denoting the channel vectors except for $\mathcal{H}_{k}$. 
In (\ref{posterior_pdf_Hk}), the pdf 
$p(\mathcal{Y}_{t'} | \mathcal{H}, \mathcal{S}_{t'}, \mathcal{U}_{t'})$ 
represents the MIMO DS-CDMA channel~(\ref{MIMO_DS_CDMA}) in symbol 
period~$t'$. 
Then, the marginal posterior pdfs $\{p(\mathcal{H}_{k}| 
\mathcal{I}_{\mathcal{T}_{t-1}}):\hbox{for all $k$}\}$ are sent towards the 
optimal detector.  
Sending $p(\mathcal{H}_{k}|\mathcal{I}_{\mathcal{T}_{t-1}})$ is equivalent 
to feeding the LMMSE estimates $\hat{\boldsymbol{h}}_{k,m}^{\mathcal{T}_{t-1}}=
\int\boldsymbol{h}_{k,m}
p(\mathcal{H}_{k}|\mathcal{I}_{\mathcal{T}_{t-1}})d\mathcal{H}_{k}$ and 
the covariances of the estimation errors 
$\Delta\boldsymbol{h}_{k,m}^{\mathcal{T}_{t-1}}=\boldsymbol{h}_{k,m}
-\hat{\boldsymbol{h}}_{k,m}^{\mathcal{T}_{t-1}}$ for all $m$, since the 
posterior pdf $p(\mathcal{H}_{k}|\mathcal{I}_{\mathcal{T}_{t-1}})$ is 
CSCG. 
%Note that sending the joint posterior pdf 
%$p(\mathcal{H}|\mathcal{I}_{\mathcal{T}_{t-1}})$ 
%is optimal in terms of performance. We have replaced the LMMSE channel 
%estimator by the suboptimal one~(\ref{posterior_pdf_Hk}) to apply the replica 
%method. 

The optimal detector uses the information about the received 
vectors $\mathcal{Y}_{t}$, the data symbols $\mathcal{B}_{t,[1,k)}$ decoded 
in the preceding substages, the spreading sequences $\mathcal{S}_{t}$, and 
the posterior pdfs $\{p(\mathcal{H}_{k}|\mathcal{I}_{\mathcal{T}_{t-1}})\}$ 
provided by the channel estimator to detect the data symbols 
$\mathcal{B}_{t,k}$. The term ``optimal detector'' indicates that  
the detector is optimal among all detectors that regard the product 
$\prod_{k=1}^{K}p(\mathcal{H}_{k}|\mathcal{I}_{\mathcal{T}_{t-1}})$ of the 
marginal posterior pdfs as the true joint posterior pdf. 
%In other words, the performance of our optimum separate receiver provides a 
%lower bound of the performance for the {\em true} optimum separated receiver 
%utilizing the joint posterior pdf $p(\mathcal{H}|\mathcal{I}_{\mathrm{tr}})$. 
The optimal detector constructs a posterior pdf 
$p(\tilde{\mathcal{B}}_{t,k} | \mathcal{Y}_{t}, \mathcal{B}_{t,[1,k)}, 
\mathcal{S}_{t}, \mathcal{I}_{\mathcal{T}_{t-1}})$ and subsequently forwards 
it to the corresponding per-user decoder. In the posterior pdf, the data 
symbols $\tilde{\mathcal{B}}_{t,k}=\{\tilde{b}_{t,k,m}\in\mathbb{C}:
\hbox{for all $m$}\}\sim \prod_{m=1}^{M}p(b_{t,k,m})$ denotes the data 
symbols in the MIMO DS-CDMA channel postulated by the optimal detector 
\begin{equation} \label{MIMO_DS_CDMA_MMSE}
\tilde{\boldsymbol{y}}_{l,t} = \frac{1}{\sqrt{L}}\sum_{k=1}^{K}
\sum_{m=1}^{M}\tilde{\boldsymbol{h}}_{k,m}s_{l,t,k,m}\tilde{b}_{t,k,m} 
+ \tilde{\boldsymbol{n}}_{l,t}, 
\end{equation} 
with $\tilde{\boldsymbol{n}}_{l,t}\sim\mathcal{CN}(\boldsymbol{0},
N_{0}\boldsymbol{I}_{N})$. In (\ref{MIMO_DS_CDMA_MMSE}), 
$\tilde{\mathcal{H}}_{k}=\{\tilde{\boldsymbol{h}}_{k,m}\in\mathbb{C}^{N}: 
\hbox{for all $m$}\}$ denotes random vectors representing the information 
about $\mathcal{H}_{k}$ provided by the channel estimator. The joint 
posterior pdf of $\tilde{\mathcal{H}}=\{\tilde{\mathcal{H}}_{k}:
\hbox{for all $k$}\}$ satisfies  
\begin{equation} \label{postulated_posterior_H}
p(\tilde{\mathcal{H}} |\mathcal{I}_{\mathcal{T}_{t-1}}) 
= \prod_{k=1}^{K}p(\mathcal{H}_{k}=\tilde{\mathcal{H}}_{k}|
\mathcal{I}_{\mathcal{T}_{t-1}}). 
\end{equation}
Note that $p(\tilde{\mathcal{H}} |\mathcal{I}_{\mathcal{T}_{t-1}})\neq 
p(\mathcal{H} |\mathcal{I}_{\mathcal{T}_{t-1}})$ since 
$p(\mathcal{H} |\mathcal{I}_{\mathcal{T}_{t-1}})$ is not decomposed into the 
product of the marginal pdfs.  The posterior pdf 
$p(\tilde{\mathcal{B}}_{t,k} | \mathcal{Y}_{t}, \mathcal{B}_{t,[1,k)}, 
\mathcal{S}_{t}, \mathcal{I}_{\mathcal{T}_{t-1}})$ is an abbreviation of 
$p(\tilde{\mathcal{B}}_{t,k} | \tilde{\mathcal{Y}}_{t}=\mathcal{Y}_{t}, 
\tilde{\mathcal{B}}_{t,[1,k)}=\mathcal{B}_{t,[1,k)}, \mathcal{S}_{t}, 
\mathcal{I}_{\mathcal{T}_{t-1}})$, with 
$\tilde{\mathcal{Y}}_{t}=\{\tilde{\boldsymbol{y}}_{l,t}\in\mathbb{C}^{N}:
\hbox{for all $l$}\}$ and $\tilde{\mathcal{B}}_{t,[1,k)}
=\{\tilde{\mathcal{B}}_{t,k'}:k'=1,\ldots,k-1\}$ denoting the received vectors 
in (\ref{MIMO_DS_CDMA_MMSE}) and the postulated data symbols in the 
preceding substages, respectively, given by  
\begin{equation} \label{posterior_MMSE_k} 
p(\tilde{\mathcal{B}}_{t,k} | \tilde{\mathcal{Y}}_{t}, 
\tilde{\mathcal{B}}_{t,[1,k)}, \mathcal{S}_{t}, 
\mathcal{I}_{\mathcal{T}_{t-1}}) = 
\frac{
 \int p(\tilde{\mathcal{Y}}_{t} | \tilde{\mathcal{B}}_{t}, 
 \mathcal{S}_{t}, \mathcal{I}_{\mathcal{T}_{t-1}})p(\tilde{\mathcal{B}}_{t})
 d\tilde{\mathcal{B}}_{t,(k,K]} 
}
{
 \int p(\tilde{\mathcal{Y}}_{t} | \tilde{\mathcal{B}}_{t}, 
 \mathcal{S}_{t}, \mathcal{I}_{\mathcal{T}_{t-1}})p(\tilde{\mathcal{B}}_{t})
 d\tilde{\mathcal{B}}_{t,[k,K]}
}, 
\end{equation}
%with $\tilde{\mathcal{B}}_{t}=\{\tilde{\boldsymbol{b}}_{t,k}:\hbox{for all 
%$k$}\}$ and $\tilde{\mathcal{B}}_{t,\backslash\{k\}}=
%\{\tilde{\boldsymbol{b}}_{t,k'}:\hbox{for all $k'\neq k$}\}$ denoting 
%all postulated data symbols and the postulated data symbols except for 
%$\tilde{\boldsymbol{b}}_{t,k}$, respectively. 
%In (\ref{posterior_MMSE_k}), $p(\tilde{\mathcal{Y}}_{t} | 
%\tilde{\mathcal{B}}_{t}, \mathcal{S}_{t}, \mathcal{I}_{\mathrm{tr}})$ is 
%defined as 
with $\tilde{\mathcal{B}}_{t,(k,K]}=\{\tilde{\mathcal{B}}_{t,k'}: 
k'=k+1,\ldots,K\}$ and $\tilde{\mathcal{B}}_{t,[k,K]}
=\{\tilde{\mathcal{B}}_{t,k'}: k'=k,\ldots,K\}$. 
In (\ref{posterior_MMSE_k}), $p(\tilde{\mathcal{Y}}_{t} | 
\tilde{\mathcal{B}}_{t}, \mathcal{S}_{t}, \mathcal{I}_{\mathcal{T}_{t-1}})$ 
is given by 
\begin{equation} \label{postulated_channel} 
p(\tilde{\mathcal{Y}}_{t} | \tilde{\mathcal{B}}_{t}, 
\mathcal{S}_{t}, \mathcal{I}_{\mathcal{T}_{t-1}}) = 
\int p(\tilde{\mathcal{Y}}_{t} | \tilde{\mathcal{H}}, \mathcal{S}_{t}, 
\tilde{\mathcal{B}}_{t})p(\tilde{\mathcal{H}} | 
\mathcal{I}_{\mathcal{T}_{t-1}})d\tilde{\mathcal{H}}, 
\end{equation}
where $p(\tilde{\mathcal{Y}}_{t} | \tilde{\mathcal{H}}, \mathcal{S}_{t}, 
\tilde{\mathcal{B}}_{t})$ represents the MIMO DS-CDMA 
channel~(\ref{MIMO_DS_CDMA_MMSE}) postulated by the optimal detector in 
symbol period~$t$. The marginal posterior pdf~(\ref{posterior_MMSE_k}) would 
reduce to the true one $p(\mathcal{B}_{t,k} | \mathcal{Y}_{t}, 
\mathcal{B}_{t,[1,k)}, \mathcal{S}_{t}, \mathcal{I}_{\mathcal{T}_{t-1}})$ if 
the channel estimator sent the joint posterior 
pdf~$p(\mathcal{H}|\mathcal{I}_{\mathcal{T}_{t-1}})$. 

The spectral efficiency~$C_{\mathrm{joint}}$ of the joint CE-MUDD based on the 
suboptimal LMMSE channel estimation is given by, 
\begin{equation} \label{target_lower_bound} 
C_{\mathrm{joint}} = \frac{1}{LT_{\mathrm{c}}} 
\sum_{t=\tau+1}^{T_{\mathrm{c}}}\sum_{k=1}^{K}C_{t,k}, 
\end{equation}
with 
\begin{equation} \label{target_each_mutual_inf} 
C_{t,k} = 
I(\mathcal{B}_{t,k};\tilde{\mathcal{B}}_{t,k} | \mathcal{B}_{t,[1,k)}, 
\mathcal{I}_{\mathcal{T}_{t-1}}, \mathcal{S}_{t}).  
\end{equation}
%with 
%\begin{equation} \label{mutual_information_opt} 
%I(\mathcal{B}_{t};\tilde{\mathcal{B}}_{t}| \mathcal{S}_{t}, 
%\mathcal{I}_{\mathcal{T}_{t-1}}) = 
%\mathbb{E}\left[
% \log\frac{
%  p(\tilde{\mathcal{B}}_{t} | \mathcal{B}_{t}, \mathcal{S}_{t}, 
%  \mathcal{I}_{\mathcal{T}_{t-1}})
% }
% {
%  \int p(\tilde{\mathcal{B}}_{t} | \mathcal{B}_{t}, \mathcal{S}_{t}, 
%  \mathcal{I}_{\mathcal{T}_{t-1}})p(\mathcal{B}_{t})d\mathcal{B}_{t} 
% }
%\right]. 
%\end{equation}
The conditional mutual information~(\ref{target_each_mutual_inf}) 
is characterized by the equivalent channel between user~$k$ and 
the corresponding decoder, given by  
\begin{equation} \label{equivalent_channel} 
p(\tilde{\mathcal{B}}_{t,k} | \mathcal{B}_{t,k}, \mathcal{B}_{t,[1,k)}, 
\mathcal{I}_{\mathcal{T}_{t-1}}, \mathcal{S}_{t}) = 
\int p(\tilde{\mathcal{B}}_{t,k} | \mathcal{Y}_{t}, \mathcal{B}_{t,[1,k)}, 
\mathcal{S}_{t}, \mathcal{I}_{\mathcal{T}_{t-1}})p(\mathcal{Y}_{t}| 
\mathcal{B}_{t,[1,k]}, 
\mathcal{S}_{t}, \mathcal{I}_{\mathcal{T}_{t-1}})d\mathcal{Y}_{t}, 
\end{equation}
where 
\begin{equation} \label{channel} 
p(\mathcal{Y}_{t}| \mathcal{B}_{t,[1,k]}, \mathcal{S}_{t}, 
\mathcal{I}_{\mathcal{T}_{t-1}}) = 
\int p(\mathcal{Y}_{t} | \mathcal{H},\mathcal{S}_{t},\mathcal{B}_{t})
p(\mathcal{B}_{t,(k,K]})p(\mathcal{H}|\mathcal{I}_{\mathcal{T}_{t-1}})
d\mathcal{H}d\mathcal{B}_{t,(k,K]}, 
\end{equation} 
with $\mathcal{B}_{t,[1,k]}=\{\mathcal{B}_{t,k'}: k'=1,\ldots,k\}$ 
and $\mathcal{B}_{t,(k,K]}=\{\mathcal{B}_{t,k'}: k'=k+1,\ldots,K\}$. 
%with $\mathcal{B}_{t}=\{\mathcal{B}_{t,k}:\hbox{for all $k$}\}$ and 
%$\mathcal{B}_{t,\backslash\{k\}}=\{\boldsymbol{b}_{t,k'}: 
%\hbox{for all $k'\neq k$}\}$ denoting all data symbols and 
%the data symbols except for $\mathcal{B}_{t,k}$. 
In (\ref{channel}), 
$p(\mathcal{Y}_{t} | \mathcal{H},\mathcal{S}_{t},\mathcal{B}_{t})$ 
represents the MIMO DS-CDMA channel~(\ref{MIMO_DS_CDMA}) in symbol 
period~$t$. 
Note that $\tilde{\mathcal{B}}_{t,k}$ in (\ref{target_each_mutual_inf}) plays 
the role of random variables representing the information about 
$\mathcal{B}_{t,k}$ provided to the per-user decoder, although we have 
introduced $\tilde{\mathcal{B}}_{t,k}$ as random variables representing the 
data symbols postulated by the optimal detector in (\ref{MIMO_DS_CDMA_MMSE}). 

The spectral efficiency~(\ref{target_lower_bound}) of the joint CE-MUDD based 
on the suboptimal LMMSE channel estimation is a lower bound for 
the spectral efficiency~(\ref{target_spectral_efficiency}) based on the 
LMMSE channel estimation. We hereafter focus on the lower 
bound~(\ref{target_lower_bound}). Thus, the joint CE-MUDD based on the 
suboptimal LMMSE channel estimation is simply referred to as the joint 
CE-MUDD. The spectral efficiency~(\ref{target_lower_bound}) of the joint 
CE-MUDD should not be confused with the spectral 
efficiency~(\ref{target_spectral_efficiency}) based on the LMMSE channel 
estimation or with the constraint capacity~(\ref{mutual_inf_tmp}).   

%Iterative CE-MUDD consists of a channel estimator, a detector, and per-user 
%decoders (See Fig.~\ref{fig1}). In terms of complexity, the channel estimator 
%and the detector for iterative CE-MUDD have to calculate their filter 
%coefficients in every iteration. Therefore, the computational complexity of 
%iterative CE-MUDD is 
%$O(n(T_{\mathrm{ch}}+T_{\mathrm{det}}+KT_{\mathrm{dec}}))$ 
%for $n$ iterations, 
%in which $T_{\mathrm{ch}}$, $T_{\mathrm{det}}$, and $T_{\mathrm{dec}}$ denote 
%the computational costs per iteration of the channel estimator, the 
%detector, and each per-user decoder. If the LMMSE filter is used, 
%$T_{\mathrm{ch}}=O(K^{3})$ and $T_{\mathrm{det}}=O(K^{3})$. 
%For multistage implementation, 
%$T_{\mathrm{det}}=O(K^{2})$~\cite{Cottatellucci07}.

\section{Receivers Based on One-Shot Channel Estimation} 
\label{section_one_shot} 
\subsection{One-Shot Channel Estimation} 
For comparison with the joint CE-MUDD, we consider three receivers based on 
one-shot channel estimation, in which the decoded data symbols are not 
used to refine the channel estimates. A first receiver performs joint MUDD 
based on one-shot LMMSE channel estimation, called one-shot CE-MUDD. This 
receiver is obtained by eliminating the feedback from the per-user decoders 
to the suboptimal LMMSE channel estimator in Fig.~\ref{fig1}. 
A second receiver performs separated decoding based on one-shot LMMSE channel 
estimation, called the optimum separated receiver. The receiver is obtained 
by eliminating the feedback from the per-user decoders to the channel 
estimator and to the optimal detector. 
%The optimal detector requires infeasible computational complexity, probably, 
%exponential computational costs in $K$. 
The last receiver is an LMMSE receiver in which the detector in the optimum 
separated receiver is replaced by an LMMSE detector.  
The four receivers considered in this paper are listed in Table~\ref{table}. 

%The computational costs of the optimum separated receiver and 
%the LMMSE receiver are given by 
%$O(T_{\mathrm{ch}}+T_{\mathrm{det}}+KT_{\mathrm{dec}})$ and 
%$O(T_{\mathrm{ch}}+T_{\mathrm{det}}^{(\mathrm{L})}+KT_{\mathrm{dec}})$, 
%respectively, in which $T_{\mathrm{det}}$ and 
%$T_{\mathrm{det}}^{(\mathrm{L})}$ denote the computational costs per iteration 
%of the optimal and LMMSE detectors.  
%In order to define receivers based on one-shot channel estimation, we consider 
%the optimal receiver with perfect CSI. A sufficient statistic of the data 
%symbols $\mathcal{B}_{t}$ depends only on the received vectors 
%$\mathcal{Y}_{t}$ in the same symbol period, and does not depend on the 
%received vectors in the other symbol periods. 
%This observation implies that it is reasonable to consider receivers in 
%which the channel estimator estimates the channel vectors only from 
%the information in the training phase.  

%The performance gap between the joint CE-MUDD and the one-shot CE-MUDD  
%corresponds to the gains obtained by using the decoded data symbols to refine 
%the channel estimates. 
%The gap between the one-shot CE-MUDD and the optimum separated receiver 
%is related to the gains obtained by using the decoded data symbols to 
%mitigate MAI. The performance gap between the 
%optimum separated receiver and the LMMSE receiver corresponds to the gains 
%obtained by performing the optimal MUD, instead of the LMMSE MUD. 

\begin{figure}[t]
\begin{center}
\includegraphics[width=0.5\hsize]{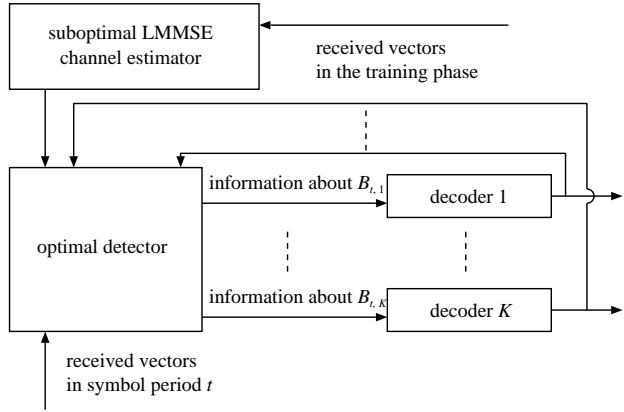}
\end{center} 
\caption{
One-shot CE-MUDD based on suboptimal LMMSE channel estimation. 
}
\label{fig2} 
\end{figure}

\subsection{One-Shot CE-MUDD} 
We define the one-shot CE-MUDD based on the suboptimal LMMSE channel 
estimation (See Fig.~\ref{fig2}).  
%while the definitions of the remaining two receivers are deferred to 
%Section~\ref{section_LMMSE}. 
In stage~$t$ of the joint CE-MUDD, the channel estimator has used the 
information $\mathcal{I}_{\mathcal{T}_{t-1}}=
\{\mathcal{Y}_{t'},\ \mathcal{S}_{t'},\  
\mathcal{U}_{t'}: t'\in\mathcal{T}_{t-1}\}$ about the received vectors 
$\{\mathcal{Y}_{t'}\}$, the spreading sequences $\{\mathcal{S}_{t'}\}$, 
and the transmitted symbols $\mathcal{U}_{t'}$ in symbol periods 
$t'\in\mathcal{T}_{t-1}$. On the other hand, the LMMSE channel estimation in 
the one-shot CE-MUDD cannot utilize the data symbols 
$\{\mathcal{B}_{t'}: t'=\tau+1,\ldots,t-1\}$ decoded in the preceding stages. 
This restriction is equivalent to assuming that the information 
$\mathcal{I}_{(\tau,t)}=\{\mathcal{Y}_{t'},\ \mathcal{S}_{t'},\  
\mathcal{U}_{t'}: t'=\tau+1,\ldots,t-1\}$ in the preceding stages is not used 
for channel estimation, as discussed in Remark~\ref{remark1}. In other words, 
the information $\mathcal{I}_{\mathcal{T}_{\tau}}=\{\mathcal{Y}_{t'},\ 
\mathcal{S}_{t'},\ \mathcal{X}_{t'}:
\hbox{for all $t'\in\mathcal{T}_{\tau}$}\}$ in 
the training phase is utilized for channel estimation. Thus, the suboptimal 
LMMSE channel estimator in the one-shot CE-MUDD provides the marginal 
posterior pdfs $\{p(\mathcal{H}_{k}|\mathcal{I}_{\mathcal{T}_{\tau}})\}$ to the 
optimal detector. Note that sending the joint posterior pdf 
$p(\mathcal{H}|\mathcal{I}_{\mathcal{T}_{\tau}})$ is of course optimal. 
Strictly speaking, the one-shot CE-MUDD should be referred to as the one-shot 
CE-MUDD based on the suboptimal LMMSE channel estimation. However, we simply 
call it the one-shot CE-MUDD, since the true LMMSE channel estimator is not 
analyzed in this paper.   

The marginal posterior pdf 
$p(\mathcal{H}_{k}|\mathcal{I}_{\mathcal{T}_{\tau}})$ 
is equal to the one constructed in the first stage of the joint CE-MUDD. 
Thus, the spectral efficiency of the one-shot CE-MUDD is given by  
\begin{equation} \label{spectral_efficiency_one} 
C_{\mathrm{one}} = \left(
 1 - \frac{\tau}{T_{\mathrm{c}}} 
\right)L^{-1}\sum_{k=1}^{K}
C_{\tau+1,k}, 
\end{equation} 
where the mutual information $C_{\tau+1,k}$ is defined as   
(\ref{target_each_mutual_inf}).

\begin{figure}[t]
\begin{center}
\includegraphics[width=0.5\hsize]{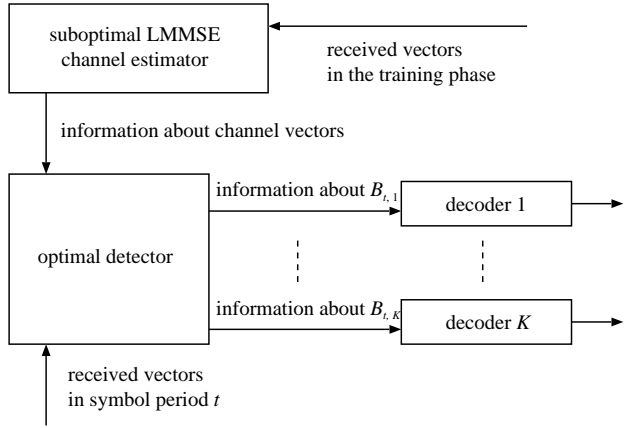} 
\end{center}
\caption{
Optimum separated receiver based on suboptimal LMMSE channel estimation. 
}
\label{receiver_fig} 
\end{figure}

\subsection{Optimum Separated Receiver} 
\label{section_Non-Iterative_Optimal_Receiver}
We define the optimum separate receiver based on the suboptimal LMMSE 
channel estimation (See Fig.~\ref{receiver_fig}). 
In substage~$k$ of the joint CE-MUDD, the optimal detector has used 
the data symbols $\mathcal{B}_{t,[1,k)}$ in the preceding substages to 
mitigate MAI. In the optimum separated receiver, on the other hand, the 
information $\mathcal{B}_{t,[1,k)}$ is not utilized for MUD. The 
posterior pdf $p(\tilde{\mathcal{B}}_{t,k} | \mathcal{Y}_{t}, \mathcal{S}_{t}, 
\mathcal{I}_{\mathcal{T}_{\tau}})$ is constructed and sent to the corresponding 
per-user decoder. The posterior pdf is equal to the one constructed in 
the first substage of the joint CE-MUDD. 
Thus, the spectral efficiency of the optimum separated 
receiver is given by   
\begin{equation} \label{spectral_efficiency_MMSE} 
C_{\mathrm{sep}} = \frac{K}{L}\left(
 1 - \frac{\tau}{T_{\mathrm{c}}}
\right)C_{\tau+1,1},   
\end{equation}
where the mutual information $C_{\tau+1,1}$ is defined as   
(\ref{target_each_mutual_inf}). 

\subsection{LMMSE Receiver} 
\label{section_LMMSE} 
The LMMSE receiver is obtained by replacing the optimal detector in the 
optimum separated receiver by the LMMSE detector. 
We first derive the LMMSE estimator of the data symbols $\mathcal{B}_{t,k}$, 
after which we define the information about $\mathcal{B}_{t,k}$ provided from 
the LMMSE detector to the corresponding decoder. 
The LMMSE estimator of $\mathcal{B}_{t,k}$ would be obtained by regarding 
$\mathcal{B}_{t,k}$ as CSCG random variables 
$\tilde{\mathcal{B}}_{t,k}^{(\mathrm{L})}=\{ 
\tilde{b}_{t,k,m}^{(\mathrm{L})}\in\mathbb{C}: \hbox{for all $m$}\}$ with 
$\mathbb{E}[\tilde{b}_{t,k,m}^{(\mathrm{L})}
(\tilde{b}_{t,k,m'}^{(\mathrm{L})})^{*}]= (P/M)\delta_{m,m'}$ 
if the receiver had perfect CSI \cite{Guo051}. 
However, the posterior mean of $\tilde{\mathcal{B}}_{t,k}^{(\mathrm{L})}$ 
is nonlinear in the received vectors $\mathcal{Y}_{t}$ 
since the channel model~(\ref{MIMO_DS_CDMA_MMSE}) includes 
multiplicative noise due to the influence of channel estimation errors. 
We approximate the channel model~(\ref{MIMO_DS_CDMA_MMSE}) by a 
channel model without multiplicative noise. 
We extract the term including channel estimation errors from the first term 
on the right-hand side of (\ref{MIMO_DS_CDMA_MMSE}), and subsequently 
approximate the extracted one by an AWGN term with the same covariance. 
Therefore, the MIMO DS-CDMA channel in symbol period $t$ postulated by the 
LMMSE detector is given as 
\begin{equation} \label{MIMO_DS_CDMA_LMMSE}
\tilde{\boldsymbol{y}}_{l,t}^{(\mathrm{L})} = 
\frac{1}{\sqrt{L}}\sum_{k=1}^{K}\sum_{m=1}^{M}s_{l,t,k,m}\left(
 \hat{\boldsymbol{h}}_{k,m}^{\mathcal{I}_{\tau}}
 \tilde{b}_{t,k,m}^{(\mathrm{L})}  
 + \tilde{\boldsymbol{w}}_{t,k,m}^{(\mathrm{L})} 
\right)
+ \tilde{\boldsymbol{n}}_{l,t}^{(\mathrm{L})}, 
\end{equation}
with $\tilde{\boldsymbol{n}}_{l,t}^{(\mathrm{L})}
\sim\mathcal{CN}(\boldsymbol{0}, N_{0}\boldsymbol{I}_{N})$. 
In (\ref{MIMO_DS_CDMA_LMMSE}), 
$\hat{\boldsymbol{h}}_{k,m}^{\mathcal{I}_{\tau}}$ denotes the 
LMMSE channel estimates $\hat{\boldsymbol{h}}_{k,m}^{\mathcal{I}_{\tau}}
=\int \boldsymbol{h}_{k,m}p(\mathcal{H}_{k}|\mathcal{I}_{\tau})
d\mathcal{H}_{k}$. The vectors 
$\{\tilde{\boldsymbol{w}}_{t,k,m}^{(\mathrm{L})}\in\mathbb{C}^{N}\}$ 
are independent CSCG random vectors with the covariance matrix 
$(P/M)\boldsymbol{\Sigma}_{k,m}$ for all $t$, $k$, and $m$, in which 
$\boldsymbol{\Sigma}_{k,m}$ denotes the covariance matrix of the 
channel estimation errors $\Delta\boldsymbol{h}_{k,m}^{\mathcal{I}_{\tau}}
=\boldsymbol{h}_{k,m}-\hat{\boldsymbol{h}}_{k,m}^{\mathcal{I}_{\tau}}$, 
i.e.,  $\boldsymbol{\Sigma}_{k,m}= \mathbb{E}[ 
\Delta\boldsymbol{h}_{k,m}^{\mathcal{I}_{\tau}}
(\Delta\boldsymbol{h}_{k,m}^{\mathcal{I}_{\tau}})^{H}| 
\mathcal{I}_{\mathcal{T}_{\tau}}]$.  
Note that 
$\mathbb{E}[\Delta \boldsymbol{h}_{k,m}^{\mathcal{I}_{\tau}}b_{t,k,m}
(\Delta \boldsymbol{h}_{k',m}^{\mathcal{I}_{\tau}}b_{t,k',m})^{H}| 
\mathcal{I}_{\tau}]=\boldsymbol{O}$ for $k\neq k'$, since the data symbols 
$\{b_{t,k,m}\}$ are independent unbiased random variables.  
Thus, the same detector would be obtained even if the 
true LMMSE channel estimator was used instead of the suboptimal one. 

Let $\tilde{\mathcal{Y}}_{t}^{(\mathrm{L})}=
\{\tilde{\boldsymbol{y}}_{l,t}^{(\mathrm{L})}\in\mathbb{C}^{N}:
\hbox{for all $l$}\}$ 
denote the received vectors postulated by the LMMSE detector 
in symbol period~$t$. Furthermore, we write the postulated data symbols 
in symbol period~$t$ and the postulated data symbols 
except for $\tilde{\mathcal{B}}_{t,k}$ as   
$\tilde{\mathcal{B}}_{t}^{(\mathrm{L})}=\{
\tilde{\mathcal{B}}_{t,k}^{(\mathrm{L})}:\hbox{for all $k$}\}$ and 
$\tilde{\mathcal{B}}_{t,\backslash\{k\}}^{(\mathrm{L})}=
\{\tilde{\mathcal{B}}_{t,k'}^{(\mathrm{L})}:
\hbox{for all $k'\neq k$}\}$, respectively. 
The linear estimator of $\mathcal{B}_{t,k}$ is given by the 
mean of $\tilde{\mathcal{B}}_{t,k}^{(\mathrm{L})}$ with respect to the 
posterior pdf 
\begin{equation} \label{posterior_pdf_LMMSE} 
p( \tilde{\mathcal{B}}_{t,k}^{(\mathrm{L})} | 
\tilde{\mathcal{Y}}_{t}^{(\mathrm{L})}=\mathcal{Y}_{t}, 
\mathcal{S}_{t}, \mathcal{I}_{\mathcal{T}_{\tau}})  = 
\frac{
 \int p(\tilde{\mathcal{Y}}_{t}^{(\mathrm{L})}=\mathcal{Y}_{t} | 
 \tilde{\mathcal{B}}_{t}^{(\mathrm{L})}, \mathcal{S}_{t}, 
 \mathcal{I}_{\mathcal{T}_{\tau}})
 p(\tilde{\mathcal{B}}_{t}^{(\mathrm{L})})
 d\tilde{\mathcal{B}}_{t,\backslash \{k\}}^{(\mathrm{L})} 
}
{
 \int p(\tilde{\mathcal{Y}}_{t}^{(\mathrm{L})}=\mathcal{Y}_{t} | 
 \tilde{\mathcal{B}}_{t}^{(\mathrm{L})}, \mathcal{S}_{t}, 
 \mathcal{I}_{\mathrm{\tau}})p(\tilde{\mathcal{B}}_{t}^{(\mathrm{L})})
 d\tilde{\mathcal{B}}_{t}^{(\mathrm{L})}
},
\end{equation}
with 
\begin{equation} 
p(\tilde{\mathcal{Y}}_{t}^{(\mathrm{L})} | 
\tilde{\mathcal{B}}_{t}^{(\mathrm{L})}, \mathcal{S}_{t}, 
\mathcal{I}_{\mathrm{tr}}) = 
\int p(\tilde{\mathcal{Y}}_{t}^{(\mathrm{L})} | 
\tilde{\mathcal{B}}_{t}^{(\mathrm{L})}, \mathcal{S}_{t}, 
\{\tilde{\boldsymbol{w}}_{t,k,m}^{(\mathrm{L})}\}, 
\{\hat{\boldsymbol{h}}_{k,m}^{\mathcal{I}_{\tau}}\})
\prod_{k=1}^{K}\prod_{m=1}^{M}\left\{ 
 p(\tilde{\boldsymbol{w}}_{t,k,m}^{(\mathrm{L})})
 d\tilde{\boldsymbol{w}}_{t,k,m}^{(\mathrm{L})}
\right\}, 
\end{equation}
where $p(\tilde{\mathcal{Y}}_{t}^{(\mathrm{L})} | 
\tilde{\mathcal{B}}_{t}^{(\mathrm{L})}, \mathcal{S}_{t}, 
\{\tilde{\boldsymbol{w}}_{t,k,m}^{(\mathrm{L})}\}, 
\{\hat{\boldsymbol{h}}_{k,m}\})$ represents the MIMO 
DS-CDMA channel~(\ref{MIMO_DS_CDMA_LMMSE}) in the $t(>\tau)$th symbol 
period postulated by the LMMSE detector. 

The LMMSE detector provides the posterior pdf~(\ref{posterior_pdf_LMMSE}) to 
the decoder of the $k$th user. 
The transfer of (\ref{posterior_pdf_LMMSE}) is equivalent to feeding the 
LMMSE estimate of $\mathcal{B}_{t,k}$ and the covariance matrix of its 
estimation errors. 
We remark that our LMMSE receiver is equivalent to the one proposed 
by Evans and Tse~\cite{Evans00} for $N=M=1$.  

The spectral efficiency of the LMMSE receiver is given by 
\begin{equation} \label{spectral_efficiency_LMMSE} 
C_{\mathrm{L}} = \frac{K}{L}\left(
 1 - \frac{\tau}{T_{\mathrm{c}}}
\right)\sum_{k=1}^{K}I(\mathcal{B}_{t,k}; 
\tilde{\mathcal{B}}_{t,k}^{(\mathrm{L})} | \mathcal{I}_{\tau}, 
\mathcal{S}_{t}). 
\end{equation} 
%\begin{equation} \label{mutual_information_LMMSE} 
%I(\mathcal{B}_{t,k};\tilde{\boldsymbol{b}}^{(\mathrm{L})}_{t,k}) = 
%\mathbb{E}\left[
% \left. 
%  \log\frac{
%   p(\tilde{\boldsymbol{b}}_{t,k}^{(\mathrm{L})} | \mathcal{B}_{t,k}, 
%   \mathcal{S}_{t}, \mathcal{I}_{\mathrm{tr}})
%  }
%  {
%   \int p(\tilde{\boldsymbol{b}}_{t,k}^{(\mathrm{L})} | \mathcal{B}_{t,k}, 
%   \mathcal{S}_{t}, \mathcal{I}_{\mathrm{tr}})
%   p(\mathcal{B}_{t,k})d\mathcal{B}_{t,k} 
%  }
% \right| \mathcal{S}_{t}, \mathcal{I}_{\mathrm{tr}} 
%\right]. 
%\end{equation}
In (\ref{spectral_efficiency_LMMSE}), the mutual information 
$I(\mathcal{B}_{t,k}; \tilde{\mathcal{B}}_{t,k}^{(\mathrm{L})} | 
\mathcal{I}_{\tau}, \mathcal{S}_{t})$ is characterized 
by the equivalent channel between user~$k$ and the corresponding decoder for 
the LMMSE receiver, 
\begin{equation}
p(\tilde{\mathcal{B}}_{t,k}^{(\mathrm{L})} | \mathcal{B}_{t,k}, 
\mathcal{S}_{t}, \mathcal{I}_{\mathcal{T}_{\tau}}) = 
\int p( \tilde{\mathcal{B}}_{t,k}^{(\mathrm{L})} | 
\tilde{\mathcal{Y}}_{t}^{(\mathrm{L})}=\mathcal{Y}_{t}, \mathcal{S}_{t}, 
\mathcal{I}_{\mathcal{T}_{\tau}}) 
p(\mathcal{Y}_{t} | \mathcal{B}_{t,k}, \mathcal{S}_{t}, 
\mathcal{I}_{\mathcal{T}_{\tau}})d\mathcal{Y}_{t}, 
\end{equation}
where $p(\mathcal{Y}_{t} | \mathcal{B}_{t,k}, \mathcal{S}_{t}, 
\mathcal{I}_{\mathcal{T}_{\tau}})$ is defined in the same manner as 
in (\ref{channel}).

\section{Main Results} \label{section_main_result}  
\subsection{Large-System Analysis}
%The use of the per-antenna spreading 
%scheme reduces the complexity of each per-user decoder significantly, since 
%joint decoding of all data streams for user~$k$ can be replaced by $M$ 
%parallel per-stream decoding without any performance loss when 
%perfect CSI is available at the receiver~\cite{Takeuchi082}.  
%We only present the results for the per-antenna spreading  
%scheme~\cite{Mantravadi03}, in which different spreading sequences are used 
%for different transmit antennas of each user. 
%The scheme is represented by taking $J_{k}=M$ and 
%$\boldsymbol{\Gamma}_{k,j}=\boldsymbol{E}_{M}^{(j)}$ for all $j$. 

In order to evaluate the spectral efficiencies of the four receivers 
listed in Table~\ref{table}, we consider the large-system limit, in which 
the number of users $K$ and the spreading factor $L$ tend to infinity while 
the system load $\beta=K/L$ and the number of transmit and receive antennas 
are kept constant. More precisely, we evaluate the spectral efficiencies by 
analyzing asymptotic properties of the equivalent channel between a finite 
number of users and their decoders. 
We write a finite number of users as $\mathcal{K}$, which is a finite 
subset of $\{1,\ldots,K\}$, and consider 
the large-system limit $\lim_{K,L\rightarrow\infty}$ in which $K$ and $L$ 
tend to infinity with $\beta$ and $\mathcal{K}$ fixed. 

We focus on substage~$k$ within stage~$t$ of the joint CE-MUDD. 
A finite subset of users $\mathcal{K}$ is chosen from the set $\{k,\ldots,K\}$ 
of users decoded in the current and following substages, i.e., 
$\mathcal{K}\cap\{1,\ldots,k-1\}=\emptyset$.  
Let $\mathcal{B}_{t,\mathcal{K}}=\{\mathcal{B}_{t,k}:k\in\mathcal{K}\}$ 
and $\tilde{\mathcal{B}}_{t,\mathcal{K}}=\{\tilde{\mathcal{B}}_{t,k}: 
k\in\mathcal{K}\}$ denote the data symbols and the postulated data symbols 
for a finite subset $\mathcal{K}$ of users in symbol period~$t$, respectively. 
The equivalent channel $p(\tilde{\mathcal{B}}_{t,\mathcal{K}} | 
\mathcal{B}_{t,\mathcal{K}}, \mathcal{B}_{t,[1,k)}, 
\mathcal{I}_{\mathcal{T}_{t-1}}, \mathcal{S}_{t},\mathcal{H})$ between the 
users in $\mathcal{K}$ and their decoders, defined in the same manner as 
in (\ref{equivalent_channel}), is expected to be self-averaging with respect 
to the spreading 
sequences $\mathcal{S}_{t}$ in the large-system limit: The equivalent channel 
converges to the one averaged over the spreading sequences for almost all 
realizations of the data symbols $\mathcal{B}_{t,\mathcal{K}}$, 
the data symbols $\mathcal{B}_{t,[1,k)}$ decoded in the preceding substages, 
the known information $I_{\mathcal{T}_{t-1}}$ in symbol 
periods~$t'\in\mathcal{T}_{t-1}$, 
and the channel vectors $\mathcal{H}$ in the large-system limit. The 
self-averaging property has been proved for 
linear receivers~\cite{Tse99,Evans00} and for the constrained capacity of CDMA 
systems with perfect CSI at the receiver~\cite{Korada08,Korada10}. However, 
it is an open challenging problem to show whether the self-averaging property 
holds for general receivers. Therefore, we postulate the self-averaging 
property. 
 
\begin{assumption} \label{assumption1} 
The conditional distribution of $\tilde{\mathcal{B}}_{t,\mathcal{K}}$ 
given $\mathcal{B}_{t,\mathcal{K}}$, $\mathcal{B}_{t,[1,k)}$, 
$\mathcal{I}_{\mathcal{T}_{t-1}}$, $\mathcal{S}_{t}$, and $\mathcal{H}$ 
converges in law to a conditional distribution that is independent of  
$\mathcal{S}_{t}$ in the large-system limit. 
\end{assumption}

Another crucial assumption is replica symmetry (RS)~\cite{Tanaka02}. 
See Appendix~\ref{derivation_lemma_channel_estimation} for a formal 
definition of the RS assumption. 
In order to present an intuitive understanding of the RS assumption, 
let us consider an iterative MUD algorithm based on belief propagation 
(BP)~\cite{Kabashima03}. Roughly speaking, the algorithm iteratively 
calculates a local minimum solution of an object function, called free energy, 
in the large-system limit. The global minimum solution of the free energy 
corresponds to the optimal one. The RS 
assumption implies that the free energy has the unique stable solution or 
at most two stable solutions. If the free energy has many stable solutions, 
replica-symmetry breaking (RSB) should be assumed~\cite{Mezard87}. 

Two necessary conditions for checking the RS assumption are known: 
de Almeida-Thouless (AT) stability~\cite{Almeida78} and the 
non-negative-entropy 
condition~\cite{Mezard87}. The AT condition is a necessary condition for the 
stability of RS solutions. The non-negative-entropy condition is a necessary 
condition under which the entropy for the posterior distribution of replicated 
random variables is non-negative when the random variables are discrete. 
Note that the non-negative-entropy condition is not defined for the no-CSI 
case, since the channel vectors are not discrete. See 
Appendix~\ref{appendix_derivation_proposition1} for details. 
The RS solution for the 
individually-optimal (IO) receiver with perfect CSI, which corresponds to the 
optimum separated receiver in this paper, has been proved to 
satisfy the AT stability condition~\cite{Tanaka02,Nishimori02}. 
Furthermore, that solution satisfies the non-negative-entropy condition derived 
in \cite[Equation~(69)]{Tanaka02} for CDMA systems with perfect CSI\footnote{
It is straightforward to prove that the non-negative-entropy 
condition~\cite[Equation~(69)]{Tanaka02} is satisfied for the IO receiver, 
although the proof was not presented in \cite{Tanaka02}.}. 
%Another freezing condition based on the perturbation of the postulated 
%variance of noise has been investigated in \cite{Yoshida07}, while 
%this type of perturbation is not considered in this paper. This freezing 
%condition for the IO receiver with perfect CSI produces a trivial condition: 
%The constrained capacity must not be larger than the transmission rate. 
These results may imply that the RS assumption is valid for the 
IO receiver. In fact, several rigorous studies 
have shown that this statement is partially correct: 
Nishimori~\cite{Nishimori02} has used a gauge theory 
to show that the free energy has no complicated structure for the IO receiver. 
See Appendix~\ref{derivation_lemma_channel_estimation} for the precise 
statement. 
Korada and Montanari~\cite{Korada10} have proved that the RS assumption 
is correct if the free energy under the RS assumption has the unique 
stable solution. Thus, we postulate the RS assumption in this paper.  

We show under these assumptions that the randomly-spread MIMO 
DS-CDMA channel with no CSI is decoupled into a bank of single-user 
single-input multiple-output (SIMO) channels with no CSI  
\begin{equation} \label{SIMO} 
\underline{\boldsymbol{y}}_{t,k,m} = \boldsymbol{h}_{k,m}u_{t,k,m} + 
\underline{\boldsymbol{n}}_{t,k,m}, 
\end{equation}
where $\underline{\boldsymbol{n}}_{t,k,m}\sim\mathcal{CN}(0,\sigma_{t}^{2}
\boldsymbol{I}_{N})$ denotes AWGN with variance $\sigma_{t}^{2}$. 
The equivalent channel between the users in $\mathcal{K}$ and their decoders 
looks like a bundle of the single-user SIMO 
channels~(\ref{SIMO}) with the {\em original} channel 
estimator~(\ref{posterior_pdf_Hk}). 
Furthermore, MAI to the users in $\mathcal{K}$  
converges towards MAI from the single-user SIMO channels~(\ref{SIMO}) 
with no CSI in the large-system limit. 
Our result is an extension of the decoupling results for randomly-spread 
MIMO DS-CDMA channels with perfect CSI at the 
receiver~\cite{Guo051,Takeuchi082} to the no-CSI case at the 
receiver. The difference from the previous decoupling results appears in 
the receiver structure for the SIMO channel~(\ref{SIMO}), which depends on the 
receiver structure of the original MIMO DS-CDMA systems. 

\begin{figure}[t]
\begin{center}
\includegraphics[width=0.5\hsize]{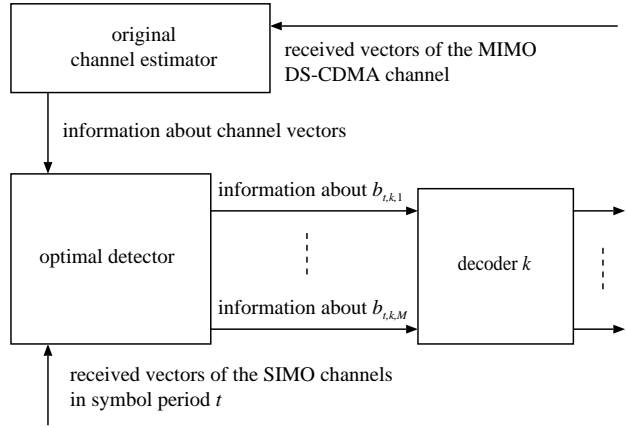} 
\end{center} 
\caption{
Receiver for the users in $\mathcal{K}$. 
}
\label{fig4} 
\end{figure}

\begin{figure}[t]
\begin{center}
\includegraphics[width=0.5\hsize]{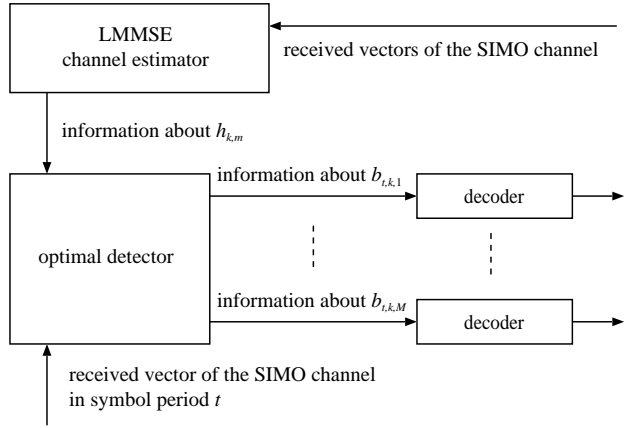} 
\end{center}
\caption{
Receiver for the users interfering to the users in $\mathcal{K}$. 
}
\label{fig5} 
\end{figure}

The equivalent channel between the users in $\mathcal{K}$ and the 
corresponding decoders is characterized by two 
receivers for the decoupled SIMO channels~(\ref{SIMO}), i.e., 
a receiver for the users in $\mathcal{K}$ and another 
receiver for the users who interfere to the users in $\mathcal{K}$. 
In the former receiver joint decoding of all data streams for each user is 
performed. The receiver consists of the original LMMSE channel 
estimator providing the marginal posterior pdfs~(\ref{posterior_pdf_Hk}), 
an optimal detector, and the decoder of user~$k$ (See Fig.~\ref{fig4}). 
The latter receiver performs per-stream decoding, and is to be used in the 
decoupled expression for quantifying the strength of MAI from outside 
$\mathcal{K}$. The receiver consists of another LMMSE channel estimator, 
the optimal detector, and the per-stream decoders (See Fig.~\ref{fig5}). 
Note that the information $\mathcal{B}_{t,[1,k)}$ in the preceding 
substages does not appear explicitly in the two receivers. It affects 
the decoupling results via the noise variance $\sigma_{t}^{2}$.  

\begin{definition}[Receiver for the users in $\mathcal{K}$] \label{definition1} 
In symbol period~$t(>\tau)$, the optimal detector uses the information about 
the received vectors $\underline{\mathcal{Y}}_{t,k}=\{
\underline{\boldsymbol{y}}_{t,k,m}: \hbox{for all $m$}\}$ of the single-user 
SIMO channel~(\ref{SIMO}) in symbol period~$t$ and about the posterior 
pdf~(\ref{posterior_pdf_Hk}) provided by the original channel estimator to  
construct the posterior pdf 
\begin{equation} \label{posterior_pdf_SIMO} 
p(\mathcal{B}_{t,k} | \underline{\mathcal{Y}}_{t,k}, 
\mathcal{I}_{\mathcal{T}_{t-1}}) = 
\frac{
 \prod_{m=1}^{M}
  p(\underline{\boldsymbol{y}}_{t,k,m}|\boldsymbol{h}_{k,m},b_{t,k,m})
  p(\mathcal{H}_{k}|\mathcal{I}_{\mathcal{T}_{t-1}})p(\mathcal{B}_{t,k})
}
{
 \int \prod_{m=1}^{M}
  p(\underline{\boldsymbol{y}}_{t,k,m}|\boldsymbol{h}_{k,m}, b_{t,k,m})
  p(\mathcal{H}_{k}|\mathcal{I}_{\mathcal{T}_{t-1}})p(\mathcal{B}_{t,k})
  d\mathcal{H}_{k}d\mathcal{B}_{t,k} 
}, 
\end{equation}
where $p(\underline{\boldsymbol{y}}_{t,k,m}|\boldsymbol{h}_{k,m},b_{t,k,m})$ 
represents the single-user SIMO channel~(\ref{SIMO}). Subsequently, the 
optimal detector sends the posterior pdf~(\ref{posterior_pdf_SIMO}) towards 
the decoder of user~$k$. 
\end{definition} 

\begin{definition}[Receiver for the interfering users] \label{definition2} 
The LMMSE channel estimator constructs the posterior pdf 
$p(\boldsymbol{h}_{k,m} | \underline{\mathcal{I}}_{\mathcal{T}_{t-1},k,m})$ of 
$\boldsymbol{h}_{k,m}$ by utilizing the information 
$\underline{\mathcal{I}}_{\mathcal{T}_{t-1},k,m}=\{
\underline{\mathcal{I}}_{t',k,m}: t'\in\mathcal{T}_{t-1}\}$ in the preceding 
stages, in which $\underline{\mathcal{I}}_{t',k,m}
=\{u_{t',k,m},\ \underline{\boldsymbol{y}}_{t',k,m}\}$ denotes the 
information about the input symbol $u_{t',k,m}$ and the received vector 
$\underline{\boldsymbol{y}}_{t',k,m}$ in the single-user SIMO 
channel~(\ref{SIMO}). Subsequently, the posterior pdf 
$p(\boldsymbol{h}_{k,m} | \underline{\mathcal{I}}_{\mathcal{T}_{t-1},k,m})$ is 
sent towards the optimal detector. 

In symbol period~$t(>\tau)$, the optimal detector utilizes the information 
about the received vector $\underline{\boldsymbol{y}}_{t,k,m}$ and 
$p(\boldsymbol{h}_{k,m} | \underline{\mathcal{I}}_{\mathcal{T}_{t-1},k,m})$ 
provided by the channel estimator for the SIMO channel~(\ref{SIMO}), and 
constructs the posterior pdf 
\begin{equation} \label{posterior_pdf} 
p(b_{t,k,m} | \underline{\boldsymbol{y}}_{t,k,m}, 
\underline{\mathcal{I}}_{\mathcal{T}_{t-1},k,m}) = 
\int p(b_{t,k,m},\boldsymbol{h}_{k,m}| 
\underline{\boldsymbol{y}}_{t,k,m}, 
\underline{\mathcal{I}}_{\mathcal{T}_{t-1},k,m})
d\boldsymbol{h}_{k,m}, 
\end{equation}
with 
\begin{equation} \label{joint_posterior_pdf} 
p(b_{t,k,m},\boldsymbol{h}_{k,m}| 
\underline{\boldsymbol{y}}_{t,k,m}, 
\underline{\mathcal{I}}_{\mathcal{T}_{t-1},k,m}) = 
\frac{
 p(\underline{\boldsymbol{y}}_{t,k,m} | \boldsymbol{h}_{k,m}, b_{t,k,m}) 
 p(b_{t,k,m})p(\boldsymbol{h}_{k,m}| 
 \underline{\mathcal{I}}_{\mathcal{T}_{t-1},k,m})
}
{
 \int p(\underline{\boldsymbol{y}}_{t,k,m} | \boldsymbol{h}_{k,m}, b_{t,k,m}) 
 p(b_{t,k,m})p(\boldsymbol{h}_{k,m}| 
 \underline{\mathcal{I}}_{\mathcal{T}_{t-1},k,m})
 db_{t,k,m}d\boldsymbol{h}_{k,m} 
}. 
\end{equation}
Subsequently, the optimal detector sends the posterior 
pdf~(\ref{posterior_pdf}) towards the corresponding per-stream decoder. 
\end{definition}

We summarize several symbols used for claiming the main results. 
The LMMSE estimate of $\boldsymbol{h}_{k,m}$ for the single-user SIMO 
channel~(\ref{SIMO}) is denoted by 
\begin{equation} \label{decoupled_LMMSE_channel_estimate} 
\underline{\hat{\boldsymbol{h}}}_{k,m}^{\mathcal{T}_{t-1}}
=\int \boldsymbol{h}_{k,m}p(\boldsymbol{h}_{k,m}|
\underline{\mathcal{I}}_{\mathcal{T}_{t-1},k,m})d\boldsymbol{h}_{k,m}. 
\end{equation} 
When $\sigma_{t'}^{2}=\sigma_{\mathrm{tr}}^{2}$ for $t'\in\mathcal{T}_{t-1}$, 
the covariance matrix of the LMMSE estimation error 
$\Delta\underline{\boldsymbol{h}}_{k,m}^{\mathcal{T}_{t-1}}=
\boldsymbol{h}_{k,m}
-\underline{\hat{\boldsymbol{h}}}_{k,m}^{\mathcal{T}_{t-1}}$ 
is given by $\xi^{2}(\sigma_{\mathrm{tr}}^{2},t-1)
\boldsymbol{I}_{N}$~\cite{Tse05}, with  
\begin{equation} \label{xi} 
\xi^{2}(\sigma_{\mathrm{tr}}^{2},t-1) = 
\frac{\sigma_{\mathrm{tr}}^{2}}{(t-1)(P/M) + \sigma_{\mathrm{tr}}^{2}}, 
\end{equation}
where $(t-1)$ corresponds to the length of the {\em training} phase 
in stage~$t$. 
The estimate $\underline{\hat{\boldsymbol{h}}}_{k,m}^{\mathcal{T}_{t-1}}$ is 
a CSCG random vector with the covariance matrix 
$(1-\xi^{2}(\sigma_{\mathrm{tr}}^{2},t-1))
\boldsymbol{I}_{N}$, since 
$\underline{\hat{\boldsymbol{h}}}_{k,m}^{\mathcal{T}_{t-1}}$
and $\Delta\underline{\boldsymbol{h}}_{k,m}^{\mathcal{T}_{t-1}}$ are 
uncorrelated. Furthermore, 
we write the posterior mean of $b_{t,k,m}$ given 
$\underline{\boldsymbol{y}}_{t,k,m}$ and 
$\underline{\mathcal{I}}_{\mathcal{T}_{t-1},k,m}$ as 
\begin{equation} \label{posterior_mean} 
\langle b_{t,k,m} \rangle=\int b_{t,k,m}p(b_{t,k,m},\boldsymbol{h}_{k,m}| 
\underline{\boldsymbol{y}}_{t,k,m}, 
\underline{\mathcal{I}}_{\mathcal{T}_{t-1},k,m})db_{t,k,m}
d\boldsymbol{h}_{k,m}. 
\end{equation}  
Finally, 
let $\underline{C}_{t}(\sigma_{\mathrm{tr}}^{2}(t-1),\sigma_{t}^{2})$ denote 
the conditional mutual information between the data symbol $b_{t,1,1}$ 
and the received vector $\underline{\boldsymbol{y}}_{t,1,1}$ for 
the SIMO channel~(\ref{SIMO}) given the information 
$\underline{\mathcal{I}}_{\mathcal{T}_{t-1},1,1}$, i.e., 
\begin{equation} \label{decoupled_mutual_inf} 
\underline{C}_{t}(\sigma_{\mathrm{tr}}^{2}(t-1),\sigma_{t}^{2}) 
= I(b_{t,1,1};\underline{\boldsymbol{y}}_{t,1,1}|
\underline{\mathcal{I}}_{\mathcal{T}_{t-1},1,1}),
\end{equation}
 with $\sigma_{t'}^{2}=\sigma_{\mathrm{tr}}^{2}(t-1)$ 
for $t'\in\mathcal{T}_{t-1}$. 
The conditional mutual information corresponds to the spectral efficiency of 
the receiver defined in Definition~\ref{definition2} for the single-user 
SIMO channel~(\ref{SIMO}). 

\begin{remark}
We shall discuss the relationship between the RSB assumption and the 
decoupling result. Under the RSB assumption, the equivalent channel would 
still be decoupled into single-user channels. However, the noise in the SIMO 
channel~(\ref{SIMO}) would be non-Gaussian. This implies that if the RS  
assumption is not correct a naive intuition is wrong: MAI should converge to  
Gaussian noise due to the central limit theorem. In fact, the RSB assumption 
should be considered if the jointly-optimal (JO) receiver is 
used~\cite{Tanaka02,Yoshida07}. 
\end{remark}

\subsection{Optimum Separated Receiver} 
\label{section_decoupled_Non-Iterative_Optimal_Receiver}
It is the main part in our derivation to analyze the spectral 
efficiency~(\ref{spectral_efficiency_MMSE}) of the optimum separated receiver 
in the large-system limit. The spectral efficiencies of the one-shot CE-MUDD 
and the joint CE-MUDD can be straightforwardly derived from the result for 
the optimum separated receiver. Thus, we first present an analytical 
expression for the spectral efficiency of the optimum separated receiver.  
Analytical formulas for the spectral efficiencies of the one-shot CE-MUDD and 
the joint CE-MUDD will be presented in Section~\ref{section_main_one_shot} 
and Section~\ref{section_main_joint}. 

Let $\underline{\tilde{\mathcal{B}}}_{t,k}=\{
\underline{\tilde{b}}_{t,k,m}\in\mathbb{C}: \hbox{for all $m$}\}\sim 
p(\mathcal{B}_{t,k}| \underline{\mathcal{Y}}_{t,k}, 
\mathcal{I}_{\mathcal{T}_{\tau}})$, defined by (\ref{posterior_pdf_SIMO}), 
denote random variables representing the information about 
$\mathcal{B}_{t,k}$, provided by the optimal detector. The equivalent channel 
between user~$k$ and the corresponding decoder is given by 
\begin{equation}
p(\underline{\tilde{\mathcal{B}}}_{t,k} | \mathcal{B}_{t,k}, 
\mathcal{H}_{k}, \mathcal{I}_{\mathcal{T}_{\tau}}) = 
\int p(\mathcal{B}_{t,k}=\underline{\tilde{\mathcal{B}}}_{t,k}| 
\underline{\mathcal{Y}}_{t,k}, \mathcal{I}_{\mathcal{T}_{\tau}})
\prod_{m=1}^{M}\left\{ 
 p(\underline{\boldsymbol{y}}_{t,k,m} | \boldsymbol{h}_{k,m}, b_{t,k,m})
 d\underline{\boldsymbol{y}}_{t,k,m}
\right\}. 
\end{equation}

\begin{proposition} \label{proposition1} 
Assume that Assumption~\ref{assumption1} and the RS assumption hold. Then, 
\begin{equation} \label{decoupling} 
\lim_{K,L\rightarrow\infty} 
p(\tilde{\mathcal{B}}_{t,\mathcal{K}} | \mathcal{B}_{t,\mathcal{K}}, 
\mathcal{I}_{\mathcal{T}_{\tau}}, \mathcal{S}_{t}, \mathcal{H}) 
= \prod_{k\in\mathcal{K}}p(\underline{\tilde{\mathcal{B}}}_{t,k}=
\tilde{\mathcal{B}}_{t,k} | \mathcal{B}_{t,k}, 
\mathcal{H}_{k}, \mathcal{I}_{\mathcal{T}_{\tau}})   
\quad\hbox{in law,}
\end{equation}
where $\sigma_{t}^{2}=\sigma_{\mathrm{c}}^{2}$ for $t\in\mathcal{C}_{\tau+1}$ 
is given by a solution to the fixed-point equation 
\begin{equation}
\sigma_{\mathrm{c}}^{2} = N_{0} + \lim_{K\rightarrow\infty}\frac{\beta}{K}
\sum_{k\notin\mathcal{K}}\Biggl\{
\frac{ P\xi^{2}\sigma_{\mathrm{c}}^{2}}
{(P/M)\xi^{2} + \sigma_{\mathrm{c}}^{2}}  
+ \frac{M}{N}\left(
 \frac{\sigma_{\mathrm{c}}^{2}}
 {(P/M)\xi^{2} + \sigma_{\mathrm{c}}^{2}} 
\right)^{2}
\mathbb{E}\left[
 \|\underline{\hat{\boldsymbol{h}}}_{k,1}^{\mathcal{T}_{\tau}}\|^{2}
 |b_{t,k,1} - \langle b_{t,k,1} \rangle|^{2} 
\right]
\Biggr\}, \label{fixed_point_data} 
\end{equation}
with $\underline{\hat{\boldsymbol{h}}}_{k,1}^{\mathcal{T}_{\tau}}$ and 
$\xi^{2}=\xi^{2}(\sigma_{\mathrm{tr}}^{2}(\tau),\tau)$ give by 
(\ref{decoupled_LMMSE_channel_estimate}) and (\ref{xi}), respectively. 
In evaluating (\ref{fixed_point_data}), 
$\sigma_{t'}^{2}=\sigma_{\mathrm{tr}}^{2}(\tau)$ for 
$t'\in\mathcal{T}_{\tau}$ is 
given as the unique solution to the fixed-point equation 
\begin{equation} \label{fixed_point_channel} 
\sigma_{\mathrm{tr}}^{2}(\tau) = N_{0} 
+ \beta P\xi^{2}(\sigma_{\mathrm{tr}}^{2}(\tau),\tau). 
\end{equation}
If (\ref{fixed_point_data}) has multiple solutions, one should choose the 
solution minimizing the following quantity 
\begin{equation} \label{free_energy} 
\lim_{K\rightarrow\infty}\frac{\beta M}{K}\sum_{k\notin\mathcal{K}}
I(b_{t,k,1},\boldsymbol{h}_{k,1};\underline{\boldsymbol{y}}_{t,k,1}| 
\underline{\mathcal{I}}_{\mathcal{T}_{\tau},k,1}) 
+ ND(N_{0}\|\sigma_{\mathrm{c}}^{2}).  
\end{equation}  
\end{proposition}

The derivation of Proposition~\ref{proposition1} is deferred to the 
end of this section. 
The fixed-point equation~(\ref{fixed_point_channel}) was originally derived 
in \cite{Evans00} by using {\em rigorous} random matrix theory, while 
we have used the replica method. 
The second term of the right-hand side of (\ref{fixed_point_data}) corresponds 
to MAI from the users who do not belong to the users in $\mathcal{K}$. 
This expression implies that the asymptotic MAI becomes the 
sum of interference from $(K-|\mathcal{K}|)M$ independent SIMO 
channels~(\ref{SIMO}). Furthermore, each interference is represented by two 
effects: The first term within the curly 
brackets in (\ref{fixed_point_data}) corresponds to contribution from channel 
estimation errors, and the second term corresponds to MAI  
from the single-user SIMO channel with perfect CSI at the receiver, 
\begin{equation} \label{SIMO_perfect} 
\underline{\boldsymbol{z}}_{t,k,m}=
\underline{\hat{\boldsymbol{h}}}_{k,m}^{\mathcal{T}_{\tau}}b_{t,k,m} 
+ \underline{\boldsymbol{v}}_{t,k,m}, 
\end{equation}
with $\underline{\boldsymbol{v}}_{t,k,m}\sim\mathcal{CN}(\boldsymbol{0}, 
[(P/M)\xi^{2}+\sigma_{\mathrm{c}}^{2}]\boldsymbol{I}_{N})$. 
The received vector $\underline{\boldsymbol{z}}_{t,k,m}$ conditioned on 
$\underline{\mathcal{I}}_{\mathcal{T}_{\tau},k,m}$ and $b_{t,k,m}$ is 
statistically equivalent to $\underline{\boldsymbol{y}}_{t,k,m}$ in 
(\ref{SIMO}) under the same conditions, 
due to $|b_{t,k,m}|^{2}=P/M$ with probability one. 
We remark that this interpretation holds only for phase-shift keying 
modulations. 

The fixed-point equation~(\ref{fixed_point_data}) can have multiple solutions. 
The criterion~(\ref{free_energy}) to select the correct solution corresponds 
to the conditional mutual information $L^{-1}I(\mathcal{B}_{t},\mathcal{H}; 
\mathcal{Y}_{t}|\mathcal{S}_{t},\mathcal{I}_{\mathcal{T}_{\tau}})$ in the 
large-system limit, although its proof is omitted.  
This phenomenon is related to the so-called phase coexistence in statistical 
mechanics. For the details in the context of wireless communications, see 
\cite{Tanaka02,Guo051}.  
The existence of multiple solutions implies that the asymptotic performance 
discontinuously changes at a critical point. 
This asymptotic result predicts that the performance sharply changes in the 
neighborhood of the critical point for finite-sized systems.  

Proposition~\ref{proposition1} implies that the equivalent channel between 
the users in $\mathcal{K}$ and their decoders is not fully decoupled into 
single-user channels, since the channel estimator in 
Definition~\ref{definition1} utilizes the information 
$\mathcal{I}_{\mathcal{T}_{\tau}}$ depending on all users. 
This is the main difference between the cases of perfect CSI and of no CSI. 
The following proposition indicates that the equivalent channel is decomposed 
in terms of the spectral efficiency. 

\begin{proposition} \label{proposition2} 
Assume that Assumption~\ref{assumption1} and the RS assumption hold. Then, 
the spectral efficiency~(\ref{spectral_efficiency_MMSE}) of the optimum 
separated receiver converges to the spectral 
efficiency of the optimum separated receiver for the single-user SIMO 
channel~(\ref{SIMO}) in the large-system limit,  
\begin{equation} \label{spectral_efficiency_MMSE_dec} 
\lim_{K,L\rightarrow\infty}C_{\mathrm{sep}} = 
\beta M\left(
 1 - \frac{\tau}{T_{\mathrm{c}}}
\right)\underline{C}_{\tau+1}(\sigma_{\mathrm{tr}}^{2}(\tau),
\sigma_{\mathrm{c}}^{2}), 
\end{equation}
where $\underline{C}_{\tau+1}(\sigma_{\mathrm{tr}}^{2}(\tau),
\sigma_{\mathrm{c}}^{2})$ is given by (\ref{decoupled_mutual_inf}). 
In evaluating the right-hand side, 
$\sigma_{\mathrm{tr}}^{2}(\tau)$ is given as the solution to the fixed-point 
equation~(\ref{fixed_point_channel}). On the other hand, 
$\sigma_{\mathrm{c}}^{2}$ is given as a solution to the fixed-point 
equation~(\ref{fixed_point_data}). If (\ref{fixed_point_data}) has 
multiple solutions, the solution minimizing (\ref{free_energy}) should be 
chosen. 
\end{proposition}
\begin{IEEEproof}[Proof of Proposition~\ref{proposition2}] 
We use a technical lemma, presented in the end of this section, to prove 
Proposition~\ref{proposition2}. See Appendix~\ref{derivation_proposition2} 
for the details. 
\end{IEEEproof} 

This result implies that the asymptotic equivalent channel between user~$k$ 
and the associated decoder looks like the SIMO channel~(\ref{SIMO}) in terms 
of the spectral efficiency. In other words, the performance loss caused by 
coding the data streams for each user separately vanishes in the large-system 
limit, as shown in \cite{Takeuchi082} for the perfect-CSI case. 
We remark that it is relatively easy to evaluate 
(\ref{spectral_efficiency_MMSE_dec}) numerically, 
by using $\underline{\boldsymbol{y}}_{t,k,m}\sim
\underline{\boldsymbol{z}}_{t,k,m}$ in (\ref{SIMO_perfect}) conditioned on 
$\underline{\mathcal{I}}_{\mathcal{T}_{\tau},k,m}$ and $b_{t,k,m}$. 

The prefactor $(1-\tau/T_{\mathrm{c}})$ corresponds to the rate loss due to 
the transmission of pilot symbols. The spectral efficiency 
$\underline{C}_{\tau+1}(\sigma_{\mathrm{tr}}^{2}(\tau),
\sigma_{\mathrm{c}}^{2})$ grows with the increase of $\tau$ since 
the channel estimation improves, while the prefactor decreases. Thus, there is 
the optimal length of the training phase to maximize the 
spectral efficiency of the optimum separated receiver, as shown in 
Section~\ref{section_numerical_result}. 

We shall present a sketch of the derivation of Proposition~\ref{proposition1}. 
The derivation of Proposition~\ref{proposition1} consists of two parts: 
the analysis of the channel estimator and the analysis of the optimal 
detector. The goal in the analysis of the channel estimator is to prove 
the following lemma. 

\begin{lemma} \label{lemma_channel_estimation} 
Let $\mathcal{H}_{k}^{\{a\}}=\{\boldsymbol{h}_{k,m}^{\{a\}}:\hbox{for all 
$m$}\}$ be replicas of the channel vectors $\mathcal{H}_{k}$ 
for $a\in\{1, \ldots, n\}$, with a natural number $n$: 
$\{\mathcal{H}_{k}^{\{a\}}\}$ are independently drawn from 
$p(\mathcal{H}_{k})$ for all $k$. 
Suppose that $\{\mathcal{A}_{k}\}$ are mutually disjoint subsets of 
$\{2,3,\ldots,n\}$ for all $k\in\mathcal{K}$. 
We define a random variable 
$X_{k}(\mathcal{H}_{k}^{\{1\}}, \mathcal{I}_{\mathcal{T}_{\tau}};\Theta)
\in\mathbb{R}$ with a set $\Theta$ of fixed parameters as 
\begin{equation} \label{X_k} 
X_{k}(\mathcal{H}_{k}^{\{1\}}, \mathcal{I}_{\mathcal{T}_{\tau}};\Theta) = 
\int f_{k}(\mathcal{H}_{k}^{\{1\}}, \mathcal{H}_{k}^{\mathcal{A}_{k}};
\Theta)\prod_{a\in \mathcal{A}_{k}}\left\{
 p( \mathcal{H}_{k}^{\{a\}} | \mathcal{I}_{\mathcal{T}_{\tau}})
 d\mathcal{H}_{k}^{\{a\}}
\right\}, 
\end{equation}
with $\mathcal{H}_{k}^{\mathcal{A}_{k}}=\{\mathcal{H}_{k}^{\{a\}}: 
a\in\mathcal{A}_{k}\}$ denoting the set of the replicated channel vectors 
associated with indices~$\mathcal{A}_{k}$. In (\ref{X_k}), 
$f_{k}(\mathcal{H}_{k}^{\{1\}}, \mathcal{H}_{k}^{\mathcal{A}_{k}};
\Theta)\in\mathbb{R}$ is a deterministic function of 
$\mathcal{H}_{k}^{\{1\}}$, $\mathcal{H}_{k}^{\mathcal{A}_{k}}$, and 
$\Theta$. Suppose that the joint moment generating 
function of $\{X_{k}(\mathcal{H}_{k}^{\{1\}},\mathcal{I}_{\mathcal{T}_{\tau}};
\Theta): k\in\mathcal{K}\}$ exists in the neighborhood of the origin. Then, 
\begin{equation} \label{decoupling_channel} 
\lim_{K,L\rightarrow\infty}p(\{X_{k}: k\in\mathcal{K}\})
= \prod_{k\in\mathcal{K}}p(\underline{X}_{k}=X_{k}), 
\end{equation} 
in which 
\begin{equation} \label{decoupled_X_k}    
\underline{X}_{k}(\mathcal{H}_{k}^{\{1\}}, 
\underline{\mathcal{I}}_{\mathcal{T}_{\tau},k};
\Theta) = \int f_{k}(\mathcal{H}_{k}^{\{1\}}, 
\mathcal{H}_{k}^{\mathcal{A}_{k}};\Theta)
\prod_{a\in \mathcal{A}_{k}}\prod_{m=1}^{M}\left\{
 p( \boldsymbol{h}_{k,m}=\boldsymbol{h}_{k,m}^{\{a\}} | 
 \underline{\mathcal{I}}_{\mathcal{T}_{\tau},k,m})d\boldsymbol{h}_{k,m}^{\{a\}}
\right\},  
\end{equation}
with $\underline{\mathcal{I}}_{\mathcal{T}_{\tau},k}
=\{\underline{\mathcal{I}}_{\mathcal{T}_{\tau},k,m}: \hbox{for all $m$}\}$.   
In (\ref{decoupling_channel}), $X_{k}$ and $\underline{X}_{k}$ are 
abbreviations of (\ref{X_k}) and (\ref{decoupled_X_k}). 
In evaluating (\ref{decoupled_X_k}), 
$\sigma_{t}^{2}=\sigma_{\mathrm{tr}}^{2}(\tau)$ for $t\in\mathcal{T}_{\tau}$ is 
given by the solution to the fixed-point equation~(\ref{fixed_point_channel}). 
\end{lemma}
\begin{IEEEproof}[Proof of Lemma~\ref{lemma_channel_estimation}] 
See Appendix~\ref{derivation_lemma1}. 
\end{IEEEproof}

Lemma~\ref{lemma_channel_estimation} is used to prove that MAI to the users 
in $\mathcal{K}$ is self-averaging 
with respect to $\mathcal{I}_{\mathcal{T}_{\tau}}$ under 
Assumption~\ref{assumption1}. The natural number $n$ corresponds to the 
number of replicas introduced in the analysis of the optimal detector. 
The details of the derivation of Proposition~\ref{proposition1} are 
summarized in Appendix~\ref{appendix_derivation_proposition1}.  

%Note that (\ref{spectral_efficiency_MMSE_dec}) is bounded from 
%above by $2\beta(1-\tau/T_{\mathrm{c}})\bar{M}$, with the average number 
%of transmit antennas 
%$\bar{M}=\lim_{K\rightarrow\infty}K^{-1}\sum_{k=1}^{K}M$ . 
%We will find in Section~\ref{section_numerical_result} that the non-iterative 
%optimal receiver cannot achieve the upper bound 
%$2\beta(1-\tau/T_{\mathrm{c}})\bar{M}$ even for high signal-to-noise 
%ratio (SNR).  

\subsection{One-Shot CE-MUDD} \label{section_main_one_shot} 
We present an analytical expression for the spectral 
efficiency~(\ref{spectral_efficiency_one}) of the one-shot CE-MUDD. 
The expression is straightforwardly obtained from 
Proposition~\ref{proposition2}. 

\begin{proposition} \label{proposition3} 
Suppose that Assumption~\ref{assumption1} and the RS assumption hold. 
Then, the spectral efficiency~(\ref{spectral_efficiency_one}) of the 
one-shot CE-MUDD is given by 
\begin{equation} \label{spectral_efficiency_one_low} 
\lim_{K,L\rightarrow\infty}C_{\mathrm{one}} = \beta M\left(
 1 - \frac{\tau}{T_{\mathrm{c}}}
\right)\int_{0}^{1}\underline{C}_{\tau+1}(\sigma_{\mathrm{tr}}^{2}(\tau),
\sigma_{\mathrm{c}}^{2}(\kappa))d\kappa,  
\end{equation}
in the large-system limit, in which 
$\underline{C}_{\tau+1}(\sigma_{\mathrm{tr}}^{2}(\tau),
\sigma_{\mathrm{c}}^{2}(\kappa))$ is defined as (\ref{decoupled_mutual_inf}). 
In evaluating the integrand, 
$\sigma_{t}^{2}=\sigma_{\mathrm{tr}}^{2}(\tau)$ for $t\in\mathcal{T}_{\tau}$ 
is given as the solution to the fixed-point 
equation~(\ref{fixed_point_channel}). On the other hand, 
$\sigma_{\mathrm{c}}^{2}(\kappa)$ satisfies the fixed-point equation 
\begin{equation}  \label{fixed_point_data_low} 
\sigma_{\mathrm{c}}^{2}(\kappa) = N_{0} 
+ \frac{ \beta P\xi^{2}\sigma_{\mathrm{c}}^{2}(\kappa)}
{(P/M)\xi^{2} + \sigma_{\mathrm{c}}^{2}(\kappa)}
+ \frac{\beta(1-\kappa)M}{N}\left(
 \frac{\sigma_{\mathrm{c}}^{2}(\kappa)}
 {(P/M)\xi^{2} + \sigma_{\mathrm{c}}^{2}(\kappa)} 
\right)^{2}
\mathbb{E}\left[
 \|\underline{\hat{\boldsymbol{h}}}_{1,1}\|^{2}
 |b_{\tau+1,1,1} - \langle {b}_{\tau+1,1,1} \rangle|^{2}  
\right],  
\end{equation}
with $\underline{\hat{\boldsymbol{h}}}_{1,1}
=\underline{\hat{\boldsymbol{h}}}_{1,1}^{\mathcal{T}_{\tau}}$ and 
$\xi^{2}=\xi^{2}(\sigma_{\mathrm{tr}}^{2}(\tau),\tau)$ given by 
(\ref{decoupled_LMMSE_channel_estimate}) and (\ref{xi}), respectively. 
If the fixed-point equation~(\ref{fixed_point_data_low}) has multiple 
solutions, one should choose the solution minimizing the following 
quantity 
\begin{equation} \label{free_energy_kappa} 
\beta M\left[
 \kappa I(\boldsymbol{h}_{1,1};\underline{\boldsymbol{y}}_{\tau+1,1,1}| 
 b_{\tau+1,1,1}, \underline{\mathcal{I}}_{\mathcal{T}_{\tau},1,1}) 
 +(1-\kappa)I(b_{\tau+1,1,1},\boldsymbol{h}_{1,1};
 \underline{\boldsymbol{y}}_{\tau+1,1,1}| 
 \underline{\mathcal{I}}_{\mathcal{T}_{\tau},1,1})
\right] + ND(N_{0}\|\sigma_{\mathrm{c}}^{2}(\kappa)). 
\end{equation}
\end{proposition}

The integrand in (\ref{spectral_efficiency_one_low}) is equal to  
the asymptotic spectral efficiency of the optimum separated receiver 
for a MIMO DS-CDMA system in which $(1-\kappa)K$ users send data symbols 
and $\kappa K$ users transmit symbols known to the receiver in symbol 
period $\tau+1$. The symbols known to the receiver corresponds to the data 
symbols decoded successfully in the preceding substages.  
The factor $1-\kappa$ appears in the last term of the right-hand side of 
(\ref{fixed_point_data_low}), since the system load decreases effectively as 
successive decoding proceeds. 

\begin{IEEEproof}[Proof of Proposition~\ref{proposition3}]
The quantity $C_{\tau+1,k}$ in the spectral 
efficiency~(\ref{spectral_efficiency_one}) of the one-shot CE-MUDD 
corresponds to the spectral efficiency of the optimum separated receiver 
for a MIMO DS-CDMA system in which the first $(k-1)$ users transmit known 
symbols $\{p_{t,k',m}\}$ for $k'=1,\ldots,k-1$ in symbol 
period~$t(>\tau)$. We have already evaluated the spectral 
efficiency in Proposition~\ref{proposition2} when all users transmit data 
symbols in symbol period~$t$. We do not use the statistical 
property of each data symbol $b_{t,k,m}$ in the derivation of 
Proposition~\ref{proposition2}. Furthermore, the derivation still 
holds even if the noise variance $\sigma_{t}^{2}$ for the single-user SIMO 
channel~(\ref{SIMO}) depends on $k$. For some natural number $K_{0}$, 
consider the large-system limit in which $K$, $L$, and $K_{0}$ tend to 
infinity while $\beta=K/L$ and $\kappa_{0}=K_{0}/K$ are kept constant. 
Replacing the prior of $b_{t,k',m}$ for $k'=1,\ldots,k-1$ 
by $\mathrm{Prob}(b_{t,k',m}=p_{t,k',m})=1$ and 
$\mathrm{Prob}(b_{t,k',m}\neq p_{t,k',m})=0$ in substage~$k\geq K_{0}$, 
we find that the fixed-point equation~(\ref{fixed_point_data}) reduces 
to (\ref{fixed_point_data_low}) with $\kappa=k/K$, since 
$\mathbb{E}[\|\boldsymbol{h}_{k',1}^{\mathcal{T}_{\tau}}\|^{2}
|b_{t,k',1} - \langle b_{t,k',1} \rangle|^{2}]=0$ for $k'=1,\ldots,k-1$. 
Similarly, the quantity~(\ref{free_energy}) for $t=\tau+1$ reduces to 
(\ref{free_energy_kappa}).  
These results imply 
\begin{equation} \label{averaged_lower_bound} 
C_{\mathrm{one}} 
= \beta \left(
 1 - \frac{\tau}{T_{\mathrm{c}}}
\right)\left[
 \frac{1}{K}\sum_{k=1}^{K_{0}-1}C_{t,k} 
 + \frac{M}{K}\sum_{k=K_{0}}^{K}\underline{C}_{\tau+1}
 (\sigma_{\mathrm{tr}}^{2}(\tau),\sigma_{\mathrm{c}}^{2}(\kappa))  
\right],   
\end{equation}
in the large-system limit. 
In evaluating the second term, $\sigma_{\mathrm{tr}}^{2}(\tau)$ is given 
as the solution to the fixed-point equation~(\ref{fixed_point_channel}). 
On the other hand, $\sigma_{\mathrm{c}}^{2}(\kappa)$ 
satisfies the fixed-point equation~(\ref{fixed_point_data_low}). 
If the latter fixed-point equation has multiple solutions, the solution 
minimizing (\ref{free_energy_kappa}) should be chosen. 

The definition of the Riemann integral implies that 
$K^{-1}\sum_{k=K_{0}}^{K}$ in the second term of (\ref{averaged_lower_bound}) 
becomes $\int_{\kappa_{0}}^{1}d\kappa$ in the large-system limit. Therefore, 
we arrive at Proposition~\ref{proposition3} by taking 
$\kappa_{0}\rightarrow0$, since the first term in (\ref{averaged_lower_bound}) 
vanishes in that limit. 
\end{IEEEproof}

\subsection{Joint CE-MUDD} \label{section_main_joint} 
We evaluate the spectral efficiency~(\ref{target_lower_bound}) of 
the joint CE-MUDD in the large-system limit. 
An analytical expression of the spectral efficiency~(\ref{target_lower_bound}) 
is immediately obtained from Proposition~\ref{proposition3}. 

\begin{proposition} \label{proposition4} 
Suppose that Assumption~\ref{assumption1} and the RS assumption hold. Then, 
the spectral efficiency~(\ref{target_lower_bound}) of the joint 
CE-MUDD is given by  
\begin{equation} \label{spectral_efficiency_low} 
\lim_{K,L\rightarrow\infty}C_{\mathrm{joint}} = \frac{\beta M}{T_{\mathrm{c}}}
\sum_{t=\tau+1}^{T_{\mathrm{c}}}\int_{0}^{1}
\underline{C}_{t}(\sigma_{\mathrm{tr}}^{2}(t-1),
\sigma_{\mathrm{c}}^{2}(\kappa))d\kappa,  
\end{equation}
in the large-system limit, in which 
$\underline{C}_{t}(\sigma_{\mathrm{tr}}^{2}(t-1),
\sigma_{\mathrm{c}}^{2}(\kappa))$ is defined as (\ref{decoupled_mutual_inf}). 
In evaluating the integrand, 
$\sigma_{\mathrm{tr}}^{2}(t-1)$ is given as the solution 
to the fixed point equation~(\ref{fixed_point_channel}) with $\tau=t-1$. 
On the other hand, $\sigma_{\mathrm{c}}^{2}(\kappa)$ satisfies 
the fixed-point equation~(\ref{fixed_point_data_low}) 
with $\underline{\hat{\boldsymbol{h}}}_{1,1}
=\underline{\hat{\boldsymbol{h}}}_{1,1}^{\mathcal{T}_{t-1}}$ and 
$\xi^{2}=\xi^{2}(\sigma_{\mathrm{tr}}^{2}(t-1),t-1)$ given by 
(\ref{decoupled_LMMSE_channel_estimate}) and (\ref{xi}), respectively. 
If the fixed-point equation~(\ref{fixed_point_data_low}) has multiple 
solutions, the solution minimizing (\ref{free_energy_kappa}) should be chosen. 
\end{proposition}

\begin{IEEEproof}[Derivation of Proposition~\ref{proposition4}]
The quantity $C_{t,k}$ in (\ref{target_lower_bound}) corresponds to the 
spectral efficiency of the one-shot CE-MUDD for the MIMO DS-CDMA 
channel~(\ref{MIMO_DS_CDMA}) with the first $t-1$ symbol periods as a 
training phase. Applying Proposition~\ref{proposition3} with $\tau=t-1$ to 
(\ref{target_lower_bound}), we obtain (\ref{spectral_efficiency_low}). 
\end{IEEEproof} 

The spectral efficiency~(\ref{spectral_efficiency_low}) does not decrease 
with each increase of $\tau$. Furthermore, the integral in 
(\ref{spectral_efficiency_low}) is equal to zero for $t=1$, since 
$t=1$ implies no pilots. Therefore, 
the spectral efficiency~(\ref{spectral_efficiency_low}) is maximized 
at $\tau=0$ and $\tau=1$ if the integral in (\ref{spectral_efficiency_low}) 
is strictly positive for $t\geq2$. This observation implies that 
the joint CE-MUDD can reduce the training overhead significantly. 

\begin{remark}
We have so far considered the equal power case. One interesting issue would be  
temporal power allocation. It is straightforward to extend 
Proposition~\ref{proposition4} to the temporally unequal power case. 
Intuitively, allocating much power to around the beginning of one fading block 
improves the accuracy of the channel estimates, while power used for  
transmission of data symbols around the end decreases. Thus, it is not 
straightforward to find the temporally optimal power allocation. 
This power allocation issue is left as future work. 
\end{remark}

\begin{remark}
It is possible in principle to evaluate the spectral efficiency of joint 
CE-MUDD based on nonlinear channel estimation in the large-system 
limit, in which the channel estimator sends the marginal posterior pdfs 
$\{p(\mathcal{H}_{k}|\mathcal{I}_{\mathcal{T}_{t-1}}, 
\overline{\mathcal{I}}_{\mathcal{C}_{t+1}})\}$ 
to the optimal detector in stage~$t$, instead of (\ref{posterior_pdf_Hk}). 
The spectral efficiency would be 
given as that of a receiver with the optimal nonlinear channel estimator for 
the single-user SIMO channel~(\ref{SIMO}). However, the obtained spectral 
efficiency is difficult to calculate in terms of the computational complexity. 
In this sense, the bound based on nonlinear channel estimation is not an 
analytical one, while the formula itself is simple. 
\end{remark}

\subsection{LMMSE Receiver}
\label{section_decouled_LMMSE}
The LMMSE receiver for $M=N=1$ has been analyzed rigorously 
in \cite{Evans00} by using random matrix theory. In this section, we show 
that the replica method can derive the same result as the rigorous one. 
 
The spectral efficiency~(\ref{spectral_efficiency_LMMSE}) of the  
LMMSE receiver is given via the spectral efficiency of a bank of the 
single-user SIMO channels~(\ref{SIMO}) with an LMMSE receiver. 
We first define the LMMSE receiver for the SIMO 
channel~(\ref{SIMO}), after which we present an analytical expression for the 
spectral efficiency~(\ref{spectral_efficiency_LMMSE}) of the LMMSE receiver 
in the large-system limit. 
The LMMSE receiver is obtained by replacing the optimal detector in 
Fig.~\ref{fig5} by the LMMSE detector. 

\begin{definition} 
The LMMSE detector estimates the data symbol $b_{t,k,m}$ from the received 
vector $\underline{\boldsymbol{y}}_{t,k,m}$ for the single-user SIMO 
channel~(\ref{SIMO}) in the same symbol period~$t$ and the information about 
$\boldsymbol{h}_{k,m}$ provided by the LMMSE channel estimator defined in 
Definition~\ref{definition2}, 
and feeds the information about $b_{t,k,m}$ to the per-stream decoders. 
In order to define the information about $b_{t,k,m}$, we consider 
the single-user SIMO channel in the $t$th symbol period postulated by the 
LMMSE detector,    
\begin{equation} \label{postulated_LMMSE_SIMO}   
\underline{\tilde{\boldsymbol{y}}}_{t,k,m}^{(\mathrm{L})} = 
\underline{\hat{\boldsymbol{h}}}_{k,m}^{\mathcal{T}_{\tau}}
\tilde{b}_{t,k,m}^{(\mathrm{L})} 
+ \underline{\tilde{\boldsymbol{w}}}_{t,k,m}^{(\mathrm{L})} 
+ \underline{\tilde{\boldsymbol{n}}}_{t,k,m}^{(\mathrm{L})},   
\end{equation}
with $\underline{\tilde{\boldsymbol{n}}}_{t,k,m}^{(\mathrm{L})}
\sim\mathcal{CN}(\boldsymbol{0},\sigma_{t}^{2}\boldsymbol{I}_{N})$. 
In (\ref{postulated_LMMSE_SIMO}), 
$\underline{\hat{\boldsymbol{h}}}_{k,m}^{\mathcal{T}_{\tau}}$ denotes 
the LMMSE channel estimate~(\ref{decoupled_LMMSE_channel_estimate}) 
for the single-user SIMO channel~(\ref{SIMO}). 
The vector 
$\underline{\tilde{\boldsymbol{w}}}_{t,k,m}^{(\mathrm{L})}\in\mathbb{C}^{N}$ 
conditioned on 
$\underline{\mathcal{I}}_{\mathcal{T}_{\tau},k,m}$ is a CSCG random vector with 
the covariance matrix $(P/M)\underline{\boldsymbol{\Sigma}}_{t,k,m}$, in 
which $\underline{\boldsymbol{\Sigma}}_{t,k,m}$ denotes the covariance matrix 
of the channel estimation errors $\Delta
\underline{\boldsymbol{h}}_{k,m}^{\mathcal{T}_{\tau}} 
= \boldsymbol{h}_{k,m} 
- \underline{\hat{\boldsymbol{h}}}_{k,m}^{\mathcal{T}_{\tau}}$, i.e.,  
$\underline{\boldsymbol{\Sigma}}_{t,k,m} = 
\mathbb{E}[ \Delta\underline{\hat{\boldsymbol{h}}}_{k,m}^{\mathcal{T}_{\tau}}
(\Delta\underline{\hat{\boldsymbol{h}}}_{k,m}^{\mathcal{T}_{\tau}})^{H}   
| \underline{\mathcal{I}}_{\mathcal{T}_{\tau},k,m}]$. 
The information about $b_{t,k,m}$ is defined as 
\begin{equation} 
p(\tilde{b}_{t,k,m}^{(\mathrm{L})} | 
\underline{\boldsymbol{y}}_{t,k,m}^{(\mathrm{L})}=
\underline{\boldsymbol{y}}_{t,k,m}, 
\underline{\mathcal{I}}_{\mathcal{T}_{\tau},k,m} ) = 
\int p(\tilde{b}_{t,k,m}^{(\mathrm{L})}, 
\underline{\tilde{\boldsymbol{w}}}_{t,k,m}^{(\mathrm{L})} | 
\underline{\boldsymbol{y}}_{t,k,m}^{(\mathrm{L})}
=\underline{\boldsymbol{y}}_{t,k,m}, 
\underline{\mathcal{I}}_{\mathcal{T}_{\tau},k,m} )
d\underline{\tilde{\boldsymbol{w}}}_{t,k,m}^{(\mathrm{L})}, 
\end{equation}
with  
\begin{equation} 
p(\tilde{b}_{t,k,m}^{(\mathrm{L})}, 
\underline{\tilde{\boldsymbol{w}}}_{t,k,m}^{(\mathrm{L})} | 
\underline{\boldsymbol{y}}_{t,k,m}^{(\mathrm{L})}, 
\underline{\mathcal{I}}_{\mathcal{T}_{\tau},k,m} ) = 
\frac{
p(\underline{\boldsymbol{y}}_{t,k,m}^{(\mathrm{L})} | 
\tilde{b}_{t,k,m}^{(\mathrm{L})}, 
\underline{\tilde{\boldsymbol{w}}}_{t,k,m}^{(\mathrm{L})}, 
\underline{\hat{\boldsymbol{h}}}_{k,m}^{\mathcal{T}_{\tau}} ) 
p(\underline{\tilde{\boldsymbol{w}}}_{t,k,m}^{(\mathrm{L})} )
p(\tilde{b}_{t,k,m}^{(\mathrm{L})})
}
{
\int p(\underline{\boldsymbol{y}}_{t,k,m}^{(\mathrm{L})} | 
\tilde{b}_{t,k,m}^{(\mathrm{L})}, 
\underline{\tilde{\boldsymbol{w}}}_{t,k,m}^{(\mathrm{L})}, 
\underline{\hat{\boldsymbol{h}}}_{k,m}^{\mathcal{T}_{\tau}} ) 
p(\underline{\tilde{\boldsymbol{w}}}_{t,k,m}^{(\mathrm{L})})
p(\tilde{b}_{t,k,m}^{(\mathrm{L})})
d\underline{\tilde{\boldsymbol{w}}}_{t,k,m}^{(\mathrm{L})}
d\tilde{b}_{t,k,m}^{(\mathrm{L})}
}, 
\end{equation}
where $p(\underline{\boldsymbol{y}}_{t,k,m}^{(\mathrm{L})} | 
\tilde{b}_{t,k,m}^{(\mathrm{L})}, 
\underline{\tilde{\boldsymbol{w}}}_{t,k,m}^{(\mathrm{L})}, 
\underline{\hat{\boldsymbol{h}}}_{k,m}^{\mathcal{T}_{\tau}} )$ denotes 
the single-user SIMO channel~(\ref{postulated_LMMSE_SIMO}) postulated by 
the LMMSE detector.  
\end{definition}

\begin{proposition} \label{proposition5} 
Under the RS assumption, the spectral 
efficiency~(\ref{spectral_efficiency_LMMSE}) of the 
LMMSE receiver converges to the 
spectral efficiency of the LMMSE receiver for the single-user SIMO 
channel~(\ref{SIMO}) with in the large-system limit: 
\begin{equation} \label{spectral_efficiency_LMMSE_dec}  
\lim_{K,L\rightarrow\infty}C_{\mathrm{L}} = 
\beta M\left(
 1-\frac{\tau}{T_{\mathrm{c}}} 
\right)\underline{C}_{\tau+1}(\sigma_{\mathrm{tr}}^{2}(\tau), 
\sigma_{\mathrm{L}}^{2}),  
\end{equation} 
where $\underline{C}_{\tau+1}(\sigma_{\mathrm{tr}}^{2}(\tau), 
\sigma_{\mathrm{L}}^{2})$ is given by (\ref{decoupled_mutual_inf}). 
In evaluating (\ref{spectral_efficiency_LMMSE_dec}), 
$\sigma_{\mathrm{tr}}^{2}(\tau)$ for $t\in\mathcal{T}_{\tau}$ is 
given as the solution to the fixed-point equation~(\ref{fixed_point_channel}). 
On the other hand, $\sigma_{\mathrm{L}}^{2}$ satisfies 
the fixed-point equation
\begin{equation}
\sigma_{\mathrm{L}}^{2} = N_{0} 
+ \frac{\beta P\xi^{2}\sigma_{\mathrm{L}}^{2}}
 {(P/M)\xi^{2}+\sigma_{\mathrm{L}}^{2}} 
 + \frac{\beta M}{N}\left(
  \frac{\sigma_{\mathrm{L}}^{2}}{(P/M)\xi^{2}+
  \sigma_{\mathrm{L}}^{2}}
 \right)^{2}\mathbb{E}\left[
  \|\underline{\hat{\boldsymbol{h}}}_{k,m}^{\mathcal{T}_{\tau}}\|^{2}
  |b_{t,k,m} - \langle \tilde{b}_{t,k,m}^{(\mathrm{L})} 
  \rangle_{\mathrm{L}}|^{2}
 \right],  \label{fixed_point_LMMSE} 
\end{equation} 
with $\xi^{2}=\xi^{2}(\sigma_{\mathrm{tr}}^{2}(\tau),\tau)$ and 
$\langle \cdots \rangle_{\mathrm{L}}= 
\int \cdots p(\tilde{b}_{t,k,m}^{(\mathrm{L})}, 
\underline{\tilde{\boldsymbol{w}}}_{t,k,m}^{(\mathrm{L})} | 
\underline{\tilde{\boldsymbol{y}}}_{t,k,m}^{(\mathrm{L})}
=\underline{\boldsymbol{y}}_{t,k,m}, 
\underline{\mathcal{I}}_{\mathcal{T}_{\tau},k,m} )
d\tilde{b}_{t,k,m}^{(\mathrm{L})}
d\underline{\tilde{\boldsymbol{w}}}_{t,k,m}^{(\mathrm{L})}$.  
\end{proposition}
%\begin{IEEEproof}[Derivation of Proposition~\ref{proposition5}]
%See Appendix~\ref{derivation_proposition5}. 
%\end{IEEEproof}

Note that the LMMSE receiver can achieve the constrained capacity of the 
single-user SIMO channel~(\ref{SIMO}). 
The fixed-point equation~(\ref{fixed_point_LMMSE}) coincides with that 
in \cite{Evans00} for $M=N=1$. This implies that the result obtained by 
using the replica method is correct for the LMMSE receiver.  
The difference between the fixed-point equations~(\ref{fixed_point_data}) 
and~(\ref{fixed_point_LMMSE}) appears in the last terms of their respective 
right-hand sides. 
The last term of the right-hand side of (\ref{fixed_point_LMMSE}) 
corresponds to the mean-squared error (MSE) of the {\em linear} MMSE estimate 
of $b_{t,k,m}$ for the single-user SIMO channel~(\ref{SIMO_perfect}) with 
perfect CSI at the receiver, whereas the last term of the right-hand side of 
(\ref{fixed_point_data}) corresponds to the MSE of the minimum mean-squared 
error (MMSE) estimate of $b_{t,k,m}$ for the same SIMO channel.

\section{Numerical Results} \label{section_numerical_result} 
The spectral efficiency of the joint CE-MUDD is compared to 
the spectral efficiencies of the one-shot CE-MUDD, the optimum 
separated receiver, and the LMMSE receiver, on the basis of 
Propositions~\ref{proposition2}--\ref{proposition5}. 
See Table~\ref{table} for the differences between the four receivers. 
The performance gap between the joint CE-MUDD and the one-shot CE-MUDD  
corresponds to the gains obtained by using the decoded data symbols to refine 
the channel estimates. 
The gap between the one-shot CE-MUDD and the optimum separated receiver 
is related to the gains obtained by using the decoded data symbols to 
mitigate MAI. The performance gap between the optimum separated receiver and 
the LMMSE receiver corresponds to the gains obtained by performing the optimal 
MUD, instead of the LMMSE MUD. 
The spectral efficiencies of the three receivers based on the one-shot channel 
estimation are well-defined for $\tau\geq0$. 
For clarity, these spectral efficiencies are calculated  
for $\tau\geq0$, while the spectral efficiency~(\ref{spectral_efficiency_low}) 
of the joint CE-MUDD is evaluated for $\tau=0,1,\ldots,T_{\mathrm{c}}$. 
In all numerical results, unbiased QPSK input symbols are used. 

\begin{figure}[t]
\begin{center}
\includegraphics[width=\hsize]{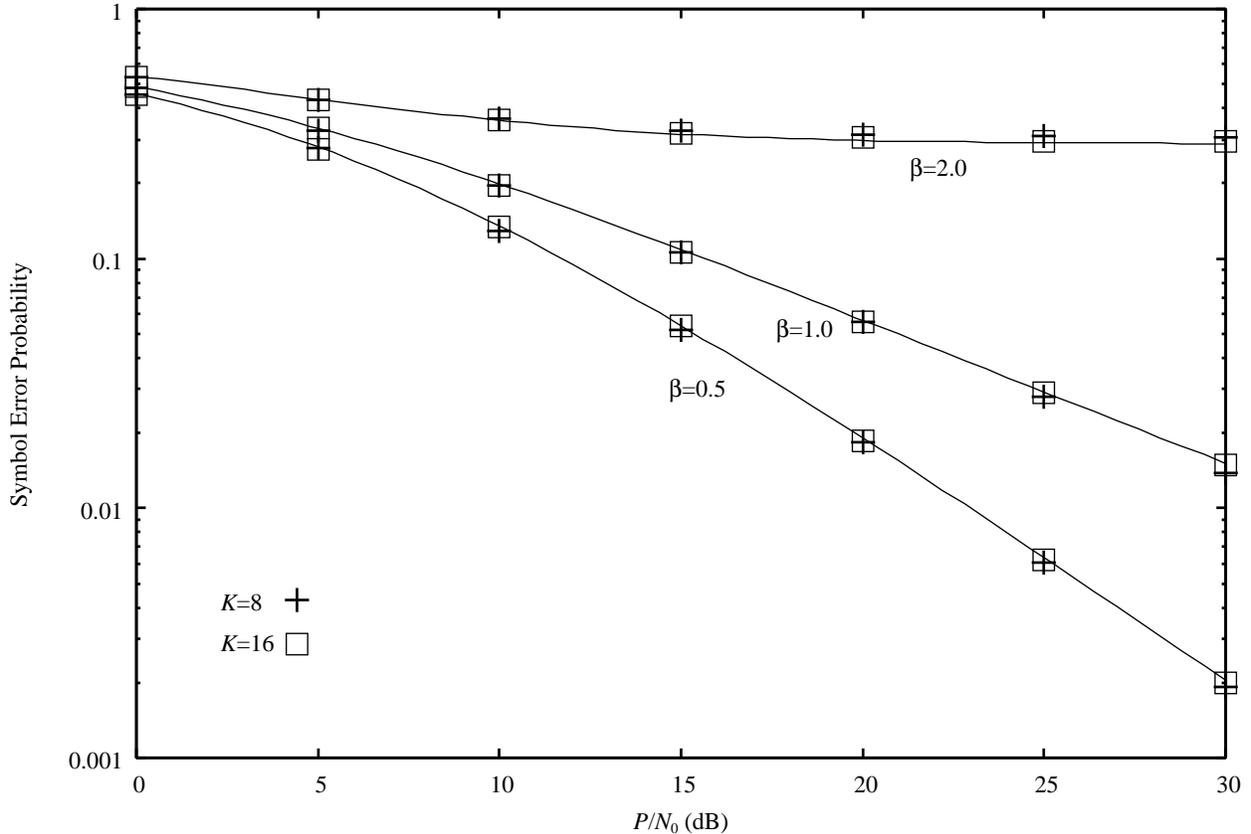}
\end{center}
\caption{
Symbol error probability of $b_{1,t,1}$ versus $P/N_{0}$ 
for the LMMSE receiver. 
From top to bottom, $\{+\}$ denote the symbol error probabilities of the 
LMMSE receiver for $(K,L)=(8,4), (8,8), (8,16)$.  
$\{\square\}$ represent the symbol error probabilities of 
the LMMSE receiver for $(K,L)=(16,8), (16,16), (16,32)$ from top 
to bottom. The solid lines represent the symbol error probabilities of the 
LMMSE receiver in the large-system limit for $\beta=2.0,1.0,0.5$ 
from top to bottom. $\tau=4$ and $M=N=1$. 
}
\label{fig6}
\end{figure}

We first present Monte Carlo simulation results for finite-sized systems 
and compare them with the analytical predictions derived in the large-system 
limit. In the Monte Carlo simulations we assume QPSK spreading, which 
is performed by using two mutually independent binary-antipodal random 
spreading sequences for in-phase and quadrature-phase channels. 
We only consider the LMMSE receiver since the computational 
complexity of the optimal detector is high.  
Figure~\ref{fig6} plots the symbol error probability of $b_{1,t,1}$  
for $M=N=1$ and $\tau=4$. 
The solid lines represent the symbol error probability of the 
LMMSE receiver for the single-user SIMO channel~(\ref{SIMO}) in which 
$\sigma_{t'}^{2}$ for $t'\in\mathcal{T}_{\tau}$ is given by the solution to  
the fixed-point equation~(\ref{fixed_point_channel}) and in which 
$\sigma_{t}^{2}$ is given as the solution to the fixed-point 
equation~(\ref{fixed_point_LMMSE}). 
We find that the analytical predictions are in agreement with the Monte Carlo 
simulation results for $K=16$, while they are slightly different from the 
simulation results for $K=8$. This result has two consequences: One is that 
the asymptotic results for the LMMSE receiver are applicable to a 
non-Gaussian distribution of $s_{l,t,k,m}$, as noted in \cite{Evans00}.  
The other is that the convergence of spectral efficiency to its asymptotic 
value is so fast that our analytical results, especially for the  
LMMSE receiver, provide reasonably good approximations of the true spectral 
efficiencies even for small-sized systems. 
The latter observation has also been made in \cite{Biglieri02,Wen07}. 

\begin{figure}[t]
\begin{center}
\includegraphics[width=\hsize]{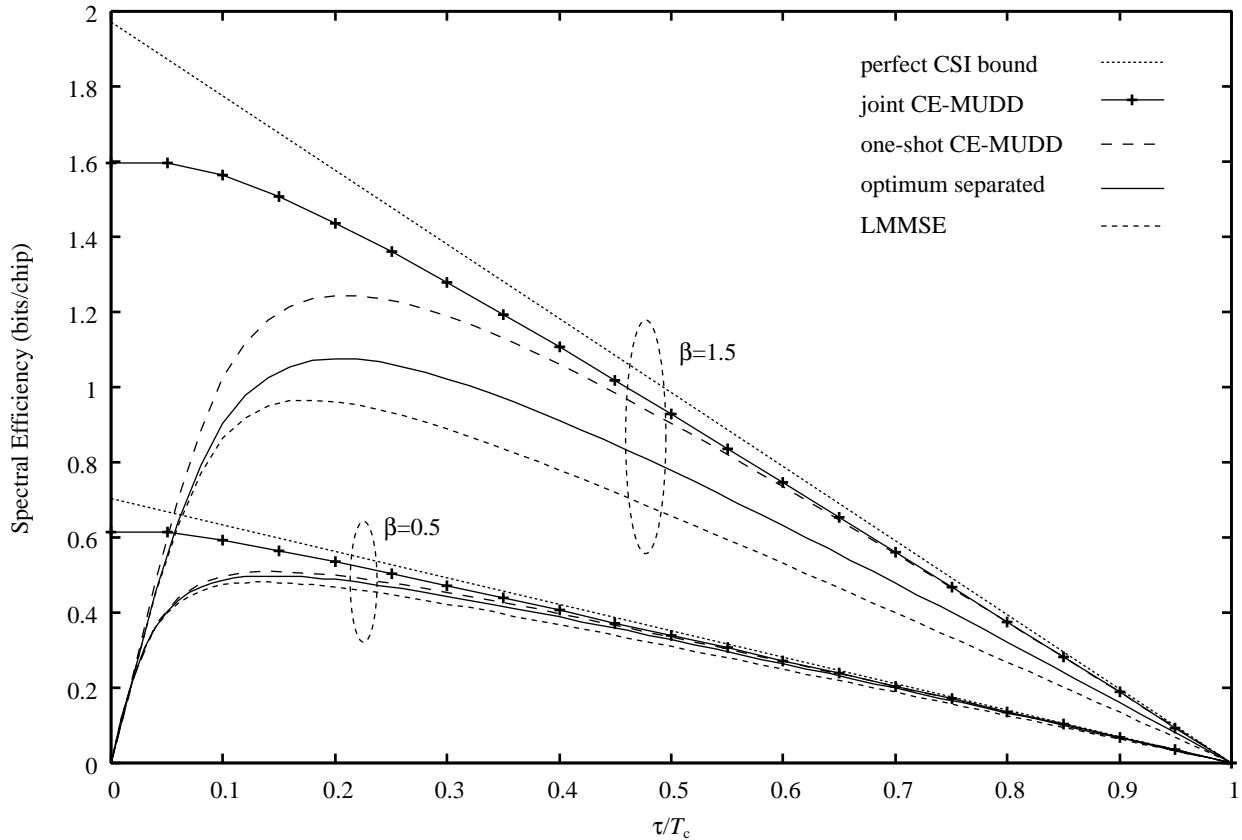} 
\end{center}
\caption{
Spectral efficiency versus $\tau/T_{\mathrm{c}}$ in the large-system limit. 
The pluses connected by straight lines display the spectral efficiency of 
the joint CE-MUDD. The dashed lines, the solid lines, and the dotted 
lines represent the spectral efficiencies of the one-shot 
CE-MUDD, the optimum separated receiver, and the LMMSE receiver, respectively. 
The fine dotted straight lines shows an upper bound of the constrained 
capacity~(\ref{spectral_efficiency}) based on the optimal receiver with 
perfect CSI. $P/N_{0}=6$~dB, $M=N=1$, and $T_{\mathrm{c}}=20$. 
}
\label{fig7} 
\end{figure}
 
We next focus on the asymptotic spectral efficiencies for single-antenna 
systems. Figure~\ref{fig7} displays the spectral 
efficiencies of the four receivers for $P/N_{0}=6$~dB. 
An upper bound $(1-\tau/T_{\mathrm{c}})C_{\mathrm{opt}}^{(\mathrm{per})}$ 
of the constrained capacity~(\ref{spectral_efficiency}) is also shown, which 
is based on the asymptotic spectral efficiency 
$C_{\mathrm{opt}}^{(\mathrm{per})}$ of the optimal receiver with 
perfect CSI \cite{Takeuchi082}.  
We find that the spectral efficiencies of the one-shot CE-MUDD and 
the two separated receivers are maximized at optimal 
$\tau=\tau_{\mathrm{opt}}$. This observation results from two effects: 
the improvement of the accuracy of the channel estimation and the decrease of 
the number of transmitted data symbols, both caused by the increase of $\tau$. 
The spectral efficiency~(\ref{spectral_efficiency_low}) 
of the joint CE-MUDD is maximized at $\tau=0$ and $\tau=1$. These 
results imply that the training overhead can be significantly reduced by 
using joint CE-MUDD. 

\begin{figure}[t]
\begin{center}
\includegraphics[width=\hsize]{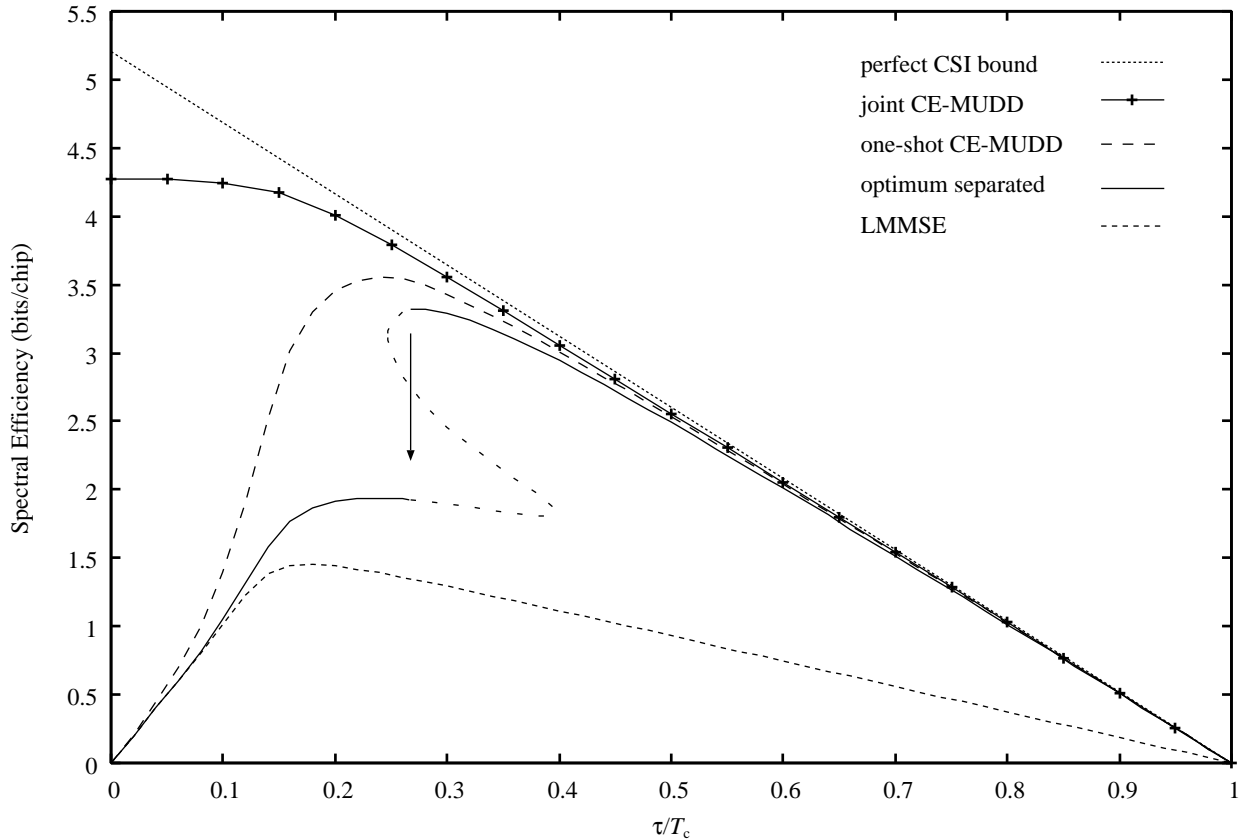} 
\end{center}
\caption{
Spectral efficiency versus $\tau/T_{\mathrm{c}}$ in the large-system limit. 
The pluses connected by straight lines display the spectral efficiency of 
the joint CE-MUDD. The dashed lines, the solid lines, and the dotted 
lines represent the spectral efficiencies of the one-shot CE-MUDD, 
the optimum separated receiver, and the LMMSE receiver, respectively. 
The fine dotted straight lines shows an upper bound of the constrained 
capacity~(\ref{spectral_efficiency}) based on the optimal receiver with 
perfect CSI. The widely-spaced dotted lines denotes solutions for the optimum 
separated receivers which are not selected by the 
criterion~(\ref{free_energy}). $\beta=2.75$, $P/N_{0}=15$~dB, $M=N=1$, and 
$T_{\mathrm{c}}=20$. 
}
\label{fig8} 
\end{figure}

The performance gaps between the one-shot CE-MUDD  
and the two separated receivers are negligibly small for $\beta=0.5$, 
while there is a noticeable gap between the spectral 
efficiencies of the joint CE-MUDD and the one-shot CE-MUDD. 
We found by numerical evaluation that 
$P/N_{0}\approx8.2$~dB is required for $\beta=0.5$ in order 
for the one-shot CE-MUDD to achieve the same spectral efficiency as that of 
the joint CE-MUDD for $P/N_{0}=6$~dB. In other words, 
the joint CE-MUDD provides a performance gain of $2.2$~dB. 
For $\beta=1.5$, the performance gaps between the optimal one-shot CE-MUDD, 
the optimum separated receiver, and the LMMSE receiver 
are large. This result implies that performance gains can be obtained 
by using an MUDD scheme with higher performance than the LMMSE receiver. 
Furthermore, there is a large performance gap between the joint 
CE-MUDD and the one-shot CE-MUDD. More precisely, 
we found that $P/N_{0}\approx8.5$~dB is needed 
for $\beta=1.5$ in order for the one-shot CE-MUDD to achieve the same 
spectral efficiency as that of the joint CE-MUDD for $P/N_{0}=6$~dB. 
These observations indicate that joint CE-MUDD can provide 
significant performance gains regardless of $\beta$, 
compared to one-shot CE-MUDD.  

Figure~\ref{fig8} shows the spectral efficiencies of the four receivers for 
$P/N_{0}=15$~dB and $\beta=2.75$. 
The upper bound $(1-\tau/T_{\mathrm{c}})C_{\mathrm{opt}}^{(\mathrm{per})}$ 
is also shown. 
One interesting observation is that the spectral efficiency of the 
optimum separated receiver is discontinuous. This result predicts that  
the spectral efficiency of the optimum separated receiver exhibits a 
waterfall behavior: The spectral efficiency rapidly degrades in the 
neighborhood of the discontinuous point 
$\tau=\tau_{\mathrm{c}}$, shown by the arrow, which corresponds to the 
threshold between interference-limited and non-limited regions. 
The system is not {\em interference-limited} for $\tau>\tau_{\mathrm{c}}$, 
i.e., the asymptotic multiuser efficiency $N_{0}/\sigma_{\mathrm{c}}^{2}$ is 
close to one. On the other hand, the system is {\em interference-limited} for 
$\tau<\tau_{\mathrm{c}}$, i.e., the asymptotic multiuser efficiency is small.

The spectral efficiency of the one-shot CE-MUDD seems to be continuous. 
This observation is explained as follows: Users decoded in the initial 
substages of successive decoding are interference-limited, while the remaining 
users are not interference-limited. 
Therefore, the achievable rate of each user changes discontinuously for the  
one-shot CE-MUDD. However, the achievable sum rate of the one-shot CE-MUDD 
changes continuously since the threshold $\kappa_{\mathrm{c}}$ between the two 
groups of interference-limited users and non-limited users 
should move continuously with the change of $\tau$. 
We remark that it might be possible to cancel out MAI successfully by 
optimizing the power allocation and the rate of each user, 
as discussed in \cite{Caire04}. 

\begin{figure}[t]
\begin{center}
\includegraphics[width=\hsize]{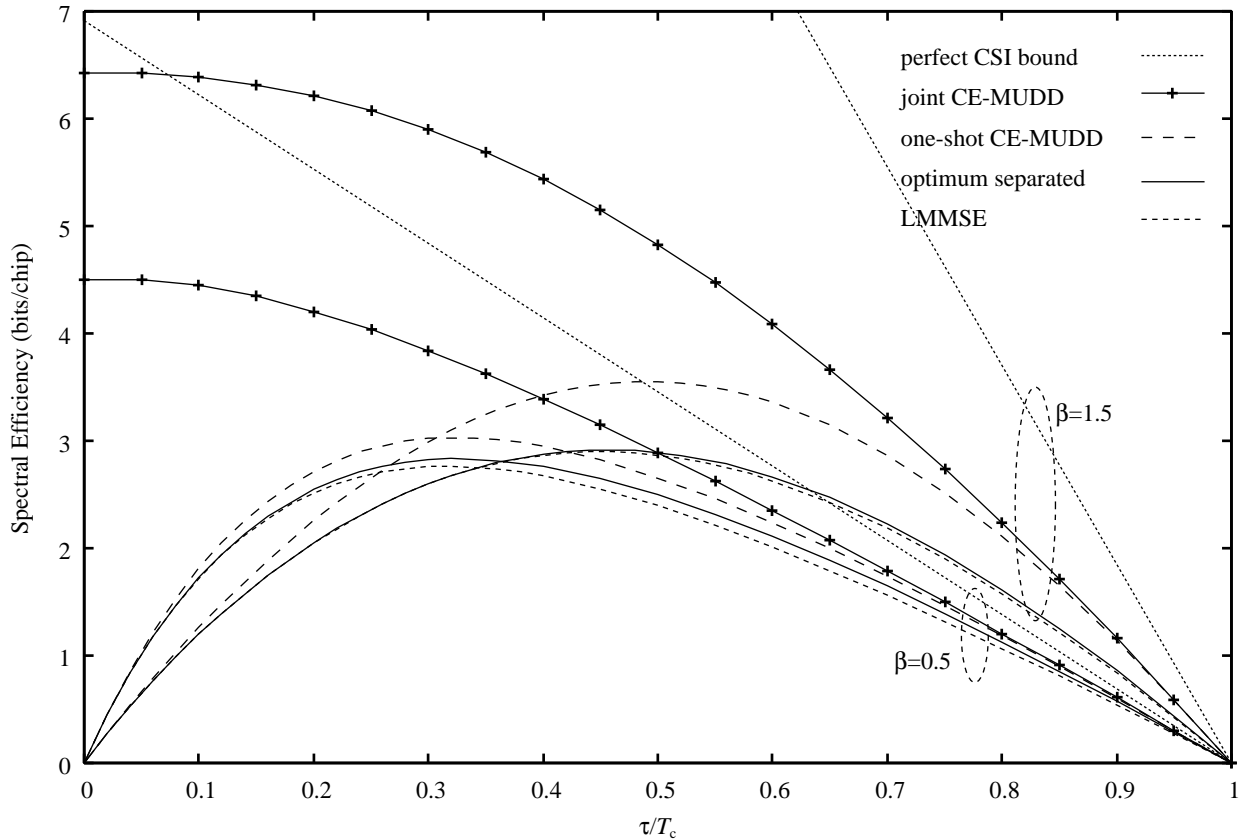} 
\end{center}
\caption{
Spectral efficiency versus $\tau/T_{\mathrm{c}}$ in the large-system limit. 
The pluses connected by straight lines display the spectral efficiency of 
the joint CE-MUDD. The dashed lines, the solid lines, and the dotted 
lines represent the spectral efficiencies of the one-shot 
CE-MUDD, the optimum separated receiver, and 
the LMMSE receiver, respectively. The fine dotted straight 
lines shows an upper bound of the constrained 
capacity~(\ref{spectral_efficiency}) based on the optimal receiver with 
perfect CSI. $P/N_{0}=6$~dB, $M=N=8$, and $T_{\mathrm{c}}=20$. 
}
\label{fig9} 
\end{figure}

Finally, we investigate multiple-antenna systems. 
Figure~\ref{fig9} shows the spectral efficiencies of the four receivers 
for $M=N=8$. 
The upper bound $(1-\tau/T_{\mathrm{c}})C_{\mathrm{opt}}^{(\mathrm{per})}$ 
is also shown. 
We find that the optimal number $\tau_{\mathrm{opt}}$ of pilot symbols is 
larger than that for $M=N=1$ (Compare Figs.~\ref{fig7} and~\ref{fig9}), 
since the number of unknown channel coefficients increases. On the other hand, 
the spectral efficiency of the joint CE-MUDD is maximized at $\tau=0$ and 
$\tau=1$ even for $M=N=8$.  
Consequently, the performance gap between the joint CE-MUDD and the 
one-shot CE-MUDD becomes larger than that for single-antenna systems. 
When $\beta=0.5$, the performance gap between the joint CE-MUDD and 
the one-shot CE-MUDD is approximately $1.47$~bits/chip for $M=N=8$ 
($0.18$~bits/chip per the number of antennas), while 
the performance gap is approximately $0.10$~bits/chip for $M=N=1$. 
We found by numerical evaluation that a performance gain of $1.47$~bits/chip 
corresponds to that of approximately $3.4$~dB.  
For $\beta=1.5$, the performance gain increases up to approximately 
$2.88$~bits/chip for $M=N=8$ ($0.36$~bits/chip per the number of 
antennas), while it is $0.35$~bits/chip for $M=N=1$. Interestingly, 
a performance gain of $2.88$~bits/chip corresponds to a gain of $7.2$~dB. 
These results imply that joint CE-MUDD can provide a significant 
performance gain for multiple-antenna systems. 

\begin{figure}[t]
\begin{center}
\includegraphics[width=\hsize]{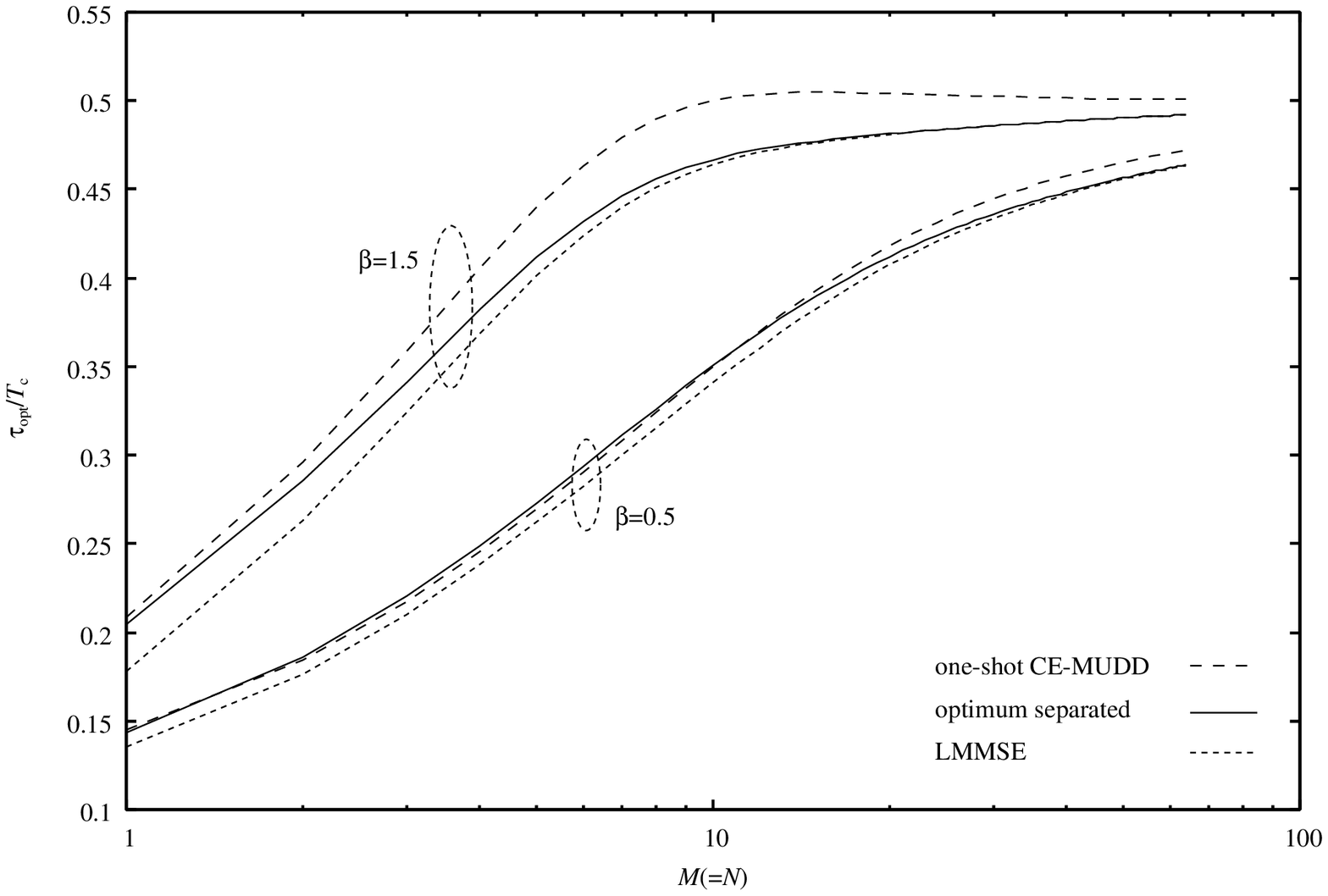} 
\end{center}
\caption{
$\tau_\mathrm{opt}/T_{\mathrm{c}}$ versus $M(=N)$ in the large-system 
limit. The dashed lines, the solid lines, and the dotted 
lines represent the optimal training overhead of the one-shot 
CE-MUDD, the optimum separated receiver, and 
the LMMSE receiver, respectively. 
$P/N_{0}=6$~dB and $T_{\mathrm{c}}=20$. 
}
\label{fig10} 
\end{figure}

Figure~\ref{fig10} shows the optimal training overhead 
$\tau_{\mathrm{opt}}/T_{\mathrm{c}}$ for $M=N$, maximizing the spectral 
efficiencies of the one-shot CE-MUDD and the two separated receivers. 
Note that the optimal training overhead for the joint CE-MUDD corresponds 
to $1/T_{\mathrm{c}}=0.05$. 
We find that the optimal training overheads increase logarithmically with the 
increase of the number of antennas for small $M$, while they tend towards  
around $0.5$ for large $M$. This observation implies that the performance 
gains obtained by using joint CE-MUDD increase as the number of antennas 
grows.

\section{Conclusions} \label{section_conclusion} 
We have analyzed the asymptotic performance of joint CE-MUDD for 
randomly-spread MIMO DS-CDMA systems with no CSI. The main contribution of 
this paper from a theoretical point of view is to derive a lower bound for the 
spectral efficiency of the optimal joint CE-MUDD on the basis of successive 
decoding with suboptimal LMMSE channel estimation, along with the spectral 
efficiencies of the one-shot CE-MUDD and the optimum separated receiver. 
The asymptotic performance of MIMO DS-CDMA systems with no CSI is 
characterized via the performance of a bank of single-user SIMO channels with 
no CSI. This decoupling result is an extension of previous 
studies~\cite{Guo051,Takeuchi082} for the case of perfect CSI at the receiver 
to the no-CSI case. 

The main contribution of this paper from a practical point of view is to 
demonstrate that joint CE-MUDD can significantly reduce the training overhead 
due to transmission of pilot signals, compared to receivers based on one-shot 
channel estimation. The results imply that joint CE-MUDD can provide 
significant performance gains for systems which require large training overhead 
for attaining accurate CSI. We conclude that the iterative refinement of 
channel estimates by utilizing soft feedback from per-user decoders may lead 
to substantial reduction of the rate loss due to transmission of pilot 
signals for MIMO DS-CDMA systems.

% use section* for acknowledgement
%\section*{Acknowledgment}

\appendices

\section{Derivations of LMMSE Estimators} \label{derivation_LMMSE} 
\subsection{Perfect-CSI Case} 
We shall present two derivations of the LMMSE estimator for vector channels 
with perfect CSI. One is based on the minimization of the MSE, and 
the other on Bayesian inference. 
Let us consider the $N\times K$ vector channel with perfect CSI 
\begin{equation} \label{imperfect_channel} 
\boldsymbol{y} = \boldsymbol{H}\boldsymbol{x} + \boldsymbol{n}
\in\mathbb{C}^{N}. 
\end{equation}
In (\ref{imperfect_channel}), $\boldsymbol{x}\in\mathbb{C}^{K}$ denotes a 
zero-mean transmitted vector with covariance matrix 
$\mathbb{E}[\boldsymbol{x}\boldsymbol{x}^{H}]=P\boldsymbol{I}_{K}$.
The noise vector $\boldsymbol{n}\in\mathbb{C}^{N}$ is a zero-mean 
random vector with covariance matrix $\boldsymbol{\Sigma}$. 
Furthermore, $\boldsymbol{H}\in\mathbb{C}^{N\times K}$ denotes a random  
channel matrix with mean $\bar{\boldsymbol{H}}\in\mathbb{C}^{N\times K}$. 
The three random variables $\boldsymbol{H}$, $\boldsymbol{x}$, and 
$\boldsymbol{n}$ are independent of each other. 

The goal is to find the LMMSE estimator, i.e., a linear estimator 
$\hat{\boldsymbol{x}}=\boldsymbol{F}_{\mathrm{per}}^{H}\boldsymbol{y}$ of 
the transmitted vector that minimizes the MSE 
$\mathbb{E}[\|\boldsymbol{x} - \hat{\boldsymbol{x}}\|^{2}|\boldsymbol{H}]$. 
Substituting $\hat{\boldsymbol{x}}=\boldsymbol{F}_{\mathrm{per}}^{H}
\boldsymbol{y}$ into the MSE 
$\mathbb{E}[\|\boldsymbol{x} - \hat{\boldsymbol{x}}\|^{2}
| \boldsymbol{H}]$ yields 
\begin{equation} \label{perfect_MSE} 
\mathbb{E}[\|\boldsymbol{x} - \hat{\boldsymbol{x}}\|^{2}
| \boldsymbol{H}] = 
P\|\boldsymbol{I}_{K} - \boldsymbol{F}_{\mathrm{per}}^{H}\boldsymbol{H}\|^{2} 
+ \mathrm{Tr}\left(
 \boldsymbol{F}_{\mathrm{per}}^{H}
 \boldsymbol{\Sigma}\boldsymbol{F}_{\mathrm{per}}
\right). 
\end{equation}
It is well known that $\boldsymbol{F}_{\mathrm{per}}$ to minimize the 
MSE~(\ref{perfect_MSE}) is given by 
\begin{equation} \label{filter} 
\boldsymbol{F}_{\mathrm{per}} = \boldsymbol{\Sigma}^{-1}\boldsymbol{H}
(P^{-1}\boldsymbol{I}_{K} + \boldsymbol{H}^{H}\boldsymbol{\Sigma}^{-1}
\boldsymbol{H})^{-1}. 
\end{equation}

In the derivation of (\ref{filter}), we have minimized the 
MSE~(\ref{perfect_MSE}) among all possible {\em linear} estimators 
$\hat{\boldsymbol{x}}=\boldsymbol{F}_{\mathrm{per}}^{H}\boldsymbol{y}$. The 
MSE~(\ref{perfect_MSE}) for the LMMSE estimator is larger than 
the MMSE in general, which is the minimum of the MSE for all possible 
estimators. However, it is known that the two MSEs coincides with each other 
when the transmitted vector $\boldsymbol{x}$ and the noise vector 
$\boldsymbol{n}$ are CSCG. This observation implies that the LMMSE estimator 
can be derived as the posterior mean estimator for a Gaussian vector channel. 
Let us define the vector channel postulated by the receiver as 
\begin{equation}
\tilde{\boldsymbol{y}}_{\mathrm{per}} = \boldsymbol{H}
\tilde{\boldsymbol{x}}_{\mathrm{per}} 
+ \tilde{\boldsymbol{n}}_{\mathrm{per}}, 
\end{equation}  
with $\tilde{\boldsymbol{x}}_{\mathrm{per}}\sim\mathcal{CN}(\boldsymbol{0},
P\boldsymbol{I}_{K})$ and $\tilde{\boldsymbol{n}}_{\mathrm{per}}
\sim\mathcal{CN}(\boldsymbol{0},\boldsymbol{\Sigma})$. 
It is straightforward to confirm that the posterior mean estimator 
$\mathbb{E}[\tilde{\boldsymbol{x}}_{\mathrm{per}} | 
\tilde{\boldsymbol{y}}_{\mathrm{per}}=\boldsymbol{y}, 
\boldsymbol{H}]$ for the postulated vector channel is equal to the 
LMMSE estimator. 

\subsection{No-CSI Case} 
We consider the no-CSI case: The channel matrix $\boldsymbol{H}$ is 
assumed to be unknown to the receiver, while the statistical properties of 
$\boldsymbol{x}$, $\boldsymbol{H}$, and $\boldsymbol{n}$ are known. 
The LMMSE estimator is defined as the linear estimator 
$\hat{\boldsymbol{x}}=\boldsymbol{F}^{H}\boldsymbol{y}$ to minimize 
the MSE $\mathbb{E}[\|\boldsymbol{x} - \hat{\boldsymbol{x}}\|^{2}]$ 
averaged over the channel matrix $\boldsymbol{H}$. 
Substituting $\hat{\boldsymbol{x}}=\boldsymbol{F}^{H}\boldsymbol{y}$ 
into this MSE gives 
\begin{equation}
\mathbb{E}[\|\boldsymbol{x} - \hat{\boldsymbol{x}}\|^{2}] 
= P\|\boldsymbol{I}_{K} - \boldsymbol{F}^{H}
\bar{\boldsymbol{H}}\|^{2} 
+ \mathrm{Tr}\left\{
 \boldsymbol{F}^{H}\left( 
  P\mathbb{E}[\Delta\boldsymbol{H}\Delta\boldsymbol{H}^{H}]
  + \boldsymbol{\Sigma}
 \right)\boldsymbol{F} 
\right\}, 
\end{equation}
with $\Delta\boldsymbol{H}=\boldsymbol{H} - \bar{\boldsymbol{H}}$. 
Comparing this expression with (\ref{perfect_MSE}), we find that the optimal 
filter $\boldsymbol{F}$ is given by 
\begin{equation}
\boldsymbol{F} = \boldsymbol{\Sigma}_{\mathrm{ef}}^{-1}\bar{\boldsymbol{H}}
(P^{-1}\boldsymbol{I}_{K} + \bar{\boldsymbol{H}}^{H}
\boldsymbol{\Sigma}_{\mathrm{ef}}^{-1}
\bar{\boldsymbol{H}})^{-1}, 
\end{equation} 
with $\boldsymbol{\Sigma}_{\mathrm{ef}} = 
P\mathbb{E}[\Delta\boldsymbol{H}\Delta\boldsymbol{H}^{H}]
+ \boldsymbol{\Sigma}$. 

In order to interpret the meaning of the effective covariance matrix 
$\boldsymbol{\Sigma}_{\mathrm{ef}}$, we re-write the vector 
channel~(\ref{imperfect_channel}) as 
\begin{equation}
\boldsymbol{y} = \bar{\boldsymbol{H}}\boldsymbol{x} + 
\Delta\boldsymbol{H}\boldsymbol{x} + \boldsymbol{n}.  
\end{equation} 
The effective covariance matrix $\boldsymbol{\Sigma}_{\mathrm{ef}}$ 
is equal to the covariance matrix of the channel estimator error plus the 
noise vector, i.e., $\Delta\boldsymbol{H}\boldsymbol{x} + \boldsymbol{n}$. 
This observation implies that the LMMSE estimator is equal to the posterior 
mean estimator $\mathbb{E}[\tilde{\boldsymbol{x}} | 
\tilde{\boldsymbol{y}}=\boldsymbol{y}]$ for the postulated vector channel, 
\begin{equation}
\tilde{\boldsymbol{y}} = \bar{\boldsymbol{H}}\tilde{\boldsymbol{x}} 
+ \tilde{\boldsymbol{w}} + \tilde{\boldsymbol{n}},  
\end{equation}
where $\tilde{\boldsymbol{x}}\in\mathbb{C}^{K}$, 
$\tilde{\boldsymbol{w}}\in\mathbb{C}^{N}$, and 
$\tilde{\boldsymbol{n}}\in\mathbb{C}^{N}$ are independent CSCG random 
vectors with covariance matrices $P\boldsymbol{I}_{K}$, 
$P\mathbb{E}[\Delta\boldsymbol{H}\Delta\boldsymbol{H}^{H}]$, 
and $\boldsymbol{\Sigma}$, respectively. If 
$\bar{\boldsymbol{H}}=\boldsymbol{O}$, the LMMSE estimator is 
obviously independent of the received vector $\boldsymbol{y}$. 

\section{Proof of Lemma~\ref{lemma_channel_estimation}} 
\label{derivation_lemma1} 
In order to prove 
Lemma~\ref{lemma_channel_estimation} we need the following lemma, which 
is derived by extending the replica analysis in \cite{Takeuchi082}. 

\begin{lemma} \label{lemma_channel_estimation_tmp}
Suppose that $\{\mathcal{A}_{k}\}$ are mutually disjoint subsets of 
$\{2,3,\ldots,n\}$ for $k\in\mathcal{K}$. Then, 
\begin{equation} \label{decoupled_measure}  
\lim_{K,L\rightarrow\infty}\mathbb{E}_{\mathcal{I}_{\mathcal{T}_{\tau}}}\left[
 p(\mathcal{H}_{\mathcal{K}}^{\{1\}} | \mathcal{I}_{\mathcal{T}_{\tau}}) 
 \prod_{k\in\mathcal{K}}\prod_{a\in\mathcal{A}_{k}}
 p(\mathcal{H}_{k}^{\{a\}} | \mathcal{I}_{\mathcal{T}_{\tau}})
\right] 
= \prod_{k\in\mathcal{K}}\prod_{m=1}^{M}  
\mathbb{E}_{\underline{\mathcal{I}}_{\mathcal{T}_{\tau},k,m}}\left[ 
 \prod_{a\in\{1\}\cup\mathcal{A}_{k}}
 p(\boldsymbol{h}_{k,m}=\boldsymbol{h}_{k,m}^{\{a\}} | 
 \underline{\mathcal{I}}_{\mathcal{T}_{\tau},k,m})
\right], 
\end{equation}
with $\mathcal{H}_{\mathcal{K}}^{\{a\}}=\{\mathcal{H}_{k}^{\{a\}}: 
k\in\mathcal{K}\}$ denoting the $a$th replicas of the channel vectors for 
the users in $\mathcal{K}$. 
In evaluating the right-hand side of (\ref{decoupled_measure}), 
$\sigma_{t}^{2}=\sigma_{\mathrm{tr}}^{2}(\tau)$ for $t\in\mathcal{T}_{\tau}$ is 
given by the solution to the fixed-point equation~(\ref{fixed_point_channel}). 
\end{lemma}
\begin{IEEEproof}[Derivation of Lemma~\ref{lemma_channel_estimation_tmp}]
See Appendix~\ref{derivation_lemma_channel_estimation}. 
\end{IEEEproof} 

Lemma~\ref{lemma_channel_estimation_tmp} implies that the information 
about $\{\boldsymbol{h}_{k,m}\}$ obtained by utilizing the information 
$\mathcal{I}_{\mathcal{T}_{\tau}}$ is mutually independent in the large-system 
limit for all $k\in\mathcal{K}$ and $m$, and that the information about each 
$\boldsymbol{h}_{k,m}$ looks like that provided by the LMMSE channel 
estimator for the single-user SIMO channel~(\ref{SIMO}).  
A result equivalent to Lemma~\ref{lemma_channel_estimation_tmp} was proved 
in \cite{Evans00} for $M=N=1$, by using random matrix theory. 
Lemma~\ref{lemma_channel_estimation_tmp} is used to evaluate the joint 
moment sequence of $\{X_{k}:k\in\mathcal{K}\}$, defined by (\ref{X_k}), 
in the following proof of Lemma~\ref{lemma_channel_estimation}.  

\begin{IEEEproof}[Proof of Lemma~\ref{lemma_channel_estimation}] 
It is sufficient to prove that the joint moment sequence of 
$\{X_{k}:k\in\mathcal{K}\}$ converges to that for the distribution defined 
by the right-hand side of (\ref{decoupling_channel}) in the large-system 
limit, since we have assumed the existence of the joint moment generating 
function. 

For non-negative integers $\{\tilde{n}_{k}\}$, 
a joint moment of $\{X_{k}:k\in\mathcal{K}\}$ is given by 
\begin{equation} \label{correlation}  
\mathbb{E}\left[
 \prod_{k\in\mathcal{K}}X_{k}^{\tilde{n}_{k}}
\right] = 
\mathbb{E}\left[
 \prod_{k\in\mathcal{K}}\prod_{i=1}^{\tilde{n}_{k}}\int 
 f_{k}(\mathcal{H}_{k}^{\{1\}},
 \mathcal{H}_{k}^{\mathcal{A}_{k}^{(i)}};\Theta)
 \prod_{a\in\mathcal{A}_{k}^{(i)}}\left\{
  p(\mathcal{H}_{k}^{\{a\}}|
  \mathcal{I}_{\mathcal{T}_{\tau}})d\mathcal{H}_{k}^{\{a\}} 
 \right\} 
\right], 
\end{equation}
where $\{\mathcal{A}_{k}^{(i)}: \hbox{for all $k\in\mathcal{K}$, $i$}\}$ are 
mutually disjoint subsets of $\{2,3,\ldots,n\}$ satisfying 
$|\mathcal{A}_{k}^{(i)}|=|\mathcal{A}_{k}|$. The right-hand side 
of (\ref{correlation}) is defined as the integration of 
$\prod_{k\in\mathcal{K}}\prod_{i=1}^{\tilde{n}_{k}}
f_{k}(\mathcal{H}_{k}^{\{1\}},
 \mathcal{H}_{k}^{\mathcal{A}_{k}^{(i)}};\Theta)$ with respect to 
the measure 
\begin{equation} \label{measure1} 
\mathbb{E}_{\mathcal{I}_{\mathcal{T}_{\tau}}}\left[ 
 p(\mathcal{H}_{\mathcal{K}}^{\{1\}} | \mathcal{I}_{\mathcal{T}_{\tau}})
 \prod_{k\in\mathcal{K}}\prod_{a\in\tilde{\mathcal{A}}_{k}} 
 p(\mathcal{H}_{k}^{\{a\}} | \mathcal{I}_{\mathcal{T}_{\tau}})  
\right]d\mathcal{H}_{\mathcal{K}}^{\{1\}} 
\prod_{k\in\mathcal{K}}\prod_{a\in\tilde{\mathcal{A}}_{k}}
d\mathcal{H}_{k}^{\{a\}}, 
\end{equation}
with $\tilde{\mathcal{A}}_{k}=\cup_{i=1}^{\tilde{n}_{k}}
\mathcal{A}_{k}^{(i)}$. 
Applying Lemma~\ref{lemma_channel_estimation_tmp} to 
(\ref{measure1}), we obtain  
\begin{equation}
\lim_{K,L\rightarrow\infty}\mathbb{E}\left[
 \prod_{k\in\mathcal{K}}X_{k}^{\tilde{n}_{k}}
\right] 
= \prod_{k\in\mathcal{K}}\mathbb{E}\left[
 \prod_{i=1}^{\tilde{n}_{k}}\int f_{k}(\mathcal{H}_{k}^{\{1\}}, 
 \mathcal{H}_{k}^{\mathcal{A}_{k}^{(i)}};\Theta) 
 \prod_{m=1}^{M}
 \prod_{a\in\mathcal{A}_{k}^{(i)}}\left\{
  p(\boldsymbol{h}_{k,m}=\boldsymbol{h}_{k,m}^{\{a\}} 
  | \underline{\mathcal{I}}_{\mathcal{T}_{\tau},k,m})  
  d\boldsymbol{h}_{k,m}^{\{a\}} 
 \right\}
\right], 
\end{equation}  
in the large-system limit, which is equal to the corresponding joint moment 
$\prod_{k\in\mathcal{K}}\mathbb{E}[\underline{X}_{k}^{\tilde{n}_{k}}]$ for 
the distribution defined as the right-hand side of (\ref{decoupling_channel}). 
\end{IEEEproof}

\section{Derivation of Lemma~\ref{lemma_channel_estimation_tmp}}  
\label{derivation_lemma_channel_estimation} 
\subsection{Replica Method} 
\label{tmp}
We present a brief introduction of the replica method. For details of the 
replica method, see \cite{Nishimori01,Fischer91,Mezard87}. 
Let $f(X,Y;N)>0$ denote a deterministic function of two random variables 
$X$ and $Y$ with a parameter $N$. Our goal is to evaluate the expectation 
$\mathbb{E}_{Y}[Z(Y;N)^{-1}]$ for the so-called partition 
function $Z(Y;N)=\mathbb{E}_{X}[f(X,Y;N)]$ in the limit $N\rightarrow\infty$. 
For that purpose, we first evaluate $\lim_{N\rightarrow\infty}
\mathbb{E}_{Y}[Z(Y;N)^{\tilde{n}-1}]$ for any natural number 
$\tilde{n}\in\mathbb{N}$, 
by utilizing the following expression 
\begin{equation} \label{special_expression} 
\mathbb{E}_{Y}\left[
 Z(Y;N)^{\tilde{n}-1}
\right] = \mathbb{E}\left[
 \prod_{a=1}^{\tilde{n}-1}f(X_{a},Y;N) 
\right], 
\end{equation} 
where $\{X_{a}:a=1,\ldots,\tilde{n}-1\}$ are 
i.i.d.\ replicated random variables following $p(X)$. Suppose that 
the analytical expression obtained via (\ref{special_expression}) is 
well-defined even for $\tilde{n}\in\mathbb{R}$.  
We take $\tilde{n}\rightarrow+0$ to obtain an analytical expression of 
$\lim_{N\rightarrow\infty}\mathbb{E}_{Y}[Z(Y;N)^{-1}]$, assuming that the 
obtained expression coincides with the correct one. 
It is a challenging problem to prove whether this assumption holds.  

\subsection{Formulation} 
In order to show Lemma~\ref{lemma_channel_estimation_tmp} by using the 
replica method, we transform the left-hand side of (\ref{decoupled_measure}) 
into a formula corresponding to $\mathbb{E}_{Y}[Z(Y;N)^{\tilde{n}-1}]$. 
The posterior pdf of the replicated channel vectors 
$\mathcal{H}_{\mathcal{K}}^{\{a\}}$ for the users in $\mathcal{K}$, 
defined in the same manner as in (\ref{posterior_pdf_Hk}), 
is given by  
\begin{equation} \label{posterior_pdf_Hk_appen} 
p(\mathcal{H}_{\mathcal{K}}^{\{a\}} | \mathcal{I}_{\mathcal{T}_{\tau}}) = 
\frac{
 \int \prod_{t=1}^{\tau}p(\mathcal{Y}_{t} | \mathcal{H}^{\{a\}}, 
 \mathcal{S}_{t}, \mathcal{U}_{t})p(\mathcal{H}^{\{a\}})
 d\mathcal{H}_{\backslash\mathcal{K}}^{\{a\}}  
}
{
 \int \prod_{t=1}^{\tau}p(\mathcal{Y}_{t} | \mathcal{H}, \mathcal{S}_{t}, 
 \mathcal{U}_{t}) p(\mathcal{H})d\mathcal{H}  
}, 
\end{equation}
with $\mathcal{H}^{\{a\}}=\{\mathcal{H}_{k}^{\{a\}}:
\hbox{for all $k$}\}$ and $\mathcal{H}_{\backslash\mathcal{K}}^{\{a\}}=\{
\mathcal{H}_{k}^{\{a\}}: \hbox{for all $k\notin\mathcal{K}$}\}$. 
In (\ref{posterior_pdf_Hk_appen}), the pdf 
$p(\mathcal{Y}_{t} | \mathcal{H}^{\{a\}}, \mathcal{S}_{t}, \mathcal{U}_{t})$ 
is an abbreviation of $p(\mathcal{Y}_{t} | \mathcal{H}=\mathcal{H}^{\{a\}}, 
 \mathcal{S}_{t}, \mathcal{U}_{t})$.  
Let $\mathcal{H}_{\mathcal{K}}^{\mathcal{A}}=
\{\mathcal{H}_{k}^{\mathcal{A}_{k}}:k\in\mathcal{K}\}$ denote the replicas of 
the channel vectors for $\mathcal{A}=\cup_{k\in\mathcal{K}}\mathcal{A}_{k}$, 
with $\mathcal{H}_{k}^{\mathcal{A}_{k}}
=\{\mathcal{H}_{k}^{\{a\}}:a\in\mathcal{A}_{k}\}$.  
Applying (\ref{posterior_pdf_Hk_appen}) to the left-hand side of 
(\ref{decoupled_measure}) and subsequently introducing a real number 
$\tilde{n}\in\mathbb{R}$, we obtain  
\begin{equation} \label{conditional_distribution_channel} 
\lim_{K,L\rightarrow\infty}
\mathbb{E}_{\mathcal{I}_{\mathcal{T}_{\tau}}}\left[
 p(\mathcal{H}_{\mathcal{K}}^{\{1\}} | \mathcal{I}_{\mathcal{T}_{\tau}}) 
 \prod_{k\in\mathcal{K}}\prod_{a\in\mathcal{A}_{k}}
 p(\mathcal{H}_{k}^{\{a\}} | \mathcal{I}_{\mathcal{T}_{\tau}})
\right] = 
\lim_{K,L\rightarrow\infty}\lim_{\tilde{n}\rightarrow+0}
\Xi_{\tilde{n}}(\mathcal{H}_{\mathcal{K}}^{\{1\}},
\mathcal{H}_{\mathcal{K}}^{\mathcal{A}}), 
\end{equation}
where $\Xi_{\tilde{n}}(\mathcal{H}_{\mathcal{K}}^{\{1\}},
\mathcal{H}_{\mathcal{K}}^{\mathcal{A}})$ is given by 
\begin{IEEEeqnarray}{r}
\Xi_{\tilde{n}}(\mathcal{H}_{\mathcal{K}}^{\{1\}}, 
\mathcal{H}_{\mathcal{K}}^{\mathcal{A}}) = 
\mathbb{E}\left[ 
 \int \left\{
  \int \prod_{t=1}^{\tau}p(\mathcal{Y}_{t} | \mathcal{H}, 
  \mathcal{S}_{t}, \mathcal{U}_{t})p(\mathcal{H})d\mathcal{H}
 \right\}^{\tilde{n}-1-|\mathcal{A}|}
\right. \nonumber \\ 
\left. 
 \left. 
  \times \prod_{a\in\{0,1\}\cup\mathcal{A}}
  \left\{
   \prod_{t=1}^{\tau} 
   p(\mathcal{Y}_{t} | \mathcal{H}^{\{a\}}, \mathcal{S}_{t}, \mathcal{U}_{t})
   p(\mathcal{H}^{\{a\}})
  \right\} 
  \prod_{t=1}^{\tau}d\mathcal{Y}_{t}d\mathcal{H}^{\{0\}}   
  d\mathcal{H}_{\backslash\mathcal{K}}^{\{1\}}  
  d\backslash\mathcal{H}_{\mathcal{K}}^{\mathcal{A}} 
 \right| \mathcal{H}_{\mathcal{K}}^{\{1\}}, 
 \mathcal{H}_{\mathcal{K}}^{\mathcal{A}}
\right]. \label{Xi_channel}  
\end{IEEEeqnarray}  
In (\ref{Xi_channel}), we have written $\mathcal{H}$ as $\mathcal{H}^{\{0\}}$. 
Furthermore, the set $\backslash\mathcal{H}_{\mathcal{K}}^{\mathcal{A}}$ 
denotes all replicas of the channel vectors for $a\geq2$ except for 
$\mathcal{H}_{\mathcal{K}}^{\mathcal{A}}$. 
The expression~(\ref{Xi_channel}) implies that 
$\prod_{t=1}^{\tau}p(\mathcal{Y}_{t} | \mathcal{H}, 
\mathcal{S}_{t}, \mathcal{U}_{t})$ corresponds to the function 
$f(X,Y;N)$ with $X=\mathcal{H}$. 

\subsection{Average over Quenched Randomness} 
We evaluate (\ref{Xi_channel}) up to $O(1)$ in the large-system limit 
for any natural number $\tilde{n}$, satisfying 
$\mathcal{A}\subset\{1,2,\ldots,\tilde{n}\}$. 
We first calculate the expectations in (\ref{Xi_channel}) with respect to 
$\{\mathcal{Y}_{t}\}$ and $\{\mathcal{S}_{t}\}$. 
For $\tilde{n}\in\mathbb{N}$, we have a special expression of 
(\ref{Xi_channel}) 
\begin{equation} \label{Xi_channel_tmp1} 
\Xi_{\tilde{n}}(\mathcal{H}_{\mathcal{K}}^{\{1\}},
\mathcal{H}_{\mathcal{K}}^{\mathcal{A}}) = 
\mathbb{E}\left[
 \left. 
  \int \prod_{a=0}^{\tilde{n}}\left\{
   \prod_{t=1}^{\tau}p(\mathcal{Y}_{t} | \mathcal{H}^{\{a\}}, 
   \mathcal{S}_{t}, \mathcal{U}_{t})p(\mathcal{H}^{\{a\}})
  \right\} \prod_{t=1}^{\tau}d\mathcal{Y}_{t}d\mathcal{H}^{\{0\}}
 d\mathcal{H}_{\backslash\mathcal{K}}^{\{1\}}
 d\backslash\mathcal{H}_{\mathcal{K}}^{\mathcal{A}}
 \right| \mathcal{H}_{\mathcal{K}}^{\{1\}}, 
 \mathcal{H}_{\mathcal{K}}^{\mathcal{A}}
\right]. 
\end{equation}
Let us re-write the MIMO DS-CDMA channel~(\ref{MIMO_DS_CDMA}) in the 
training phase as 
\begin{equation} \label{vector_channel_tr} 
\boldsymbol{y}_{l} = \frac{1}{\sqrt{L}}
\sum_{k=1}^{K}\sum_{t=1}^{\tau}\sum_{m=1}^{M} 
(\boldsymbol{e}_{\tau}^{(t)}\otimes\boldsymbol{h}_{k,m})
s_{l,t,k,m}x_{t,k,m} + \boldsymbol{n}_{l}, 
\end{equation}
with $\boldsymbol{y}_{l}=(\boldsymbol{y}_{l,1}^{T},\ldots,
\boldsymbol{y}_{l,\tau}^{T})^{T}\in\mathbb{C}^{N\tau}$ and 
$\boldsymbol{n}_{l}=(\boldsymbol{n}_{l,1}^{T},\ldots,
\boldsymbol{n}_{l,\tau}^{T})^{T}\in\mathbb{C}^{N\tau}$. 
By evaluating the expectation in (\ref{Xi_channel_tmp1}) with respect to 
$\{\mathcal{S}_{t}\}$, from the independency of 
$\{s_{l,t,k,m}\}$ for all $l$, (\ref{Xi_channel_tmp1}) yields 
\begin{equation} \label{Xi_channel_tmp2} 
\Xi_{\tilde{n}}(\mathcal{H}_{\mathcal{K}}^{\{1\}},
\mathcal{H}_{\mathcal{K}}^{\mathcal{A}}) =  
p(\mathcal{H}_{\mathcal{K}}^{\{1\}})p(\mathcal{H}_{\mathcal{K}}^{\mathcal{A}})
\mathbb{E}\left[
 \left. 
  \left\{ 
   \mathbb{E}\left[
    \int\prod_{a=0}^{\tilde{n}}\left\{
     \frac{1}{(\pi N_{0})^{N\tau}}\mathrm{e}^{
      -\frac{1}{N_{0}}\|\boldsymbol{y}_{1} 
      - \sqrt{\beta}\boldsymbol{v}^{\{a\}}\|^{2}
     }d\boldsymbol{y}_{1}
    \right\}
   \right]
  \right\}^{L} 
 \right| \mathcal{H}_{\mathcal{K}}^{\{1\}}, 
 \mathcal{H}_{\mathcal{K}}^{\mathcal{A}}
\right], 
\end{equation} 
where the inner expectation is taken over 
$\{s_{1,t,k,m}:\hbox{for all $t$, $k$, $m$}\}$. In (\ref{Xi_channel_tmp2}), 
$\boldsymbol{v}^{\{a\}}\in\mathbb{C}^{N\tau}$ is defined as 
\begin{equation}
\boldsymbol{v}^{\{a\}} = 
\frac{1}{\sqrt{K}}\sum_{k=1}^{K}\sum_{t=1}^{\tau}\sum_{m=1}^{M}
s_{1,t,k,m}\boldsymbol{\omega}_{t,k,m}^{\{a\}}, 
\end{equation} 
where $\boldsymbol{\omega}_{t,k,m}^{\{a\}}\in\mathbb{C}^{N\tau}$ is given by 
\begin{equation} \label{omega} 
\boldsymbol{\omega}_{t,k,m}^{\{a\}}=
(\boldsymbol{e}_{\tau}^{(t)}\otimes\boldsymbol{h}_{k,m}^{\{a\}})x_{t,k,m}, 
\end{equation}
with $\boldsymbol{h}_{k,m}^{\{0\}}=\boldsymbol{h}_{k,m}$. 
Let us define $\boldsymbol{v}\in\mathbb{C}^{(\tilde{n}+1)N\tau}$ as 
$\boldsymbol{v}=((\boldsymbol{v}^{\{0\}})^{T},
\ldots,(\boldsymbol{v}^{\{\tilde{n}\}})^{T})^{T}$. 
We perform the Gaussian integration in (\ref{Xi_channel_tmp2}) 
with respect to $\boldsymbol{y}_{1}$ to obtain 
\begin{equation} \label{Xi_channel_tmp3} 
\Xi_{\tilde{n}}(\mathcal{H}_{\mathcal{K}}^{\{1\}},
\mathcal{H}_{\mathcal{K}}^{\mathcal{A}}) =  
p(\mathcal{H}_{\mathcal{K}}^{\{1\}})p(\mathcal{H}_{\mathcal{K}}^{\mathcal{A}})
\mathbb{E}\left[
 \left. 
  \left\{
   \frac{
    \mathbb{E}\left[ 
     \mathrm{e}^{-\boldsymbol{v}^{H}\boldsymbol{A}_{\tau}(\tilde{n})
     \boldsymbol{v}}
    \right]
   }{(\pi N_{0})^{\tilde{n}N\tau}(1+\tilde{n})^{N\tau}}
  \right\}^{L}
 \right| \mathcal{H}_{\mathcal{K}}^{\{1\}}, 
 \mathcal{H}_{\mathcal{K}}^{\mathcal{A}}
\right], 
\end{equation} 
with 
\begin{equation} \label{A}
\boldsymbol{A}_{\tau}(\tilde{n}) = 
\frac{\beta}{N_{0}+\tilde{n}N_{0}}
\begin{pmatrix}
\tilde{n} & -\boldsymbol{1}_{\tilde{n}}^{T} \\ 
-\boldsymbol{1}_{\tilde{n}} & (1 + \tilde{n})\boldsymbol{I}_{\tilde{n}} 
- \boldsymbol{1}_{\tilde{n}}\boldsymbol{1}_{\tilde{n}}^{T}  
\end{pmatrix}
\otimes \boldsymbol{I}_{N\tau}.  
\end{equation}
The CSCG assumption of $\{s_{l,t,k,m}\}$ implies that 
$\boldsymbol{v}$ in (\ref{Xi_channel_tmp3}) conditioned on 
$\{\mathcal{H}^{\{a\}}: \hbox{for all $a$}\}$ 
and $\mathcal{X}$ is a CSCG random vector with the covariance matrix 
\begin{equation} \label{Q_tr}
\boldsymbol{\mathcal{Q}} = 
\frac{1}{K}\sum_{k=1}^{K}\sum_{t=1}^{\tau}\sum_{m=1}^{M}
\boldsymbol{\omega}_{k,t,m}\boldsymbol{\omega}_{k,t,m}^{H}, 
\end{equation}
with  $\boldsymbol{\omega}_{k,t,m} 
=((\boldsymbol{\omega}_{k,t,m}^{\{0\}})^{T},\ldots,  
(\boldsymbol{\omega}_{k,t,m}^{\{\tilde{n}\}})^{T})^{T}$. 
Let us assume that $\boldsymbol{\mathcal{Q}}$ is non-singular. Then, 
(\ref{Xi_channel_tmp3}) yields  
\begin{equation} \label{Xi_channel_tmp4}
\Xi_{\tilde{n}}(\mathcal{H}_{\mathcal{K}}^{\{1\}},
\mathcal{H}_{\mathcal{K}}^{\mathcal{A}}) = 
p(\mathcal{H}_{\mathcal{K}}^{\{1\}})p(\mathcal{H}_{\mathcal{K}}^{\mathcal{A}})
\mathbb{E}\left[
 \left. 
  \mathrm{e}^{ LG_{\tau}(\boldsymbol{\mathcal{Q}};\tilde{n}) } 
 \right| \mathcal{H}_{\mathcal{K}}^{\{1\}}, 
 \mathcal{H}_{\mathcal{K}}^{\mathcal{A}} 
\right], 
\end{equation}  
with 
\begin{equation} \label{G}
G_{\tau}(\boldsymbol{\mathcal{Q}};\tilde{n}) = 
-\ln\det(\boldsymbol{I} + \boldsymbol{A}_{\tau}(\tilde{n})  
\boldsymbol{\mathcal{Q}}) - \tilde{n}N\tau\ln(\pi N_{0}) 
- N\tau\ln\left(
 1 + \tilde{n}
\right). 
\end{equation}

Note that (\ref{Xi_channel_tmp4}) holds for any $K$ and $L$. 
Without the assumption of $s_{l,t,k,m}\sim\mathcal{CN}(0,1)$, it would be 
necessary to evaluate $\ln\mathbb{E}[\exp(- 
\boldsymbol{v}^{H}\boldsymbol{A}_{\tau}(\tilde{n})\boldsymbol{v})]$ up to 
$O(K^{-1})$ in the large-system limit by using the Edgeworth expansion, 
since (\ref{decoupled_measure}) is a quantity of $O(1)$. The vector 
$\boldsymbol{v}$ conditioned on $\{\mathcal{H}^{\{a\}}: \hbox{for all $a$}\}$ 
and $\mathcal{X}$ converges in law to a CSCG random vector with the covariance 
matrix~(\ref{Q_tr}) in the large-system limit 
if one assumes that $\{\mathrm{Re}[s_{l,t,k,m}],\ 
\mathrm{Im}[s_{l,t,k,m}]\}$ are i.i.d.\ zero-mean random variables with 
variance $1/2$ and finite moments for all $l$, $t$, $k$, and $m$. Therefore, 
the expansion coefficient for $O(1)$ should coincide with (\ref{G}). 
Furthermore, it was shown that the expansion coefficient for $O(K^{-1})$ 
does not affect the result in the large-system limit 
for conventional DS-CDMA channels~\cite{Nakamura08}. Therefore, it is 
expected that the results in this paper holds for a general distribution 
of $s_{l,t,k,m}$. 

\subsection{Average over Replicated Randomness} 
The conditional expectation in (\ref{Xi_channel_tmp4}) is given as  
the integration of $\exp[LG_{\tau}(\boldsymbol{\mathcal{Q}};\tilde{n})]$ with 
respect to the measure $\mu_{K}(\boldsymbol{\mathcal{Q}};\tilde{n})
d\boldsymbol{\mathcal{Q}}$ on the space $\mathcal{M}_{(\tilde{n}+1)N\tau}^{+}$ 
of all $(\tilde{n}+1)N\tau\times(\tilde{n}+1)N\tau$ positive definite 
Hermitian matrices: 
\begin{equation} \label{Xi_channel_tmp5} 
\Xi_{\tilde{n}}(\mathcal{H}_{\mathcal{K}}^{\{1\}},
\mathcal{H}_{\mathcal{K}}^{\mathcal{A}}) = 
p(\mathcal{H}_{\mathcal{K}}^{\{1\}})p(\mathcal{H}_{\mathcal{K}}^{\mathcal{A}}) 
\int_{\mathcal{M}_{(\tilde{n}+1)N\tau}^{+}}\mathrm{e}^{ 
 LG_{\tau}(\boldsymbol{\mathcal{Q}};\tilde{n}) 
}\mu_{K}(\boldsymbol{\mathcal{Q}};\tilde{n})
d\boldsymbol{\mathcal{Q}}, 
\end{equation}
where $\mu_{K}(\boldsymbol{\mathcal{Q}};\tilde{n})$ denotes the pdf of 
$\boldsymbol{\mathcal{Q}}$ conditioned on 
$\mathcal{H}_{\mathcal{K}}^{\{1\}}$ and 
$\mathcal{H}_{\mathcal{K}}^{\mathcal{A}}$, induced from (\ref{Q_tr}). 
We shall calculate $\mu_{K}(\boldsymbol{\mathcal{Q}};\tilde{n})$ up to 
$O(1)$ since our goal is to evaluate (\ref{Xi_channel}) up to $O(1)$ in the 
large-system limit. 

We first evaluate the the moment generating function of (\ref{Q_tr}), 
defined as 
\begin{equation} \label{moment_generating}
M_{K}(\tilde{\boldsymbol{\mathcal{Q}}};\tilde{n}) =   
\mathbb{E}\left[
 \left. 
 \mathrm{e}^{K\mathrm{Tr}(\boldsymbol{\mathcal{Q}}
 \tilde{\boldsymbol{\mathcal{Q}}})}
 \right| \mathcal{H}_{\mathcal{K}}^{\{1\}}, 
 \mathcal{H}_{\mathcal{K}}^{\mathcal{A}}
\right], 
\end{equation}
where an $(\tilde{n}+1)N\tau\times (\tilde{n}+1)N\tau$ non-singular Hermitian 
matrix $\tilde{\boldsymbol{\mathcal{Q}}}$ is given by 
\begin{equation} \label{tilde_Q} 
\tilde{\boldsymbol{\mathcal{Q}}} = 
\begin{pmatrix}
\tilde{\boldsymbol{Q}}_{0,0} & \frac{1}{2}\tilde{\boldsymbol{Q}}_{0,1} & 
\cdots & \frac{1}{2}\tilde{\boldsymbol{Q}}_{0,\tilde{n}} \\ 
\frac{1}{2}\tilde{\boldsymbol{Q}}_{0,1}^{H} & \ddots & \ddots & \vdots \\ 
\vdots & \ddots & \ddots & 
\frac{1}{2}\tilde{\boldsymbol{Q}}_{\tilde{n}-1,\tilde{n}} \\
\frac{1}{2}\tilde{\boldsymbol{Q}}_{0,\tilde{n}}^{H} & \cdots & 
\frac{1}{2}\tilde{\boldsymbol{Q}}_{\tilde{n}-1,\tilde{n}}^{H} & 
\tilde{\boldsymbol{Q}}_{\tilde{n},\tilde{n}} 
\end{pmatrix}, 
\end{equation}
with $N\tau\times N\tau$ complex matrices 
$\{\tilde{\boldsymbol{Q}}_{a,b}: 0\leq a<b\leq \tilde{n}\}$, in which the 
$(j,\ k)$-element of $\tilde{\boldsymbol{Q}}_{a,b}$ is given as 
$\tilde{q}_{a,b}^{j,k}\in\mathbb{C}$, and with 
$N\tau\times N\tau$ Hermitian matrices 
\begin{equation}
\tilde{\boldsymbol{Q}}_{a,a} =  
\begin{pmatrix}
\tilde{q}^{1,1}_{a,a} & \frac{1}{2}\tilde{q}^{1,2}_{a,a} & 
\cdots & \frac{1}{2}\tilde{q}^{1,N\tau}_{a,a} \\ 
\frac{1}{2}(\tilde{q}^{1,2}_{a,a})^{*} & \ddots & \ddots & \vdots \\ 
\vdots & \ddots & \ddots & \frac{1}{2}\tilde{q}^{N\tau-1,N\tau}_{a,a} \\
\frac{1}{2}(\tilde{q}^{1,N\tau}_{a,a})^{*} & \cdots & 
\frac{1}{2}(\tilde{q}^{N\tau-1,N\tau}_{a,a})^{*} & 
\tilde{q}^{N\tau,N\tau}_{a,a} 
\end{pmatrix} 
\quad \hbox{for all $a$.} 
\end{equation}
Since $\{\mathcal{X}_{k}=\{x_{t,k,m}:\hbox{for all $t$, $m$}\}\}$ and 
$\{\mathcal{H}_{k}^{\{a\}}:a=0,\ldots,\tilde{n}\}$ are mutually 
independent for all $k$, substituting (\ref{Q_tr}) into 
(\ref{moment_generating}) implies that (\ref{moment_generating}) is 
decomposed into the product of $K$ terms:  
\begin{equation} \label{moment_generating_K_tr} 
M_{K}(\tilde{\boldsymbol{\mathcal{Q}}};\tilde{n}) = 
\prod_{k\in\mathcal{K}}M_{k} 
(\tilde{\boldsymbol{\mathcal{Q}}};\tilde{n})\prod_{k\notin \mathcal{K}}
\mathbb{E}\left[ 
 \mathrm{e}^{
  \Lambda_{k}(\tilde{\boldsymbol{\mathcal{Q}}};\tilde{n})
 }
\right], 
\end{equation}
with  
\begin{equation} \label{moment_generating_k}
M_{k}(\tilde{\boldsymbol{\mathcal{Q}}};\tilde{n}) 
= \mathbb{E}\left[
 \left. 
  \mathrm{e}^{
   \Lambda_{k}(\tilde{\boldsymbol{\mathcal{Q}}};\tilde{n})
  }
 \right| \mathcal{H}_{k}^{\{1\}}, \mathcal{H}_{k}^{\mathcal{A}_{k}} 
\right]. 
\end{equation}  
In these expressions, 
$\Lambda_{k}(\tilde{\boldsymbol{\mathcal{Q}}};\tilde{n})$ is given by 
\begin{equation} \label{Lambda}
\Lambda_{k}(\tilde{\boldsymbol{\mathcal{Q}}};\tilde{n})=
\sum_{t=1}^{\tau}\sum_{m=1}^{M}
\mathrm{Tr}(\boldsymbol{\omega}_{t,k,m}\boldsymbol{\omega}_{t,k,m}^{H}
\tilde{\boldsymbol{\mathcal{Q}}}). 
\end{equation}
It will be shown later that the moment generating 
function~(\ref{moment_generating_k}) for the users in $\mathcal{K}$ 
reduces to the right-hand side of 
(\ref{decoupled_measure}). 

We next calculate the pdf $\mu_{K}(\boldsymbol{\mathcal{Q}};\tilde{n})$ 
up to $O(1)$. The inversion formula of the moment generating 
function~(\ref{moment_generating_K_tr}) implies  
\begin{equation} \label{mu_tmp1} 
\mu_{K}(\boldsymbol{\mathcal{Q}};\tilde{n}) = 
\left(
 \frac{K}{2\pi\mathrm{i}}
\right)^{[(\tilde{n}+1)N\tau]^{2}}\int \left\{
 \prod_{k\in\mathcal{K}}M_{k}(\tilde{\boldsymbol{\mathcal{Q}}};\tilde{n}) 
\right\}\mathrm{e}^{
 -KI_{K}(\boldsymbol{\mathcal{Q}}, \tilde{\boldsymbol{\mathcal{Q}}};\tilde{n}) 
}d\tilde{\boldsymbol{\mathcal{Q}}}, 
\end{equation}
where the integrations with respect to $d\mathrm{Re}[\tilde{q}^{j,k}_{a,b}]$, 
$d\mathrm{Im}[\tilde{q}^{j,k}_{a,b}]$, and $d\tilde{q}^{j,j}_{a,a}$ 
are taken along imaginary axes, respectively. In (\ref{mu_tmp1}), 
$I_{K}(\boldsymbol{\mathcal{Q}}, \tilde{\boldsymbol{\mathcal{Q}}};\tilde{n})$ 
is defined as 
\begin{equation} \label{I_K} 
I_{K}(\boldsymbol{\mathcal{Q}}, \tilde{\boldsymbol{\mathcal{Q}}};\tilde{n}) = 
\mathrm{Tr}(\boldsymbol{\mathcal{Q}}\tilde{\boldsymbol{\mathcal{Q}}}) 
 - \frac{1}{K}\sum_{k\notin\mathcal{K}}\ln\mathbb{E}\left[
 \mathrm{e}^{
  \Lambda_{k}(\tilde{\boldsymbol{\mathcal{Q}}};\tilde{n})
 }
\right],  
\end{equation}
where $\Lambda_{k}(\tilde{\boldsymbol{\mathcal{Q}}};\tilde{n})$ is given by 
(\ref{Lambda}). Note that the limit 
$\lim_{K\rightarrow\infty}I_{K}
(\boldsymbol{\mathcal{Q}}, \tilde{\boldsymbol{\mathcal{Q}}};\tilde{n})
\equiv I(\boldsymbol{\mathcal{Q}}, \tilde{\boldsymbol{\mathcal{Q}}};\tilde{n})$ 
exists obviously. 
Applying the saddle-point method to (\ref{mu_tmp1}) in the large-system limit, 
we obtain 
\begin{equation} \label{mu_tmp2} 
\mu_{K}(\boldsymbol{\mathcal{Q}};\tilde{n}) =  
\left\{
 \prod_{k\in\mathcal{K}}M_{k}
 (\tilde{\boldsymbol{\mathcal{Q}}}_{\mathrm{s}};\tilde{n}) 
\right\} 
\left(
 \frac{K}{2\pi}
\right)^{[(\tilde{n}+1)N\tau]^{2}/2}\left|
 \det\nabla_{\tilde{\boldsymbol{\mathcal{Q}}}}^{2}
 I(\boldsymbol{\mathcal{Q}},\tilde{\boldsymbol{\mathcal{Q}}}_{\mathrm{s}};
 \tilde{n})
\right|^{-1/2}
\mathrm{e}^{
 -KI(\boldsymbol{\mathcal{Q}}, \tilde{\boldsymbol{\mathcal{Q}}}_{\mathrm{s}};
 \tilde{n})
}[1+o(K)], 
\end{equation} 
where $\tilde{\boldsymbol{\mathcal{Q}}}_{\mathrm{s}}$ is implicitly given as 
the solution to the fixed-point equation 
\begin{equation}  \label{equ1} 
\boldsymbol{\mathcal{Q}} = \lim_{K\rightarrow\infty}
\frac{1}{K}\sum_{k\notin \mathcal{K}}
\sum_{t=1}^{\tau}\sum_{m=1}^{M}\mathbb{E}\left\{
 \frac{
 \mathbb{E}\left[
  \boldsymbol{\omega}_{k,t,m}
  \boldsymbol{\omega}_{k,t,m}^{H}
  \mathrm{e}^{\Lambda_{k}(\tilde{\boldsymbol{\mathcal{Q}}};\tilde{n})}
 \right]
 }
 {
 \mathbb{E}\left[ 
  \mathrm{e}^{\Lambda_{k}(\tilde{\boldsymbol{\mathcal{Q}}};\tilde{n})}
 \right]
 }
\right\}. 
\end{equation}
In (\ref{mu_tmp2}), $\nabla_{\tilde{\boldsymbol{\mathcal{Q}}}}^{2}
I(\boldsymbol{\mathcal{Q}},\tilde{\boldsymbol{\mathcal{Q}}}_{\mathrm{s}};
\tilde{n})$ denotes the Hesse matrix of 
$I(\boldsymbol{\mathcal{Q}},\tilde{\boldsymbol{\mathcal{Q}}};\tilde{n})$ 
with respect to $\tilde{\boldsymbol{\mathcal{Q}}}$ at the saddle point 
$\tilde{\boldsymbol{\mathcal{Q}}}_{\mathrm{s}}$. 
In the derivation of (\ref{mu_tmp2}), we have used the fact that the Hesse 
matrix $\nabla_{\tilde{\boldsymbol{\mathcal{Q}}}}^{2}
I(\boldsymbol{\mathcal{Q}},
\tilde{\boldsymbol{\mathcal{Q}}}_{\mathrm{s}};\tilde{n})$ is negative definite, 
due to the assumption of the non-singularity of $\boldsymbol{\mathcal{Q}}$. 
Note that the concavity of (\ref{I_K}) with respect to 
$\tilde{\boldsymbol{\mathcal{Q}}}$ at 
$\tilde{\boldsymbol{\mathcal{Q}}}=
\tilde{\boldsymbol{\mathcal{Q}}}_{\mathrm{s}}$ is required for the use of 
the saddle-point method, since the integration in (\ref{mu_tmp1}) is taken 
over imaginary axes. 

Finally, we evaluate (\ref{Xi_channel_tmp5}) up to $O(1)$ in the large-system 
limit. Substituting (\ref{mu_tmp2}) into (\ref{Xi_channel_tmp5}) and 
subsequently using the saddle-point method, we obtain   
\begin{equation} 
\Xi_{\tilde{n}}(\mathcal{H}_{\mathcal{K}}^{\{1\}},
\mathcal{H}_{\mathcal{K}}^{\mathcal{A}}) =
D_{\tilde{n}}p(\mathcal{H}_{\mathcal{K}}^{\{1\}})
p(\mathcal{H}_{\mathcal{K}}^{\mathcal{A}}) 
\prod_{k\in\mathcal{K}}M_{k} 
 (\tilde{\boldsymbol{\mathcal{Q}}}_{\mathrm{s}};\tilde{n}) 
\mathrm{e}^{
 -K\Phi(\boldsymbol{\mathcal{Q}}_{\mathrm{s}};\tilde{n})
}[1+o(K)], \label{Xi_channel_tmp6} 
\end{equation}
where $\boldsymbol{\mathcal{Q}}_{\mathrm{s}}$ is a solution to satisfy 
\begin{equation} \label{saddle_point_condition} 
\boldsymbol{\mathcal{Q}}_{\mathrm{s}} = 
\arginf_{\boldsymbol{\mathcal{Q}}\in\mathcal{M}_{(\tilde{n}+1)N\tau}^{+}}
\Phi(\boldsymbol{\mathcal{Q}};\tilde{n}), 
\end{equation}
with 
\begin{equation} \label{Phi}
\Phi(\boldsymbol{\mathcal{Q}};\tilde{n})=
I(\boldsymbol{\mathcal{Q}}, 
\tilde{\boldsymbol{\mathcal{Q}}}_{\mathrm{s}};\tilde{n})
- \beta^{-1}G_{\tau}(\boldsymbol{\mathcal{Q}};\tilde{n}). 
\end{equation}
In (\ref{Xi_channel_tmp6}), 
$D_{\tilde{n}}=[\det\nabla_{\boldsymbol{\mathcal{Q}}}^{2}
\Phi(\boldsymbol{\mathcal{Q}}_{\mathrm{s}};\tilde{n})]^{-1/2}
|\det\nabla_{\tilde{\boldsymbol{\mathcal{Q}}}}^{2}
I(\boldsymbol{\mathcal{Q}}_{\mathrm{s}},
\tilde{\boldsymbol{\mathcal{Q}}}_{\mathrm{s}};\tilde{n})|^{-1/2}$,  
with $\nabla_{\boldsymbol{\mathcal{Q}}}^{2}
\Phi(\boldsymbol{\mathcal{Q}}_{\mathrm{s}};\tilde{n})$ denoting the Hesse 
matrix of $\Phi(\boldsymbol{\mathcal{Q}};\tilde{n})$ with respect to 
$\boldsymbol{\mathcal{Q}}$ at the saddle-point 
$\boldsymbol{\mathcal{Q}}_{\mathrm{s}}$. 
We have assumed that the Hesse matrix $\nabla_{\boldsymbol{\mathcal{Q}}}^{2}
\Phi(\boldsymbol{\mathcal{Q}}_{\mathrm{s}};\tilde{n})$ is positive definite. 
Calculating the solution~(\ref{saddle_point_condition}) with (\ref{G}) and 
(\ref{I_K}), we find that $\boldsymbol{\mathcal{Q}}_{\mathrm{s}}$ satisfies 
the stationarity condition 
\begin{equation} \label{equ2}
\tilde{\boldsymbol{\mathcal{Q}}} = -\beta^{-1}
(\boldsymbol{I} + \boldsymbol{A}_{\tau}(\tilde{n})
\boldsymbol{\mathcal{Q}})^{-1}
\boldsymbol{A}_{\tau}(\tilde{n}).
\end{equation}

\subsection{Evaluation of Fixed-Point Equations} 
It is generally difficult to solve the coupled fixed-point 
equations~(\ref{equ1}) and (\ref{equ2}). At this point, 
we assume that the solution $(\boldsymbol{\mathcal{Q}}_{\mathrm{s}},\ 
\tilde{\boldsymbol{\mathcal{Q}}}_{\mathrm{s}})$ satisfies RS. 
Note that it depends on models whether the RS assumption holds. Fortunately, 
It is empirically known that the RS assumption is correct when 
replicated random variables are CSCG. 

\begin{assumption}[Replica Symmetry] \label{RS_assumption} 
$(\boldsymbol{\mathcal{Q}}_{\mathrm{s}},\ 
\tilde{\boldsymbol{\mathcal{Q}}}_{\mathrm{s}})$ is 
invariant under all permutations of replica indices: 
\begin{equation} \label{RS}
\boldsymbol{\mathcal{Q}}_{\mathrm{s}} = 
\begin{pmatrix}
 \boldsymbol{Q}_{0} & \boldsymbol{1}_{\tilde{n}}^{T}\otimes \boldsymbol{M} \\
 \boldsymbol{1}_{\tilde{n}}\otimes\boldsymbol{M}^{H} 
 & \boldsymbol{I}_{\tilde{n}}\otimes (\boldsymbol{Q}_{1}-\boldsymbol{Q}) 
 + \boldsymbol{1}_{\tilde{n}}\boldsymbol{1}_{\tilde{n}}^{T}\otimes\boldsymbol{Q}
\end{pmatrix}, 
\end{equation}
\begin{equation} \label{RS_tilde}
\tilde{\boldsymbol{\mathcal{Q}}}_{\mathrm{s}} = 
\begin{pmatrix}
 \tilde{\boldsymbol{Q}}_{0} & 
 \boldsymbol{1}_{\tilde{n}}^{T}\otimes\tilde{\boldsymbol{M}} \\
 \boldsymbol{1}_{\tilde{n}}\otimes\tilde{\boldsymbol{M}}^{H} 
 & \boldsymbol{I}_{\tilde{n}}\otimes (\tilde{\boldsymbol{Q}}_{1}
 -\tilde{\boldsymbol{Q}}) 
 + \boldsymbol{1}_{\tilde{n}}\boldsymbol{1}_{\tilde{n}}^{T}\otimes
 \tilde{\boldsymbol{Q}}
\end{pmatrix},
\end{equation}
with $\boldsymbol{M}, \tilde{\boldsymbol{M}}
\in\mathbb{C}^{N\tau\times N\tau}$, and with $N\tau\times N\tau$ Hermitian 
matrices $\boldsymbol{Q}_{0}$, $\tilde{\boldsymbol{Q}}_{0}$, 
$\boldsymbol{Q}_{1}$, $\tilde{\boldsymbol{Q}}_{1}$, 
$\boldsymbol{Q}$, and $\tilde{\boldsymbol{Q}}$. 
\end{assumption}

We first solve the fixed-point equation~(\ref{equ2}) under 
Assumption~\ref{RS_assumption}. Evaluating the right-hand side of (\ref{equ2}) 
with (\ref{A}) and subsequently comparing both sides, we obtain 
\begin{equation} \label{equ2_rs}
\tilde{\boldsymbol{Q}}_{0} = 
-\tilde{n}(\boldsymbol{\Sigma} + \tilde{n}\boldsymbol{\Sigma}_{0})^{-1}, \quad 
\tilde{\boldsymbol{M}} = 
(\boldsymbol{\Sigma} + \tilde{n}\boldsymbol{\Sigma}_{0})^{-1}, \quad 
\tilde{\boldsymbol{Q}} = 
(\boldsymbol{\Sigma} + \tilde{n}\boldsymbol{\Sigma}_{0})^{-1}
\boldsymbol{\Sigma}_{0}\boldsymbol{\Sigma}^{-1}, \quad 
\tilde{\boldsymbol{Q}}_{1} = 
\tilde{\boldsymbol{Q}} - \boldsymbol{\Sigma}^{-1},   
\end{equation}
where $\boldsymbol{\Sigma}_{0}$ and $\boldsymbol{\Sigma}$ are given by 
\begin{equation} \label{Sigma_0_rs}
\boldsymbol{\Sigma}_{0} = N_{0}\boldsymbol{I}_{N\tau} + 
\beta(\boldsymbol{Q}_{0} - \boldsymbol{M} - \boldsymbol{M}^{H} 
+ \boldsymbol{Q}), 
\end{equation}
\begin{equation} \label{Sigma_rs}
\boldsymbol{\Sigma} = N_{0}\boldsymbol{I}_{N\tau} + 
\beta(\boldsymbol{Q}_{1} - \boldsymbol{Q}).  
\end{equation}
Note that $\boldsymbol{\Sigma}$ and 
$\boldsymbol{\Sigma}+\tilde{n}\boldsymbol{\Sigma}_{0}$ must be invertible since 
$(\boldsymbol{I}+\boldsymbol{A}_{\tau}(\tilde{n}) 
\boldsymbol{\mathcal{Q}}_{\mathrm{s}})^{-1}$ exists due to the assumption 
of the positive definiteness of $\boldsymbol{\mathcal{Q}}$. 
Nishimori~\cite{Nishimori02} proved $\boldsymbol{\Sigma}_{0}
=\boldsymbol{\Sigma}$ for unfaded CDMA systems. 
The condition $\boldsymbol{\Sigma}_{0}=\boldsymbol{\Sigma}$ implies that 
$\boldsymbol{\Sigma}_{0}$ must be invertible, since $\boldsymbol{\Sigma}^{-1}$ 
exists. In this case we can re-write $\tilde{\boldsymbol{Q}}$ as 
$\tilde{\boldsymbol{Q}}=\boldsymbol{\Sigma}^{-1}
(\tilde{n}\boldsymbol{\Sigma}^{-1}+
\boldsymbol{\Sigma}_{0}^{-1})^{-1}\boldsymbol{\Sigma}^{-1}$.  

We next evaluate $\mathrm{e}^{
\Lambda_{k}(\tilde{\boldsymbol{\mathcal{Q}}}_{\mathrm{s}};\tilde{n})}$, 
assuming that $\boldsymbol{\Sigma}_{0}$ is invertible. 
Substituting (\ref{Lambda}) and (\ref{equ2_rs}) into $\mathrm{e}^{
\Lambda_{k}(\tilde{\boldsymbol{\mathcal{Q}}}_{\mathrm{s}};\tilde{n})}$ yields  
\begin{equation} \label{mgf_tmp1}
\mathrm{e}^{\Lambda_{k}
(\tilde{\boldsymbol{\mathcal{Q}}}_{\mathrm{s}};\tilde{n})}  = 
\prod_{t=1}^{\tau}\prod_{m=1}^{M}
\mathrm{e}^{
 \boldsymbol{a}_{t,k,m}^{H}\tilde{\boldsymbol{\Sigma}}_{0}  
 \boldsymbol{a}_{t,k,m}  
 - \sum_{a=0}^{\tilde{n}}(\boldsymbol{\omega}_{t,k,m}^{\{a\}})^{H}
 \boldsymbol{\Sigma}_{a}^{-1}\boldsymbol{\omega}_{t,k,m}^{\{a\}} 
}, 
\end{equation}
with $\tilde{\boldsymbol{\Sigma}}_{0}
=(\tilde{n}\boldsymbol{\Sigma}^{-1} + \boldsymbol{\Sigma}_{0}^{-1})^{-1}$ and 
$\boldsymbol{\Sigma}_{a}=\boldsymbol{\Sigma}$ for $a=1,\ldots,\tilde{n}$.  
In (\ref{mgf_tmp1}), $\boldsymbol{a}_{t,k,m}\in\mathbb{C}^{N\tau}$ 
is defined as $\boldsymbol{a}_{t,k,m} = \sum_{a=0}^{\tilde{n}}
\boldsymbol{\Sigma}_{a}^{-1}\boldsymbol{\omega}_{t,k,m}^{\{a\}}$. 
In order to linearize the quadratic form 
$\boldsymbol{a}_{t,k,m}^{H}\tilde{\boldsymbol{\Sigma}}_{0}  
\boldsymbol{a}_{t,k,m}$ in (\ref{mgf_tmp1}), we apply the identity 
\begin{equation}
\mathrm{e}^{\boldsymbol{a}_{t,k,m}^{H}\tilde{\boldsymbol{\Sigma}}_{0}
\boldsymbol{a}_{t,k,m}} 
= \int \frac{1}{\pi^{N\tau}\det\tilde{\boldsymbol{\Sigma}}_{0}}
\mathrm{e}^{
 -\vec{\underline{\boldsymbol{y}}}_{t,k,m}^{H}
 \tilde{\boldsymbol{\Sigma}}_{0}^{-1}
 \vec{\underline{\boldsymbol{y}}}_{t,k,m}  
 + \boldsymbol{a}_{t,k,m}^{H}\vec{\underline{\boldsymbol{y}}}_{t,k,m}  
 + \vec{\underline{\boldsymbol{y}}}_{t,k,m}^{H}\boldsymbol{a}_{t,k,m}  
}d\vec{\underline{\boldsymbol{y}}}_{t,k,m}, 
\end{equation}
to (\ref{mgf_tmp1}), with 
$\vec{\underline{\boldsymbol{y}}}_{t,k,m}\in\mathbb{C}^{N\tau}$. 
Then, we obtain   
\begin{equation} \label{mgf_tmp2}
\mathrm{e}^{\Lambda_{k}
(\tilde{\boldsymbol{\mathcal{Q}}}_{\mathrm{s}};\tilde{n})} 
= C_{\tilde{n}}^{\tau M}
\int \prod_{a=0}^{\tilde{n}} q(\vec{\underline{\mathcal{Y}}}_{k} | 
\Omega_{k}^{\{a\}} )d\vec{\underline{\mathcal{Y}}}_{k}, 
\end{equation}
with 
\begin{equation}
C_{\tilde{n}} = \left[
 \pi^{\tilde{n}N\tau}\det\{\boldsymbol{\Sigma}^{\tilde{n}-1}
 (\boldsymbol{\Sigma} + \tilde{n}\boldsymbol{\Sigma}_{0})\}
\right]. 
\end{equation}
In (\ref{mgf_tmp2}), 
$q(\vec{\underline{\mathcal{Y}}}_{k} | \Omega_{k}^{\{a\}} )$, 
with $\vec{\underline{\mathcal{Y}}}_{k}
=\{\vec{\underline{\boldsymbol{y}}}_{t,k,m}:
\hbox{for all $t\in\mathcal{T}_{\tau}$,  $m$}\}$ and 
$\Omega_{k}^{\{a\}}=\{\boldsymbol{\omega}_{t,k,m}^{\{a\}}:
\hbox{for all $t\in\mathcal{T}_{\tau}$, $m$}\}$, is defined as 
\begin{equation} \label{function_q} 
q(\vec{\underline{\mathcal{Y}}}_{k} | \Omega_{k}^{\{a\}} )
= \prod_{t=1}^{\tau}\prod_{m=1}^{M}
g_{\tau N}\left(
 \vec{\underline{\boldsymbol{y}}}_{t,k,m} 
 - \boldsymbol{\omega}_{t,k,m}^{\{a\}};\boldsymbol{\Sigma}_{a}
\right), 
\end{equation}
where $g_{\tau N}(\vec{\underline{\boldsymbol{y}}}_{t,k,m} 
- \boldsymbol{\omega}_{t,k,m}^{\{a\}};\boldsymbol{\Sigma}_{a})$ is given by 
(\ref{Gauss}), with $\boldsymbol{\omega}_{t,k,m}^{\{a\}}$ defined 
as (\ref{omega}). 

We solve the fixed-point equation~(\ref{equ1}) by substituting 
(\ref{mgf_tmp2}) into (\ref{equ1}). It is sufficient to evaluate 
$\boldsymbol{Q}_{0} - \boldsymbol{M} - \boldsymbol{M}^{H} + \boldsymbol{Q}$ 
and $\boldsymbol{Q}_{1} - \boldsymbol{Q}$ owing to the fact that 
$\boldsymbol{\Sigma}_{0}$ 
and $\boldsymbol{\Sigma}$ depend on $\boldsymbol{\mathcal{Q}}_{\mathrm{s}}$ 
only through them. Comparing both sides of (\ref{equ1}), we have  
\begin{IEEEeqnarray}{r}
\boldsymbol{Q}_{0} - \boldsymbol{M} - \boldsymbol{M}^{H} + \boldsymbol{Q} = 
\lim_{K\rightarrow\infty}\frac{1}{K}\sum_{k\notin\mathcal{K}}
\sum_{t=1}^{\tau}\sum_{m=1}^{M}\mathbb{E}\left[ 
 \frac{C_{\tilde{n}}^{\tau M}}{
 \mathbb{E}\left\{
  \mathrm{e}^{\Lambda_{k}
  (\tilde{\boldsymbol{\mathcal{Q}}}_{\mathrm{s}};\tilde{n})}  
 \right\}
 }\int (\boldsymbol{\omega}_{t,k,m}^{\{0\}}  
 - \langle \boldsymbol{\omega}_{t,k,m}^{\{1\}} \rangle )
\right. \nonumber \\ 
\times ( \boldsymbol{\omega}_{t,k,m}^{\{0\}} 
- \langle \boldsymbol{\omega}_{t,k,m}^{\{1\}} \rangle )^{H}
q(\vec{\underline{\mathcal{Y}}}_{k} | \Omega_{k}^{\{0\}} )
\left\{
 \mathbb{E}_{\mathcal{H}_{k}^{\{1\}}}\left[
  \left. 
   q(\vec{\underline{\mathcal{Y}}}_{k} | \Omega_{k}^{\{1\}} )
  \right| \mathcal{X}_{k} 
 \right]
\right\}^{\tilde{n}}d\vec{\underline{\mathcal{Y}}}_{k}
\Biggl], \label{equ1_rs1} 
\end{IEEEeqnarray}
\begin{IEEEeqnarray}{r}
\boldsymbol{Q}_{1} - \boldsymbol{Q} = 
\lim_{K\rightarrow\infty}\frac{1}{K}\sum_{k\notin\mathcal{K}}
\sum_{t=1}^{\tau}\sum_{m=1}^{M}\mathbb{E}\left[ 
 \frac{C_{\tilde{n}}^{\tau M}}{
  \mathbb{E}\left\{
   \mathrm{e}^{\Lambda_{k}
   (\tilde{\boldsymbol{\mathcal{Q}}}_{\mathrm{s}};\tilde{n})}  
  \right\}
 }\int \langle (\boldsymbol{\omega}_{t,k,m}^{\{1\}}   
 - \langle \boldsymbol{\omega}_{t,k,m}^{\{1\}} \rangle )
\right. \nonumber \\ 
\times  ( \boldsymbol{\omega}_{t,k,m}^{\{1\}}
- \langle \boldsymbol{\omega}_{t,k,m}^{\{1\}} \rangle )^{H} 
\rangle q(\vec{\underline{\mathcal{Y}}}_{k} | \Omega_{k}^{\{0\}} )
\left\{
 \mathbb{E}_{ \mathcal{H}_{k}^{\{1\}}}\left[
  \left. 
   q(\vec{\underline{\mathcal{Y}}}_{k} | \Omega_{k}^{\{1\}} )
  \right| \mathcal{X}_{k} 
 \right]
\right\}^{\tilde{n}}d\vec{\underline{\mathcal{Y}}}_{k}
\Biggr], \label{equ1_rs2}
\end{IEEEeqnarray}
In (\ref{equ1_rs1}) and (\ref{equ1_rs2}), $\mathbb{E}[\exp\{ \Lambda_{k}(
\tilde{\boldsymbol{\mathcal{Q}}}_{\mathrm{s}};\tilde{n}) \} ]$ 
is explicitly given as 
\begin{equation} 
\mathbb{E}\left[
 \mathrm{e}^{\Lambda_{k}
 (\tilde{\boldsymbol{\mathcal{Q}}}_{\mathrm{s}};\tilde{n})}
\right] 
= C_{\tilde{n}}^{\tau M}\mathbb{E}\left[
 \int q(\vec{\underline{\mathcal{Y}}}_{k} | \Omega_{k}^{\{0\}} )
 \left\{
  \mathbb{E}_{\mathcal{H}_{k}^{\{1\}}}\left[
   \left. 
    q(\vec{\underline{\mathcal{Y}}}_{k} | \Omega_{k}^{\{1\}} )
   \right| \mathcal{X}_{k}  
  \right]
 \right\}^{\tilde{n}}d\vec{\underline{\mathcal{Y}}}_{k}
\right]. \label{moment_generating_k'_rs} 
\end{equation}
Furthermore, 
$\langle f(\boldsymbol{\omega}_{t,k,m}^{\{1\}}) \rangle$ for a function 
$f(\boldsymbol{\omega}_{t,k,m}^{\{1\}})$ of (\ref{omega}) denotes the 
mean of $f(\boldsymbol{\omega}_{t,k,m}^{\{1\}})$ with respect to 
the posterior measure 
$q(\mathcal{H}_{k}^{\{1\}} | \vec{\underline{\mathcal{Y}}}_{k}, \mathcal{X}_{k})
d\mathcal{H}_{k}^{\{1\}}$, 
defined as 
\begin{equation} \label{posterior_pdf_rs} 
q(\mathcal{H}_{k}^{\{a\}} | \vec{\underline{\mathcal{Y}}}_{k}, 
\mathcal{X}_{k}) = 
\frac{
 q(\vec{\underline{\mathcal{Y}}}_{k} | \Omega_{k}^{\{a\}} )
 p(\mathcal{H}_{k}^{\{a\}})
} 
{
 \int 
 q(\vec{\underline{\mathcal{Y}}}_{k} | \Omega_{k}^{\{a\}} )
 p(\mathcal{H}_{k}^{\{a\}})d\mathcal{H}_{k}^{\{a\}} 
}, 
\end{equation}
where $q(\vec{\underline{\mathcal{Y}}}_{k} | \Omega_{k}^{\{a\}} )$ is given by 
(\ref{function_q}). 
Equations~(\ref{Sigma_0_rs}), (\ref{Sigma_rs}), (\ref{equ1_rs1}), 
and~(\ref{equ1_rs2}) form closed equations for $\boldsymbol{\Sigma}_{0}$ and 
$\boldsymbol{\Sigma}$, and are well-defined for $\tilde{n}\in\mathbb{R}$. 
Regarding $\tilde{n}$ in these equations as a real number and 
taking $\tilde{n}\rightarrow+0$, we obtain the coupled fixed-point equations 
\begin{equation} \label{Sigma_0_rs_n0}
\boldsymbol{\Sigma}_{0} = N_{0}\boldsymbol{I}_{N\tau} + 
\lim_{K\rightarrow\infty}\frac{\beta}{K}\sum_{k\notin \mathcal{K}}
\sum_{t=1}^{\tau}\sum_{m=1}^{M}
\mathbb{E}\left[ 
 (\boldsymbol{\omega}_{t,k,m}^{\{0\}}  
 - \langle \boldsymbol{\omega}_{t,k,m}^{\{1\}} \rangle )
 ( \boldsymbol{\omega}_{t,k,m}^{\{0\}} 
 - \langle \boldsymbol{\omega}_{t,k,m}^{\{1\}} \rangle )^{H} 
\right], 
\end{equation}
\begin{equation} \label{Sigma_rs_n0} 
\boldsymbol{\Sigma} = N_{0}\boldsymbol{I}_{N\tau} + 
\lim_{K\rightarrow\infty}\frac{\beta}{K}\sum_{k\notin \mathcal{K}}
\sum_{t=1}^{\tau}\sum_{m=1}^{M} 
\mathbb{E}\left[ 
 \left\langle 
  (\boldsymbol{\omega}_{t,k,m}^{\{1\}}  
  - \langle \boldsymbol{\omega}_{t,k,m}^{\{1\}} \rangle )
  ( \boldsymbol{\omega}_{t,k,m}^{\{1\}} 
  - \langle \boldsymbol{\omega}_{t,k,m}^{\{1\}} \rangle )^{H} 
 \right\rangle 
\right], 
\end{equation}
where the expectations in (\ref{Sigma_0_rs_n0}) and (\ref{Sigma_rs_n0}) are 
taken with respect to the measure 
$q(\vec{\underline{\mathcal{Y}}}_{k} | \Omega_{k}^{\{0\}})
p(\mathcal{H}_{k}^{\{0\}})
p(\mathcal{X}_{k})d\vec{\underline{\mathcal{Y}}}_{k}$ $d\mathcal{H}_{k}^{\{0\}}
d\mathcal{X}_{k}$. Note that (\ref{Sigma_0_rs_n0}) and 
(\ref{Sigma_rs_n0}) are coupled since their second terms depend on both 
$\boldsymbol{\Sigma}_{0}$ and $\boldsymbol{\Sigma}$. 
 
Let us assume $\boldsymbol{\Sigma}_{0}=\boldsymbol{\Sigma}$. 
Then, it is straightforward to show that 
the coupled fixed-point equations~(\ref{Sigma_0_rs_n0}) 
and~(\ref{Sigma_rs_n0}) reduce to the single 
fixed-point equation 
\begin{equation} \label{single_Sigma_rs} 
\boldsymbol{\Sigma}_{0} = N_{0}\boldsymbol{I}_{N\tau} + 
\lim_{K\rightarrow\infty}\frac{\beta}{K}\sum_{k\notin\mathcal{K}}
\sum_{t=1}^{\tau}\sum_{m=1}^{M}\mathbb{E}\left[
 (\boldsymbol{\omega}_{t,k,m}^{\{0\}}  
 - \langle \boldsymbol{\omega}_{t,k,m}^{\{0\}} \rangle )
 ( \boldsymbol{\omega}_{t,k,m}^{\{0\}} 
 - \langle \boldsymbol{\omega}_{t,k,m}^{\{0\}} \rangle )^{H} 
\right]. 
\end{equation}
In (\ref{single_Sigma_rs}), the expectation  is taken with respect to 
$q(\vec{\underline{\mathcal{Y}}}_{k} | \Omega_{k}^{\{0\}})
p(\mathcal{H}_{k}^{\{0\}})
p(\mathcal{X}_{k})d\vec{\underline{\mathcal{Y}}}_{k}d\mathcal{H}_{k}^{\{0\}}
d\mathcal{X}_{k}$. Furthermore, 
$\langle \boldsymbol{\omega}_{t,k,m}^{\{0\}} \rangle$ denotes the 
mean of $\boldsymbol{\omega}_{t,k,m}^{\{0\}}$ 
with respect to the posterior measure 
$q(\mathcal{H}_{k}^{\{0\}} | \vec{\underline{\mathcal{Y}}}_{k}, \mathcal{X}_{k})
d\mathcal{H}_{k}^{(\{0\}}$, given by (\ref{posterior_pdf_rs}).  
Substituting (\ref{omega}) into (\ref{single_Sigma_rs}) after assuming 
$\boldsymbol{\Sigma}_{0}=
\sigma_{\mathrm{tr}}^{2}(\tau)\boldsymbol{I}_{N\tau}$, 
we find that (\ref{single_Sigma_rs}) reduces to 
(\ref{fixed_point_channel}).

\subsection{Replica Continuity} 
We evaluate (\ref{G}), (\ref{moment_generating_k}), and~(\ref{I_K}) 
under Assumption~\ref{RS_assumption}. Substituting (\ref{A}) and 
(\ref{RS}) into (\ref{G}) and subsequently using 
(\ref{Sigma_0_rs}) and (\ref{Sigma_rs}), we obtain 
\begin{equation} \label{G_rs}
G_{\tau}(\boldsymbol{\mathcal{Q}}_{\mathrm{s}};\tilde{n}) 
= - (\tilde{n}-1)\ln\det\boldsymbol{\Sigma} 
- \ln\det(\boldsymbol{\Sigma}+\tilde{n}\boldsymbol{\Sigma}_{0}) 
- \tilde{n}N\tau\ln\pi. 
\end{equation}
Next, from (\ref{mgf_tmp2}), the moment generating 
function~(\ref{moment_generating_k}) for the users in  
$\mathcal{K}$ reduces to 
\begin{equation}  
M_{k}(\tilde{\boldsymbol{\mathcal{Q}}}_{\mathrm{s}};\tilde{n}) 
= \frac{C_{\tilde{n}}^{\tau M}}{
p(\mathcal{H}_{k}^{\{1\}\cup\mathcal{A}_{k}})}
\mathbb{E}\left[
 \int q(\vec{\underline{\mathcal{Y}}}_{k} | \Omega_{k}^{\{0\}} )
 \prod_{a\in\{1\}\cup\mathcal{A}_{k}}q(\mathcal{H}_{k}^{\{a\}} | 
 \vec{\underline{\mathcal{Y}}}_{k}, \mathcal{X}_{k}) 
 \left\{
  \mathbb{E}_{\mathcal{H}_{k}^{\{1\}}}\left[
   \left. 
    q(\vec{\underline{\mathcal{Y}}}_{k} | \Omega_{k}^{\{1\}} )
   \right| \mathcal{X}_{k}   
  \right]
 \right\}^{\tilde{n}}d\vec{\underline{\mathcal{Y}}}_{k}
\right], \label{moment_generating_k_rs} 
\end{equation}
where $q(\mathcal{H}_{k}^{\{a\}} |  \vec{\underline{\mathcal{Y}}}_{k}, 
\mathcal{X}_{k})$ id given by (\ref{posterior_pdf_rs}). 
Finally, calculating the first term on the right-hand side of (\ref{I_K}), 
we have  
\begin{IEEEeqnarray}{lcr}
I(\boldsymbol{\mathcal{Q}}_{\mathrm{s}},
\tilde{\boldsymbol{\mathcal{Q}}}_{\mathrm{s}};\tilde{n}) &=& 
-\frac{\tilde{n}}{\beta}\mathrm{Tr}\left[
 \boldsymbol{I}_{N\tau} 
 - N_{0}(\boldsymbol{\Sigma}+\tilde{n}\boldsymbol{\Sigma}_{0})^{-1} 
 - N_{0}\boldsymbol{\Sigma}^{-1} 
 + N_{0}(\boldsymbol{\Sigma}+\tilde{n}\boldsymbol{\Sigma}_{0})^{-1}
 \boldsymbol{\Sigma}_{0}\boldsymbol{\Sigma}^{-1}
\right] \nonumber \\ 
&& - \lim_{K\rightarrow\infty}\frac{1}{K}\sum_{k\notin\mathcal{K}}
\ln\mathbb{E}\left\{ 
 \mathrm{e}^{
  \Lambda_{k}(\tilde{\boldsymbol{\mathcal{Q}}}_{\mathrm{s}};\tilde{n})
 } 
\right\}, \label{I_rs}
\end{IEEEeqnarray}
in the limit $K\rightarrow\infty$, in which 
$\mathbb{E}[\exp\{ \Lambda_{k}(
\tilde{\boldsymbol{\mathcal{Q}}}_{\mathrm{s}};\tilde{n}) \}  ]$ 
is given by (\ref{moment_generating_k'_rs}). 

Equations~(\ref{G_rs}), (\ref{moment_generating_k_rs}), and~(\ref{I_rs}) 
are well-defined for $\tilde{n}\in\mathbb{R}$. Let us assume that they 
coincide with the true ones for $\tilde{n}\in[0,n_{\mathrm{c}})$ with some 
$n_{\mathrm{c}}>0$. 
Then, substituting (\ref{G_rs}), (\ref{moment_generating_k_rs}), and 
(\ref{I_rs}) into (\ref{Xi_channel_tmp6}) and taking $\tilde{n}\rightarrow+0$, 
we arrive at 
\begin{equation} \label{Xi_tmp7} 
\lim_{\tilde{n}\rightarrow+0}\Xi_{\tilde{n}}(\mathcal{H}_{\mathcal{K}}^{\{1\}},
\mathcal{H}_{\mathcal{K}}^{\mathcal{A}})  = 
\left(
 \lim_{\tilde{n}\rightarrow+0}D_{\tilde{n}}
\right)\prod_{k\in\mathcal{K}}\mathbb{E}\left[
 q(\mathcal{H}_{k}^{\{1\}} | \vec{\underline{\mathcal{Y}}}_{k}, \mathcal{X}_{k})
 \prod_{a\in\mathcal{A}_{k}}q(\mathcal{H}_{k}^{\{a\}} | 
 \vec{\underline{\mathcal{Y}}}_{k}, \mathcal{X}_{k})
\right] + o(K),  
\end{equation}
where the expectation is taken with respect to the measure 
$q(\vec{\underline{\mathcal{Y}}}_{k} | \Omega_{k}^{\{0\}})
p(\mathcal{H}_{k}^{\{0\}})
p(\mathcal{X}_{k})d\vec{\underline{\mathcal{Y}}}_{k}d\mathcal{H}_{k}^{\{0\}}
d\mathcal{X}_{k}$. 
We apply (\ref{Xi_tmp7}) to (\ref{conditional_distribution_channel}) 
to obtain 
\begin{equation} \label{conditional_distribution_rs} 
\lim_{K,L\rightarrow\infty}
\mathbb{E}_{\mathcal{I}_{\mathcal{T}_{\tau}}}\left[
 p(\mathcal{H}_{\mathcal{K}}^{\{1\}} | \mathcal{I}_{\mathcal{T}_{\tau}}) 
 \prod_{k\in\mathcal{K}}\prod_{a\in\mathcal{A}_{k}}
 p(\mathcal{H}_{k}^{\{a\}} | \mathcal{I}_{\mathcal{T}_{\tau}})
\right] = \prod_{k\in\mathcal{K}}\mathbb{E}\left[
 q(\mathcal{H}_{k}^{\{1\}} | \vec{\underline{\mathcal{Y}}}_{k}, \mathcal{X}_{k})
 \prod_{a\in\mathcal{A}_{k}}q(\mathcal{H}_{k}^{\{a\}} | 
 \vec{\underline{\mathcal{Y}}}_{k}, \mathcal{X}_{k})
\right], 
\end{equation}
where $q(\mathcal{H}_{k}^{\{a\}} | \vec{\underline{\mathcal{Y}}}_{k}, 
\mathcal{X}_{k})$ is given by (\ref{posterior_pdf_rs}). 
It is straightforward to find that the right-hand side of 
(\ref{conditional_distribution_rs}) is equal to that of  
(\ref{decoupled_measure}).  
In the derivation of (\ref{conditional_distribution_rs}), we have 
used the fact that $\lim_{\tilde{n}\rightarrow+0}D_{\tilde{n}}$ should be 
equal to $1$  due to the normalization of the 
pdf~(\ref{conditional_distribution_rs}). 
We omit the proof of $\lim_{\tilde{n}\rightarrow+0}D_{\tilde{n}}=1$ since it 
requires complicated calculations and is beyond the scope of this paper.  
Furthermore, the proof of the convexity of (\ref{Phi}) at the saddle point 
is also omitted for the same reason. For details, see \cite{Tanaka02}.

\section{Derivation of Proposition~\ref{proposition1}} 
\label{appendix_derivation_proposition1} 
\subsection{Formulation} 
We analyze the equivalent channel between the users in $\mathcal{K}$ and 
their decoders in symbol period~$t(>\tau)$ 
to derive Proposition~\ref{proposition1}. Our analysis is based on the 
replica method, which is basically the same as that for 
Lemma~\ref{lemma_channel_estimation_tmp}, presented in 
Appendix~\ref{derivation_lemma_channel_estimation}. However, there are two 
differences between the two replica analyses. One is that we replicate not 
only the channel vectors but also the data symbols. The other is that 
the self-averaging property of MAI with respect to 
$\mathcal{I}_{\mathcal{T}_{\tau}}$ is shown by using 
Lemma~\ref{lemma_channel_estimation}. 

It is sufficient from Assumption~\ref{assumption1} to show that 
the distribution of $\tilde{\mathcal{B}}_{t,\mathcal{K}}$ conditioned on 
$\mathcal{B}_{t,\mathcal{K}}$, $\mathcal{H}$, and 
$\mathcal{I}_{\mathcal{T}_{\tau}}$ 
converges in law to the right-hand side of (\ref{decoupling}) in the 
large-system limit. 
We first transform $p(\tilde{\mathcal{B}}_{t,\mathcal{K}}| 
\mathcal{B}_{t,\mathcal{K}}, \mathcal{H}, \mathcal{I}_{\mathcal{T}_{\tau}})$ 
into a formula corresponding to (\ref{Xi_channel}). 
The posterior pdf of $\tilde{\mathcal{B}}_{t,\mathcal{K}}$ 
postulated by the optimal detector, defined in the same manner as in 
(\ref{posterior_MMSE_k}), is given by 
\begin{equation} \label{posterior_MMSE} 
p(\tilde{\mathcal{B}}_{t,\mathcal{K}} | \mathcal{Y}_{t}, 
\mathcal{S}_{t}, \mathcal{I}_{\mathcal{T}_{\tau}}) = 
\frac{
 \int p(\tilde{\mathcal{Y}}_{t}=\mathcal{Y}_{t} | \tilde{\mathcal{B}}_{t}, 
 \mathcal{S}_{t}, \mathcal{I}_{\mathcal{T}_{\tau}})p(\tilde{\mathcal{B}}_{t})
 d\tilde{\mathcal{B}}_{t,\backslash\mathcal{K}}
}
{
 \int p(\tilde{\mathcal{Y}}_{t}=\mathcal{Y}_{t} | \tilde{\mathcal{B}}_{t}, 
 \mathcal{S}_{t}, \mathcal{I}_{\mathcal{T}_{\tau}})p(\tilde{\mathcal{B}}_{t})
 d\tilde{\mathcal{B}}_{t}
}, 
\end{equation}
with $\tilde{\mathcal{B}}_{t,\backslash\mathcal{K}}=
\{\tilde{\mathcal{B}}_{t,k}:\hbox{for all $k\notin\mathcal{K}$}\}$. 
In (\ref{posterior_MMSE}), the pdf 
$p(\tilde{\mathcal{Y}}_{t} | \tilde{\mathcal{B}}_{t},  
\mathcal{S}_{t}, \mathcal{I}_{\mathcal{T}_{\tau}})$ 
is given by (\ref{postulated_channel}) 
with $\mathcal{I}_{t-1}=\mathcal{I}_{\tau}$. 
The equivalent channel between the users in $\mathcal{K}$ and 
their decoders is represented as 
\begin{equation} \label{appen_equivalent_channel} 
p(\tilde{\mathcal{B}}_{t,\mathcal{K}}| \mathcal{B}_{t,\mathcal{K}}, 
\mathcal{H}, \mathcal{I}_{\mathcal{T}_{\tau}}) = 
\mathbb{E}_{\mathcal{S}_{t}}\left[
 \int p(\tilde{\mathcal{B}}_{t,\mathcal{K}} | \mathcal{Y}_{t}, 
 \mathcal{S}_{t}, \mathcal{I}_{\mathcal{T}_{\tau}})p(\mathcal{Y}_{t} | 
 \mathcal{H}, \mathcal{S}_{t}, \mathcal{B}_{t})
 p(\mathcal{B}_{t,\backslash\mathcal{K}})d\mathcal{Y}_{t}
 d\mathcal{B}_{t,\backslash\mathcal{K}}
\right], 
\end{equation} 
with $\mathcal{B}_{t,\backslash\mathcal{K}}=
\{\mathcal{B}_{t,k}:\hbox{for all $k\notin\mathcal{K}$}\}$. 
In (\ref{appen_equivalent_channel}), the pdf 
$p(\mathcal{Y}_{t} | \mathcal{H}, \mathcal{S}_{t}, \mathcal{B}_{t})$
represents the true MIMO DS-CDMA channel~(\ref{MIMO_DS_CDMA}). 
Introducing a real number $n$, we obtain 
\begin{equation}
\lim_{K,L\rightarrow\infty}
p(\tilde{\mathcal{B}}_{t,\mathcal{K}}| \mathcal{B}_{t,\mathcal{K}}, 
\mathcal{H}, \mathcal{I}_{\mathcal{T}_{\tau}}) = 
\lim_{K,L\rightarrow\infty}\lim_{n\rightarrow+0}
\Xi_{n}(\tilde{\mathcal{B}}_{t,\mathcal{K}}, 
\mathcal{B}_{t,\mathcal{K}}, \mathcal{H}, \mathcal{I}_{\mathcal{T}_{\tau}}), 
\end{equation} 
with 
\begin{IEEEeqnarray}{r}
\Xi_{n}(\tilde{\mathcal{B}}_{t,\mathcal{K}}, 
\mathcal{B}_{t,\mathcal{K}}, \mathcal{H}, \mathcal{I}_{\mathcal{T}_{\tau}}) = 
\mathbb{E}_{\mathcal{S}_{t}}\left[
 \int \left\{
  \int p(\tilde{\mathcal{Y}}_{t}=\mathcal{Y}_{t} | 
  \tilde{\mathcal{H}}, \mathcal{S}_{t}, \tilde{\mathcal{B}}_{t})
  p(\tilde{\mathcal{B}}_{t})d\tilde{\mathcal{B}}_{t}
  p(\tilde{\mathcal{H}} | \mathcal{I}_{\mathcal{T}_{\tau}})
  d\tilde{\mathcal{H}}  
 \right\}^{n-1}
\right. \nonumber \\ 
\times p(\tilde{\mathcal{Y}}_{t}=\mathcal{Y}_{t} | 
\tilde{\mathcal{H}}, \mathcal{S}_{t}, \tilde{\mathcal{B}}_{t}) 
p(\tilde{\mathcal{H}} | \mathcal{I}_{\mathcal{T}_{\tau}})d\tilde{\mathcal{H}} 
p(\tilde{\mathcal{B}}_{t})d\tilde{\mathcal{B}}_{t,\backslash\mathcal{K}}
p(\mathcal{Y}_{t} | \mathcal{H}, \mathcal{S}_{t}, \mathcal{B}_{t})
p(\mathcal{B}_{t,\backslash\mathcal{K}})d\mathcal{Y}_{t}
d\mathcal{B}_{t,\backslash\mathcal{K}}
\Biggr], \label{Xi} 
\end{IEEEeqnarray}
where the posterior pdf 
$p(\tilde{\mathcal{H}} | \mathcal{I}_{\mathcal{T}_{\tau}})$ is given by 
(\ref{postulated_posterior_H}) with $\mathcal{I}_{t-1}=\mathcal{I}_{\tau}$. 

\subsection{Average over Quenched Randomness} 
We evaluate (\ref{Xi}) only for $n\in\mathbb{N}$ in the large-system limit. 
Let $\mathcal{B}_{t,k}^{\{a\}}=\{b_{t,k,m}^{\{a\}}\in\mathbb{C}
:\hbox{for all $m$}\}$ and $\tilde{\mathcal{H}}_{k}^{\{a\}}
=\{\tilde{\boldsymbol{h}}_{k,m}^{\{a\}}\in\mathbb{C}^{N}
:\hbox{for all $m$}\}$ denote 
replicas of $\tilde{\mathcal{B}}_{t,k}$ and $\tilde{\mathcal{H}}_{k}$ 
for $a=2,3,\ldots$, respectively: $\{\mathcal{B}_{t,k}^{\{a\}}\}$ are 
independently drawn from $p(\mathcal{B}_{t,k})$ for all $k$ and $a$, and 
$\{\tilde{\mathcal{H}}_{k}^{\{a\}}\}$ conditioned on 
$\mathcal{I}_{\mathcal{T}_{\tau}}$ are mutually independent random vectors 
following $p(\tilde{\mathcal{H}}_{k} | \mathcal{I}_{\mathcal{T}_{\tau}})$, 
defined by (\ref{posterior_pdf_Hk}), for 
all $k$ and $a$. For notational convenience, we introduce 
$\mathcal{B}_{t,k}^{\{0\}}=\mathcal{B}_{t,k}$,  
$\mathcal{B}_{t,k}^{\{1\}}=\tilde{\mathcal{B}}_{t,k}$,  
$\tilde{\mathcal{H}}_{k}^{\{0\}}=\mathcal{H}_{k}$, and 
$\tilde{\mathcal{H}}_{k}^{\{1\}}=\tilde{\mathcal{H}}_{k}$. 
Note that $\tilde{\mathcal{H}}_{k}^{\{a\}}$ and 
$\mathcal{H}_{k}^{\{a\}}$ for $a=1,\ldots,n$ should not be 
confused with each other. 
$\{\tilde{\mathcal{H}}_{k}^{\{a\}}\}$ 
conditioned on $\mathcal{I}_{\mathcal{T}_{\tau}}$ are mutually independent for 
all $k$, while $\{\mathcal{H}_{k}^{\{a\}}\}$ conditioned on 
$\mathcal{I}_{\mathcal{T}_{\tau}}$ have dependencies for all $k$.  
Taking the averages in (\ref{Xi}) over $\mathcal{Y}_{t}$ and 
$\mathcal{S}_{t}$ in the same manner as in the derivation of 
(\ref{Xi_channel_tmp4}), we have 
\begin{equation} \label{Xi_tmp1} 
\Xi_{n}(\tilde{\mathcal{B}}_{t,\mathcal{K}}, 
\mathcal{B}_{t,\mathcal{K}}, \mathcal{H}, \mathcal{I}_{\mathcal{T}_{\tau}}) = 
p(\tilde{\mathcal{B}}_{t,\mathcal{K}})
\mathbb{E}\left[
 \left. 
 \mathrm{e}^{LG_{1}(\boldsymbol{\mathcal{Q}}_{t};n)} 
 \right| \tilde{\mathcal{B}}_{t,\mathcal{K}}, 
\mathcal{B}_{t,\mathcal{K}}, \mathcal{H}, \mathcal{I}_{\mathcal{T}_{\tau}} 
\right], 
\end{equation}
where $G_{1}(\boldsymbol{\mathcal{Q}}_{t},n)$ is given by (\ref{G}).  
In (\ref{Xi_tmp1}), the positive definite Hermitian matrix 
$\boldsymbol{\mathcal{Q}}_{t}\in\mathcal{M}_{(n+1)N}^{+}$ is given by 
\begin{equation} \label{Q}
\boldsymbol{\mathcal{Q}}_{t} = 
\frac{1}{K}\sum_{k=1}^{K}\sum_{m=1}^{M}
\boldsymbol{\omega}_{t,k,m}^{(\mathrm{c})}
(\boldsymbol{\omega}_{t,k,m}^{(\mathrm{c})})^{H}, 
\end{equation}
with $\boldsymbol{\omega}_{t,k,m}^{(\mathrm{c})} 
=((\boldsymbol{\omega}_{t,k,m}^{(\mathrm{c}),\{0\}})^{T},\ldots,  
(\boldsymbol{\omega}_{t,k,m}^{(\mathrm{c}),\{n\}})^{T})^{T}$, in which 
$\boldsymbol{\omega}_{t,k,m}^{(\mathrm{c}),\{a\}}\in\mathbb{C}^{N}$ 
is given by 
\begin{equation} \label{omega_c}
\boldsymbol{\omega}_{t,k,m}^{(\mathrm{c}),\{a\}}=
\tilde{\boldsymbol{h}}_{k,m}^{\{a\}}b_{t,k,m}^{\{a\}}.
\end{equation}

\subsection{Average over Replicated Randomness} 
We next evaluate the expectation in (\ref{Xi_tmp1}) with respect to 
$\boldsymbol{\mathcal{Q}}_{t}$. In the same manner as in the derivation 
of (\ref{mu_tmp1}), the pdf of $\boldsymbol{\mathcal{Q}}_{t}$ conditioned 
on $\tilde{\mathcal{B}}_{t,\mathcal{K}}$,  $\mathcal{B}_{t,\mathcal{K}}$, 
$\mathcal{H}$, and $\mathcal{I}_{\mathcal{T}_{\tau}}$ is evaluated as  
\begin{equation} \label{mu_MMSE} 
\mu_{K}^{(\mathrm{c})}(\boldsymbol{\mathcal{Q}}_{t};n) = \left(
 \frac{K}{2\pi\mathrm{i}} 
\right)^{[(n+1)N]^{2}}
\int\left\{
 \prod_{k\in\mathcal{K}}M_{k}^{(\mathrm{c})}
 (\tilde{\boldsymbol{\mathcal{Q}}}_{t};n) 
\right\}
\mathrm{e}^{-KI_{K}^{(\mathrm{c})}(\boldsymbol{\mathcal{Q}}_{t}, 
\tilde{\boldsymbol{\mathcal{Q}}}_{t};n)}
d\tilde{\boldsymbol{\mathcal{Q}}}_{t}, 
\end{equation}
where we have used the fact that $\{\tilde{\mathcal{H}}_{k}^{\{a\}}: 
a=1,\ldots,n\}$ conditioned on $\mathcal{I}_{\mathcal{T}_{\tau}}$ are mutually 
independent for all $k$. 
In (\ref{mu_MMSE}), $\tilde{\boldsymbol{\mathcal{Q}}}_{t}$ denotes an 
$(n+1)N\times(n+1)N$ non-singular Hermitian matrix, defined in the same manner 
as in (\ref{tilde_Q}). The integration in 
(\ref{mu_MMSE}) with respect to 
each element of $\tilde{\boldsymbol{\mathcal{Q}}}_{t}$ is taken along 
an imaginary axis. The moment generating function 
$M_{k}^{(\mathrm{c})}(\tilde{\boldsymbol{\mathcal{Q}}}_{t};n)$ for  
the users in $\mathcal{K}$ is defined as 
\begin{equation}
M_{k}^{(\mathrm{c})}(\tilde{\boldsymbol{\mathcal{Q}}}_{t};n) 
= \mathbb{E}\left[
 \left. 
  \mathrm{e}^{
   \Lambda_{k}^{(\mathrm{c})}(\tilde{\boldsymbol{\mathcal{Q}}}_{t};n)
  }
 \right| \mathcal{B}_{t,k}, \tilde{\mathcal{B}}_{t,k}, 
 \mathcal{H}_{k}, \mathcal{I}_{\mathcal{T}_{\tau}} 
\right],
\end{equation}
with 
\begin{equation} \label{Lambda_c} 
\Lambda_{k}^{(\mathrm{c})}(\tilde{\boldsymbol{\mathcal{Q}}}_{t};n)
= \sum_{m=1}^{M}\mathrm{Tr}[\boldsymbol{\omega}_{t,k,m}^{(\mathrm{c})}
(\boldsymbol{\omega}_{t,k,m}^{(\mathrm{c})})^{H}
\tilde{\boldsymbol{\mathcal{Q}}}_{t}]. 
\end{equation}
Furthermore, the function 
$I_{K}^{(\mathrm{c})}(\boldsymbol{\mathcal{Q}}_{t}, 
\tilde{\boldsymbol{\mathcal{Q}}}_{t};n)$ is given by 
\begin{equation} \label{I_K_MMSE} 
I_{K}^{(\mathrm{c})}(\boldsymbol{\mathcal{Q}}_{t}, 
\tilde{\boldsymbol{\mathcal{Q}}}_{t};n) = 
\mathrm{Tr}(\boldsymbol{\mathcal{Q}}_{t}\tilde{\boldsymbol{\mathcal{Q}}}_{t}) 
 - \frac{1}{K}\sum_{k\notin\mathcal{K}}\ln\mathbb{E}\left[
  \left. 
   \mathrm{e}^{
    \Lambda_{k}^{(\mathrm{c})}(\tilde{\boldsymbol{\mathcal{Q}}}_{t};n)
   }
  \right| \mathcal{H}_{k}, \mathcal{I}_{\mathcal{T}_{\tau}} 
 \right]. 
\end{equation} 
Note that the second term of the right-hand side of (\ref{I_K_MMSE}) depends 
on $\mathcal{H}$ and $\mathcal{I}_{\mathcal{T}_{\tau}}$, whereas that of 
(\ref{I_K}) is a deterministic value. We use 
Lemma~\ref{lemma_channel_estimation} to show that the second term on the 
right-hand side of (\ref{I_K_MMSE}) converges in probability to a 
deterministic value in the large-system limit. We re-write 
$\mathbb{E}[ \exp\{
 \Lambda_{k}^{(\mathrm{c})}(\tilde{\boldsymbol{\mathcal{Q}}}_{t};n)
\} | \mathcal{H}_{k}, \mathcal{I}_{\mathcal{T}_{\tau}}]$ as 
$X_{k}(\mathcal{H}_{k}, \mathcal{I}_{\mathcal{T}_{\tau}})$, given by 
\begin{equation} 
X_{k}(\mathcal{H}_{k}, \mathcal{I}_{\mathcal{T}_{\tau}}) = 
\int f_{k}(\mathcal{H}_{k},\{\tilde{\mathcal{H}}_{k}^{\{a\}}\})
\prod_{a=1}^{n}\left\{
 p(\tilde{\mathcal{H}}_{k}^{\{a\}} | 
 \mathcal{I}_{\mathcal{T}_{\tau}})d\tilde{\mathcal{H}}_{k}^{\{a\}}
\right\}, 
\end{equation} 
with 
\begin{equation}
f_{k}(\mathcal{H}_{k},\{\tilde{\mathcal{H}}_{k}^{\{a\}}\}) = \mathbb{E}\left[
 \left. 
  \mathrm{e}^{
   \Lambda_{k}^{(\mathrm{c})}(\tilde{\boldsymbol{\mathcal{Q}}}_{t};n)
  }
 \right| \mathcal{H}_{k}, \{\tilde{\mathcal{H}}_{k}^{\{a\}}\} 
\right], 
\end{equation}
where $\{\tilde{\mathcal{H}}_{k}^{\{a\}}\}
=\{\tilde{\mathcal{H}}_{k}^{\{a\}}: a=1,\ldots,n\}$ denotes all replicas 
of $\tilde{\mathcal{H}}_{k}^{\{a\}}$ for user~$k$. 
Lemma~\ref{lemma_channel_estimation} implies that 
$\{X_{k}(\mathcal{H}_{k}, \mathcal{I}_{\mathcal{T}_{\tau}}):
\hbox{for all $k$}\}$ converges in law to uncorrelated random variables 
$\underline{X}_{k}
(\mathcal{H}_{k},\underline{\mathcal{I}}_{\mathcal{T}_{\tau},k})$ 
in the large-system limit, given by 
\begin{equation}
\underline{X}_{k}(\mathcal{H}_{k}, 
\underline{\mathcal{I}}_{\mathcal{T}_{\tau},k}) 
= \int f_{k}(\mathcal{H}_{k},\{\tilde{\mathcal{H}}_{k}^{\{a\}}\})
\prod_{a=1}^{n}\prod_{m=1}^{M}\left\{
 p(\boldsymbol{h}_{k,m}=\tilde{\boldsymbol{h}}_{k,m}^{\{a\}} | 
 \underline{\mathcal{I}}_{\mathcal{T}_{\tau},k,m})
 d\tilde{\boldsymbol{h}}_{k,m}^{\{a\}}
\right\}, 
\end{equation}
with $\underline{\mathcal{I}}_{\mathcal{T}_{\tau},k}=
\{\underline{\mathcal{I}}_{\mathcal{T}_{\tau},k,m}:\hbox{for all $m$}\}$.  
From the weak law of large numbers, we find that 
(\ref{I_K_MMSE}) converges in probability to 
\begin{equation} \label{I_MMSE} 
I^{(\mathrm{c})}
(\boldsymbol{\mathcal{Q}}_{t}, \tilde{\boldsymbol{\mathcal{Q}}}_{t};n) = 
\mathrm{Tr}(\boldsymbol{\mathcal{Q}}_{t}
\tilde{\boldsymbol{\mathcal{Q}}}_{t}) 
- \lim_{K\rightarrow\infty}\frac{1}{K}\sum_{k\notin\mathcal{K}}
\mathbb{E}\left[
 \ln\mathbb{E}\left\{
  \left. 
   \mathrm{e}^{
    \Lambda_{k}^{(\mathrm{c})}(\tilde{\boldsymbol{\mathcal{Q}}}_{t};n)
   }
  \right| \mathcal{H}_{k}, \underline{\mathcal{I}}_{\mathcal{T}_{\tau},k}  
 \right\}
\right], 
\end{equation} 
in the large-system limit.  
%\begin{equation}
%\mathbb{E}\left[
%  \left. 
%   \mathrm{e}^{
%    \Lambda_{k}^{(\mathrm{c})}(\tilde{\boldsymbol{\mathcal{Q}}}_{t};n)
%   }
%  \right| \mathcal{H}_{k}, \underline{\mathcal{I}}_{\mathcal{T}_{\tau},k}  
%\right] 
%= \int \mathrm{e}^{
% \Lambda_{k}^{(\mathrm{c})}(\tilde{\boldsymbol{\mathcal{Q}}}_{t};n)
%}\prod_{a=0}^{n}\left\{
% p(\mathcal{B}_{t,k}^{\{a\}})d\mathcal{B}_{t,k}^{\{a\}}  
%\right\} 
%\left\{
% \prod_{a=1}^{n}\prod_{m=1}^{M}p(\tilde{\boldsymbol{h}}_{k,m}^{\{a\}} | 
% \underline{\mathcal{I}}_{\mathcal{T}_{\tau},k,m})
% d\tilde{\boldsymbol{h}}_{k,m}^{\{a\}} 
%\right\}. 
%\end{equation}
The convergence in probability of (\ref{I_K_MMSE}) to (\ref{I_MMSE}) allows us 
to use the same method as in the derivation of (\ref{Xi_channel_tmp6}). 
Consequently, (\ref{Xi_tmp1}) yields 
\begin{equation} \label{Xi_tmp2} 
\Xi_{n}(\tilde{\mathcal{B}}_{t,\mathcal{K}}, 
\mathcal{B}_{t,\mathcal{K}}, \mathcal{H}, \mathcal{I}_{\mathcal{T}_{\tau}}) = 
D_{n}^{(\mathrm{c})}p( \tilde{\mathcal{B}}_{t,\mathcal{K}}) 
\left\{
 \prod_{k\in\mathcal{K}}M_{k}^{(\mathrm{c})}
 (\tilde{\boldsymbol{\mathcal{Q}}}_{t}^{(\mathrm{s})};n) 
\right\} 
\mathrm{e}^{
 -K\Phi^{(\mathrm{c})}(\boldsymbol{\mathcal{Q}}_{t}^{(\mathrm{s})},
\tilde{\boldsymbol{\mathcal{Q}}}_{t}^{(\mathrm{s})};n)
}[1+o(K)], 
\end{equation}
with 
\begin{equation} \label{Phi_MMSE} 
\Phi^{(\mathrm{c})}(\boldsymbol{\mathcal{Q}}_{t},
\tilde{\boldsymbol{\mathcal{Q}}}_{t};n)=
I^{(\mathrm{c})}(\boldsymbol{\mathcal{Q}}_{t},
\tilde{\boldsymbol{\mathcal{Q}}}_{t};n)
- \beta^{-1}G_{1}(\boldsymbol{\mathcal{Q}}_{t};n).  
\end{equation}
In (\ref{Xi_tmp2}), $D_{n}^{(\mathrm{c})}$ is given by $D_{n}^{(\mathrm{c})}=
[\det\nabla_{\boldsymbol{\mathcal{Q}}_{t}}^{2}
\Phi^{(\mathrm{c})}(\boldsymbol{\mathcal{Q}}_{t}^{(\mathrm{s})},
\tilde{\boldsymbol{\mathcal{Q}}}_{t}^{(\mathrm{s})};n)]^{-1/2}
|\det\nabla_{\tilde{\boldsymbol{\mathcal{Q}}}_{t}}^{2}
I^{(\mathrm{c})}(\boldsymbol{\mathcal{Q}}_{t}^{(\mathrm{s})},
\tilde{\boldsymbol{\mathcal{Q}}}_{t}^{(\mathrm{s})};n)|^{-1/2}$,  
with the Hesse matrices $\nabla_{\boldsymbol{\mathcal{Q}}_{t}}^{2}
\Phi^{(\mathrm{c})}(\boldsymbol{\mathcal{Q}}_{t}^{(\mathrm{s})},
\tilde{\boldsymbol{\mathcal{Q}}}_{t}^{(\mathrm{s})};n)$ and 
$\nabla_{\tilde{\boldsymbol{\mathcal{Q}}}_{t}}^{2}
I^{(\mathrm{c})}(\boldsymbol{\mathcal{Q}}_{t}^{(\mathrm{s})},
\tilde{\boldsymbol{\mathcal{Q}}}_{t}^{(\mathrm{s})};n)$  of 
(\ref{Phi_MMSE}) and (\ref{I_MMSE}) with respect to 
$\boldsymbol{\mathcal{Q}}_{t}$ and $\tilde{\boldsymbol{\mathcal{Q}}}_{t}$ 
at the saddle-point 
$(\boldsymbol{\mathcal{Q}}_{t},\tilde{\boldsymbol{\mathcal{Q}}}_{t})=
(\boldsymbol{\mathcal{Q}}_{t}^{(\mathrm{s})},
\tilde{\boldsymbol{\mathcal{Q}}}_{t}^{(\mathrm{s})})$, respectively, 
which is a solution to the coupled fixed-point equations  
\begin{equation} \label{equ1_MMSE} 
\boldsymbol{\mathcal{Q}}_{t} = \lim_{K\rightarrow\infty}
\frac{1}{K}\sum_{k\notin \mathcal{K}}\sum_{m=1}^{M}\mathbb{E}\left[
 \frac{
  \mathbb{E}\left\{ 
   \left. 
    \boldsymbol{\omega}_{t,k,m}^{(\mathrm{c})}
    (\boldsymbol{\omega}_{t,k,m}^{(\mathrm{c})})^{H}
    \mathrm{e}^{\Lambda_{k}^{(\mathrm{c})}
    (\tilde{\boldsymbol{\mathcal{Q}}}_{t};n)}
   \right| \mathcal{H}_{k}, \underline{\mathcal{I}}_{\mathcal{T}_{\tau},k}  
  \right\} 
 }
 {
  \mathbb{E}\left\{  
   \left. 
    \mathrm{e}^{\Lambda_{k}^{(\mathrm{c})}
    (\tilde{\boldsymbol{\mathcal{Q}}}_{t};n)}
   \right| \mathcal{H}_{k}, \underline{\mathcal{I}}_{\mathcal{T}_{\tau},k}  
  \right\} 
 }
\right], 
\end{equation}
\begin{equation} \label{equ2_MMSE} 
\tilde{\boldsymbol{\mathcal{Q}}}_{t} = -\beta^{-1}
(\boldsymbol{I} + \boldsymbol{A}_{1}(n)
\boldsymbol{\mathcal{Q}}_{t})^{-1}\boldsymbol{A}_{1}(n).
\end{equation}
If the coupled fixed-point equations~(\ref{equ1_MMSE}) 
and~(\ref{equ2_MMSE}) have multiple solutions, the solution to minimize  
(\ref{Phi_MMSE}) is chosen. 

\begin{remark}
A non-negative-entropy condition would be defined via the entropy of 
(\ref{mu_MMSE}) if (\ref{Q}) is discrete or if the data symbols and the 
channel vectors were discrete random variables~\cite{Tanaka02}. However, 
it is unrealistic to assume that the channel vectors are discrete. Thus, 
the {\em conventional} non-negative-entropy condition is not defined for the 
no-CSI case. A non-negative-entropy condition might be defined via the 
entropy for the pdf of $\boldsymbol{\mathcal{Q}}_{t}$ marginalized over 
the channel vectors. However, its calculation is not straightforward. 
\end{remark}

\subsection{Replica Continuity} 
The evaluation of (\ref{Xi_tmp2}), (\ref{equ1_MMSE}), and (\ref{equ2_MMSE}) 
under the RS assumption is almost the same as in the derivations of 
(\ref{conditional_distribution_rs}), (\ref{Sigma_0_rs_n0}), and 
(\ref{Sigma_rs_n0}).  
Therefore, we omit the details and only present the results: 
\begin{equation} \label{conditional_pdf} 
\lim_{K,L\rightarrow\infty}
p(\tilde{\mathcal{B}}_{t,\mathcal{K}}| \mathcal{B}_{t,\mathcal{K}}, 
\mathcal{H}, \mathcal{I}_{\mathcal{T}_{\tau}}) = 
\prod_{k\in\mathcal{K}}\int q(\tilde{\mathcal{B}}_{t,k} | 
\underline{\mathcal{Y}}_{t,k}, \mathcal{I}_{\mathcal{T}_{\tau}} )
\prod_{m=1}^{M}\left\{
 g_{N}\left(
  \underline{\boldsymbol{y}}_{t,k,m} - \boldsymbol{h}_{k,m}b_{t,k,m};
  \boldsymbol{\Sigma}_{0}^{(t)}
 \right)
 d\underline{\boldsymbol{y}}_{t,k,m}
\right\}, 
\end{equation}
for $\underline{\boldsymbol{y}}_{t,k,m}\in
\mathbb{C}^{N}$. In (\ref{conditional_pdf}), 
$g_{N}(\underline{\boldsymbol{y}}_{t,k,m}  
- \boldsymbol{h}_{k,m}b_{t,k,m};\boldsymbol{\Sigma}_{0}^{(t)})$ is 
defined as (\ref{Gauss}). Furthermore, 
$q(\tilde{\mathcal{B}}_{t,k} | \underline{\mathcal{Y}}_{t,k}, 
\mathcal{I}_{\mathcal{T}_{\tau}} )$, with 
$\underline{\mathcal{Y}}_{t,k} 
=\{\underline{\boldsymbol{y}}_{t,k,m}:\hbox{for all $m$}\}$, 
is given by 
\begin{equation}
q(\tilde{\mathcal{B}}_{t,k} | \underline{\mathcal{Y}}_{t,k}, 
\mathcal{I}_{\mathcal{T}_{\tau}} ) = 
\frac{
 \int \prod_{m=1}^{M}g_{N}(\underline{\boldsymbol{y}}_{t,k,m} -  
  \tilde{\boldsymbol{h}}_{k,m}\tilde{b}_{t,k,m};\boldsymbol{\Sigma}^{(t)} )
 p(\tilde{\mathcal{B}}_{t,k})
 p(\tilde{\mathcal{H}}_{k} | \mathcal{I}_{\mathcal{T}_{\tau}})
 d\tilde{\mathcal{H}}_{k} 
}
{
 \int \prod_{m=1}^{M}g_{n}(\underline{\boldsymbol{y}}_{t,k,m} -  
 \tilde{\boldsymbol{h}}_{k,m}\tilde{b}_{t,k,m};\boldsymbol{\Sigma}^{(t)} )
 p(\tilde{\mathcal{B}}_{t,k})
 p(\tilde{\mathcal{H}}_{k} | \mathcal{I}_{\mathcal{T}_{\tau}}) 
 d\tilde{\mathcal{B}}_{t,k}d\tilde{\mathcal{H}}_{k}
}. 
\end{equation}
In these expressions, 
$(\boldsymbol{\Sigma}_{0}^{(t)}, \boldsymbol{\Sigma}^{(t)})$ is a 
solution to the coupled fixed-point equations 
\begin{equation} \label{Sigma_0_rs_n0_MMSE} 
\boldsymbol{\Sigma}_{0}^{(t)} = N_{0}\boldsymbol{I}_{N} + 
\lim_{K\rightarrow\infty}\frac{\beta}{K}\sum_{k\notin \mathcal{K}}
\sum_{m=1}^{M}
\mathbb{E}\left[ 
 (\tilde{\boldsymbol{h}}_{k,m}^{\{0\}}b_{t,k,m}^{\{0\}}
 - \langle \tilde{\boldsymbol{h}}_{k,m}^{\{1\}}b_{t,k,m}^{\{1\}} 
 \rangle )( \tilde{\boldsymbol{h}}_{k,m}^{\{0\}}b_{t,k,m}^{\{0\}}
 - \langle \tilde{\boldsymbol{h}}_{k,m}^{\{1\}}b_{t,k,m}^{\{1\}} 
 \rangle )^{H} 
\right], 
\end{equation}
\begin{equation} \label{Sigma_rs_n0_MMSE} 
\boldsymbol{\Sigma}^{(t)} = N_{0}\boldsymbol{I}_{N} + 
\lim_{K\rightarrow\infty}\frac{\beta}{K}\sum_{k\notin \mathcal{K}}
\sum_{m=1}^{M}
\mathbb{E}\left[ 
 \left\langle 
  (\tilde{\boldsymbol{h}}_{k,m}^{\{1\}}b_{t,k,m}^{\{1\}}
  - \langle \tilde{\boldsymbol{h}}_{k,m}^{\{1\}}b_{t,k,m}^{\{1\}}
  \rangle )( \tilde{\boldsymbol{h}}_{k,m}^{\{1\}}b_{t,k,m}^{\{1\}}
  - \langle \tilde{\boldsymbol{h}}_{k,m}^{\{1\}}b_{t,k,m}^{\{1\}} 
  \rangle )^{H} 
 \right\rangle 
\right], 
\end{equation}
where $\langle \tilde{\boldsymbol{h}}_{k,m}^{\{a\}}b_{t,k,m}^{\{a\}} \rangle$ 
denotes the mean of $\tilde{\boldsymbol{h}}_{k,m}^{\{a\}}b_{t,k,m}^{\{a\}}$ 
with respect to $q(b_{t,k,m}^{\{a\}}, \tilde{\boldsymbol{h}}_{k,m}^{\{a\}} | 
\underline{\boldsymbol{y}}_{t,k,m}, 
\underline{\mathcal{I}}_{\mathcal{T}_{\tau},k,m})d\tilde{b}_{t,k,m}
d\tilde{\boldsymbol{h}}_{k,m}$, given by    
\begin{equation} 
q(b_{t,k,m}^{\{a\}}, \tilde{\boldsymbol{h}}_{k,m}^{\{a\}} | 
\underline{\boldsymbol{y}}_{t,k,m}, 
\underline{\mathcal{I}}_{\mathcal{T}_{\tau},k,m}) = 
\frac{
 g_{N}(\underline{\boldsymbol{y}}_{t,k,m} 
 - \tilde{\boldsymbol{h}}_{k,m}^{\{a\}}b_{t,k,m}^{\{a\}}; 
 \boldsymbol{\Sigma}_{a}^{(t)} )p(b_{t,k,m}^{\{a\}})
 p(\tilde{\boldsymbol{h}}_{k,m}^{\{a\}} | 
 \underline{\mathcal{I}}_{\mathcal{T}_{\tau},k,m}) 
}
{
 \int g_{N}(\underline{\boldsymbol{y}}_{t,k,m} -  
 \tilde{\boldsymbol{h}}_{k,m}^{\{a\}}b_{t,k,m}^{\{a\}}; 
 \boldsymbol{\Sigma}_{a}^{(t)})
 p(b_{t,k,m}^{\{a\}})p(\tilde{\boldsymbol{h}}_{k,m}^{\{a\}} | 
 \underline{\mathcal{I}}_{\mathcal{T}_{\tau},k,m}) 
 db_{t,k,m}^{\{a\}}d\tilde{\boldsymbol{h}}_{k,m}^{\{a\}}
}, 
\end{equation}
with $\boldsymbol{\Sigma}_{1}^{(t)}=\boldsymbol{\Sigma}^{(t)}$. 
In the right-hand sides of (\ref{Sigma_0_rs_n0_MMSE}) and 
(\ref{Sigma_rs_n0_MMSE}), the expectations 
are taken with respect to the measure  
$g_{N}(\underline{\boldsymbol{y}}_{t,k,m} - \boldsymbol{h}_{k,m}b_{t,k,m};
\boldsymbol{\Sigma}_{0}^{(t)})
d\underline{\boldsymbol{y}}_{t,k,m} p(b_{t,k,m})
db_{t,k,m}
p(\boldsymbol{h}_{k,m} | \underline{\mathcal{I}}_{\mathcal{T}_{\tau},k,m}) 
p(\underline{\mathcal{I}}_{\mathcal{T}_{\tau},k,m})d\boldsymbol{h}_{k,m}
d\underline{\mathcal{I}}_{\mathcal{T}_{\tau},k,m}$. 

Let us assume $\boldsymbol{\Sigma}_{0}^{(t)}=\boldsymbol{\Sigma}^{(t)}$. 
Then, the coupled fixed-point equations~(\ref{Sigma_0_rs_n0_MMSE}) and 
(\ref{Sigma_rs_n0_MMSE}) reduce to the single fixed-point equation 
\begin{equation} \label{single_Sigma_rs_n0_MMSE} 
\boldsymbol{\Sigma}_{0}^{(t)} = N_{0}\boldsymbol{I}_{N} + 
\lim_{K\rightarrow\infty}\frac{\beta}{K}\sum_{k\notin \mathcal{K}}
\sum_{m=1}^{M}
\mathbb{E}\left[ 
 (\boldsymbol{h}_{k,m}b_{t,k,m}
 - \langle \boldsymbol{h}_{k,m}b_{t,k,m} \rangle )
 ( \boldsymbol{h}_{k,m}b_{t,k,m}
 - \langle \boldsymbol{h}_{k,m}b_{t,k,m} \rangle )^{H} 
\right]. 
\end{equation} 
In (\ref{single_Sigma_rs_n0_MMSE}), 
$\langle \boldsymbol{h}_{k,m}b_{t,k,m} \rangle$ denotes the 
mean of $\boldsymbol{h}_{k,m}b_{t,k,m}$ with respect to the posterior 
pdf~(\ref{joint_posterior_pdf}). Furthermore, 
the expectation is taken with respect to the same measure as that for 
(\ref{Sigma_0_rs_n0_MMSE}). It is shown in 
Appendix~\ref{appendix_fixed_point} that 
(\ref{single_Sigma_rs_n0_MMSE}) reduces to (\ref{fixed_point_data}) under 
the assumption of 
$\boldsymbol{\Sigma}_{0}^{(t)}=\sigma_{\mathrm{c}}^{2}\boldsymbol{I}_{N}$.  
Furthermore, it is straightforward to find that (\ref{conditional_pdf}) is 
equivalent to (\ref{decoupling}) under Assumption~\ref{assumption1}. 

\subsection{Multiple Solutions} 
\label{section_multiple_solutions} 
We consider the case in which the coupled fixed-point 
equations~(\ref{Sigma_0_rs_n0_MMSE}) and 
(\ref{Sigma_rs_n0_MMSE}) have multiple solutions. In this case, 
we should choose the solution minimizing (\ref{Phi_MMSE}) under the RS 
assumption for $n\in[0,\epsilon)$, with a sufficiently small $\epsilon>0$. 
Since $\lim_{n\rightarrow+0}\Phi^{(\mathrm{c})}
(\boldsymbol{\mathcal{Q}}_{t}^{(\mathrm{s})},
\tilde{\boldsymbol{\mathcal{Q}}}_{t}^{(\mathrm{s})};n)=0$, 
that solution is given as the solution minimizing the derivative 
of (\ref{Phi_MMSE}) with respect to $n$ in the limit $n\rightarrow+0$:  
\begin{equation} \label{free_energy_rs} 
F_{\mathrm{rs}}\equiv \lim_{n\rightarrow +0}
\frac{\partial}{\partial n}\Phi^{(\mathrm{c})}
(\boldsymbol{\mathcal{Q}}_{t}^{(\mathrm{s})},
\tilde{\boldsymbol{\mathcal{Q}}}_{t}^{(\mathrm{s})};n) = 
\lim_{n\rightarrow+0}\frac{\partial}{\partial n}
\Phi^{(\mathrm{c})}\left(
 \underline{\boldsymbol{\mathcal{Q}}}_{t}^{(\mathrm{s})},
 \underline{\tilde{\boldsymbol{\mathcal{Q}}}}_{t}^{(\mathrm{s})};n
\right), 
\end{equation}
with $\underline{\boldsymbol{\mathcal{Q}}}_{t}^{(\mathrm{s})}=
\lim_{n\rightarrow+0}\boldsymbol{\mathcal{Q}}_{t}^{(\mathrm{s})}$ and 
$\underline{\tilde{\boldsymbol{\mathcal{Q}}}}_{t}^{(\mathrm{s})}=
\lim_{n\rightarrow+0}\tilde{\boldsymbol{\mathcal{Q}}}_{t}^{(\mathrm{s})}$, 
in which we have used the stationarity condition~(\ref{equ2_MMSE}) to obtain 
the last expression.  

We calculate (\ref{G}) and (\ref{I_MMSE}) in (\ref{Phi_MMSE}) under the RS 
assumption in the same manner as in the derivations of (\ref{G_rs}) and 
(\ref{I_rs}) 
and subsequently differentiate the obtained formula with respect to 
$n$ in $n\rightarrow+0$, to have 
\begin{equation} \label{deriv_G_n}
\lim_{n\rightarrow+0}\frac{\partial G_{1}}{\partial n}\left(
 \underline{\boldsymbol{\mathcal{Q}}}_{t}^{(\mathrm{s})};n
\right) = 
- \ln\det\boldsymbol{\Sigma}^{(t)} 
- \mathrm{Tr}\left[
 (\boldsymbol{\Sigma}^{(t)})^{-1}\boldsymbol{\Sigma}_{0}^{(t)}
\right] - N\ln\pi, 
\end{equation}
\begin{IEEEeqnarray}{rl} 
&\lim_{n\rightarrow+0}\frac{\partial I}{\partial n}\left(
 \underline{\boldsymbol{\mathcal{Q}}}_{t}^{(\mathrm{s})},
 \underline{\tilde{\boldsymbol{\mathcal{Q}}}}_{t}^{(\mathrm{s})};n
\right) \nonumber \\ 
=&  \lim_{K\rightarrow\infty}\frac{1}{K}\sum_{k\notin\mathcal{K}}
\sum_{m=1}^{M}\mathbb{E}[\tilde{C}_{k,m}]\ln2  
-\frac{1}{\beta}\mathrm{Tr}\left[
 \boldsymbol{I}_{N} - 2N_{0}(\boldsymbol{\Sigma}^{(t)})^{-1} 
 + N_{0}(\boldsymbol{\Sigma}^{(t)})^{-1}
 \boldsymbol{\Sigma}_{0}^{(t)}(\boldsymbol{\Sigma}^{(t)})^{-1}
\right], \label{deriv_I_n}
\end{IEEEeqnarray}  
with 
\begin{equation}
\tilde{C}_{k,m} = 
\int g_{N}(\underline{\boldsymbol{y}}_{t,k,m} -  
\boldsymbol{h}_{k,m}b_{t,k,m};\boldsymbol{\Sigma}_{0}^{(t)})
p(b_{t,k,m})\log\frac{
 g_{N}(\underline{\boldsymbol{y}}_{t,k,m} - \boldsymbol{h}_{k,m}b_{t,k,m}
 ;\boldsymbol{\Sigma}^{(t)})
}
{
 \mathbb{E}\left[
  \left. 
   g_{N}(\underline{\boldsymbol{y}}_{t,k,m} - \boldsymbol{h}_{k,m}b_{t,k,m}; 
   \boldsymbol{\Sigma}^{(t)} )
  \right| \underline{\mathcal{I}}_{\mathcal{T}_{\tau},k,m}   
 \right]
}d\underline{\boldsymbol{y}}_{t,k,m}db_{t,k,m}, 
\end{equation}
where the conditional expectation is taken with respect to 
$\boldsymbol{h}_{k,m}$ and $b_{t,k,m}$. 
Substituting (\ref{deriv_G_n}) and (\ref{deriv_I_n}) into 
(\ref{free_energy_rs}), we obtain 
\begin{IEEEeqnarray}{rl} 
\frac{\beta}{\ln2} F_{\mathrm{rs}} =&  
\lim_{K\rightarrow\infty}\frac{\beta}{K}\sum_{k\notin\mathcal{K}}
\sum_{m=1}^{M}\mathbb{E}[\tilde{C}_{k,m}] 
+ 2D(N_{0}\boldsymbol{I}_{N}\|\boldsymbol{\Sigma}^{(t)}) 
+ D(\boldsymbol{\Sigma}_{0}^{(t)}\|\boldsymbol{\Sigma}^{(t)}) 
- D(N_{0}\boldsymbol{I}_{N} \| 
\boldsymbol{\Sigma}^{(t)}(\boldsymbol{\Sigma}_{0}^{(t)})^{-1}
\boldsymbol{\Sigma}^{(t)}) \nonumber \\ 
&+ N\log(\pi\mathrm{e}N_{0}). \label{free_energy_rs_result} 
\end{IEEEeqnarray}
It is straightforward to find that (\ref{free_energy_rs_result}) 
reduces to (\ref{free_energy}) with the exception of the constant 
$N\log(\pi\mathrm{e}N_{0})$, by substituting  
$\boldsymbol{\Sigma}_{0}^{(t)}=\boldsymbol{\Sigma}^{(t)}=
\sigma_{\mathrm{c}}^{2}\boldsymbol{I}_{N}$ into (\ref{free_energy_rs_result}).

\section{Derivation of (\ref{fixed_point_data})}
\label{appendix_fixed_point}   
We assume 
$\boldsymbol{\Sigma}_{0}^{(t)}=\sigma_{\mathrm{c}}^{2}\boldsymbol{I}_{N}$. 
Taking the traces for both sides of (\ref{single_Sigma_rs_n0_MMSE}) divided 
by $N$, we obtain 
\begin{equation} \label{appen_fixed_point} 
\sigma_{\mathrm{c}}^{2} = N_{0} + \lim_{K\rightarrow\infty}\frac{\beta}{NK} 
\sum_{k\notin\mathcal{K}}\sum_{m=1}^{M}\mathrm{MMSE}_{t,k,m}, 
\end{equation}
with 
\begin{equation} \label{MMSE} 
\mathrm{MMSE}_{t,k,m}=\mathbb{E}\left[
 \|\boldsymbol{h}_{k,m}b_{t,k,m} - \langle \boldsymbol{h}_{k,m}b_{t,k,m} 
 \rangle\|^{2} 
\right]. 
\end{equation} 
In (\ref{MMSE}), $\langle \boldsymbol{h}_{k,m}b_{t,k,m} \rangle$ denotes the 
mean of $\boldsymbol{h}_{k,m}b_{t,k,m}$ with respect to the posterior 
pdf~(\ref{joint_posterior_pdf}). 

We first calculate the conditional joint pdf  
$p(\underline{\boldsymbol{y}}_{t,k,m}, \boldsymbol{h}_{k,m} | 
b_{t,k,m}, \underline{\mathcal{I}}_{\mathcal{T}_{\tau},k,m})$ 
to evaluate the posterior mean of $\boldsymbol{h}_{k,m}b_{t,k,m}$. 
The channel vector $\boldsymbol{h}_{k,m}$ conditioned on 
$\underline{\mathcal{I}}_{\mathcal{T}_{\tau},k,m}$ 
follows $\mathcal{CN}(
\underline{\hat{\boldsymbol{h}}}_{k,m}^{\mathcal{T}_{\tau}}, 
\xi^{2}\boldsymbol{I}_{N})$, with 
$\underline{\hat{\boldsymbol{h}}}_{k,m}^{\mathcal{T}_{\tau}}$ and 
$\xi^{2}=\xi^{2}(\sigma_{\mathrm{tr}}^{2}(\tau),\tau)$ given by 
(\ref{decoupled_LMMSE_channel_estimate}) and 
(\ref{xi}), respectively. 
On the other hand, (\ref{SIMO}) yields  
\begin{equation}
p(\underline{\boldsymbol{y}}_{t,k,m} | \boldsymbol{h}_{k,m}, b_{t,k,m}) = 
\frac{1}{(\pi\sigma_{\mathrm{c}}^{2})^{N}}\exp\left(
 -\frac{\|\underline{\boldsymbol{y}}_{t,k,m} 
 - \boldsymbol{h}_{k,m}b_{t,k,m}\|^{2}}
 {\sigma_{\mathrm{c}}^{2}} 
\right). 
\end{equation} 
Calculating $p(\underline{\boldsymbol{y}}_{t,k,m}, \boldsymbol{h}_{k,m} | 
b_{t,k,m}, \underline{\mathcal{I}}_{\mathcal{T}_{\tau},k,m}) 
=p(\underline{\boldsymbol{y}}_{t,k,m} | \boldsymbol{h}_{k,m}, b_{t,k,m})
p(\boldsymbol{h}_{k,m} | \underline{\mathcal{I}}_{\mathcal{T}_{\tau},k,m})$, 
we obtain 
\begin{equation}
p(\underline{\boldsymbol{y}}_{t,k,m}, \boldsymbol{h}_{k,m} | b_{t,k,m}, 
\underline{\mathcal{I}}_{\mathcal{T}_{\tau},k,m}) = 
p(\boldsymbol{h}_{k,m} | \underline{\boldsymbol{y}}_{t,k,m}, b_{t,k,m}, 
\underline{\mathcal{I}}_{\mathcal{T}_{\tau},k,m})
p(\underline{\boldsymbol{y}}_{t,k,m} | b_{t,k,m}, 
\underline{\mathcal{I}}_{\mathcal{T}_{\tau},k,m}),  
\end{equation} 
with
\begin{IEEEeqnarray}{r}
p(\boldsymbol{h}_{k,m} | \underline{\boldsymbol{y}}_{t,k,m}, b_{t,k,m},  
\underline{\mathcal{I}}_{\mathcal{T}_{\tau},k,m}) = 
\frac{(P/M)\xi^{2} + \sigma_{\mathrm{c}}^{2}}
{\pi\xi^{2}\sigma_{\mathrm{c}}^{2}}
\exp\left\{ 
 -\frac{(P/M)\xi^{2} + \sigma_{\mathrm{c}}^{2}}
 {\xi^{2}\sigma_{\mathrm{c}}^{2}}
\right. \nonumber \\ 
\left.  
 \times\left\|
  \boldsymbol{h}_{k,m}
  - \frac{\xi^{2}}{(P/M)\xi^{2} 
  + \sigma_{\mathrm{c}}^{2}}
  (b_{t,k,m})^{*}\underline{\boldsymbol{y}}_{t,k,m}
  - \frac{\sigma_{\mathrm{c}}^{2}}{(P/M)\xi^{2} + 
  \sigma_{\mathrm{c}}^{2}}
  \underline{\hat{\boldsymbol{h}}}_{k,m}^{\mathcal{T}_{\tau}}  
 \right\|^{2}
\right\}, 
\end{IEEEeqnarray}
where we have used the fact that $|b_{t,k,m}|^{2}$ equals $P/M$ with 
probability one. By definition, the posterior mean of 
$\boldsymbol{h}_{k,m}b_{t,k,m}$ is given by 
\begin{equation}
\langle \boldsymbol{h}_{k,m}b_{t,k,m} \rangle 
= \frac{(P/M)\xi^{2}}{(P/M)\xi^{2} + 
\sigma_{\mathrm{c}}^{2}}\underline{\boldsymbol{y}}_{t,k,m} 
+ \frac{\sigma_{\mathrm{c}}^{2}}{(P/M)\xi^{2} 
+ \sigma_{\mathrm{c}}^{2}}
\underline{\hat{\boldsymbol{h}}}_{k,m}^{\mathcal{T}_{\tau}}   
\langle b_{t,k,m} \rangle.  
\label{posterior_mean_red} 
\end{equation} 

We next evaluate (\ref{MMSE}). Substituting (\ref{SIMO}) and 
(\ref{posterior_mean_red}) into (\ref{MMSE}) yields 
\begin{equation} \label{eq1}
\mathrm{MMSE}_{t,k,m}
= \mathbb{E}\left[
 \left\|
  \boldsymbol{c}_{t,k,m} 
  + \frac{\sigma_{\mathrm{c}}^{2}}{(P/M)\xi^{2} 
  + \sigma_{\mathrm{c}}^{2}}
  \underline{\hat{\boldsymbol{h}}}_{k,m}^{\mathcal{T}_{\tau}}  
  (b_{t,k,m} - \langle b_{t,k,m} \rangle) 
 \right\|^{2} 
\right], 
\end{equation} 
where $\boldsymbol{c}_{t,k,m}$ is defined as 
\begin{equation}
\boldsymbol{c}_{t,k,m} = 
\frac{\sigma_{\mathrm{c}}^{2}}{(P/M)\xi^{2} 
+ \sigma_{\mathrm{c}}^{2}}
( \boldsymbol{h}_{k,m} 
- \underline{\hat{\boldsymbol{h}}}_{k,m}^{\mathcal{T}_{\tau}} )b_{t,k,m}  
- \frac{(P/M)\xi^{2}}{(P/M)\xi^{2} 
+ \sigma_{\mathrm{c}}^{2}}\underline{\boldsymbol{n}}_{t,k,m}. 
\end{equation}
The fact that $\boldsymbol{c}_{t,k,m}$ and 
$\underline{\boldsymbol{y}}_{t,k,m}$ conditioned on $b_{t,k,m}$ and 
$\underline{\mathcal{I}}_{\mathcal{T}_{\tau},k,m}$ are 
jointly CSCG is useful for showing that the two 
terms on the right-hand side of (\ref{eq1}) are mutually independent 
under the same conditions.  
The means of $\boldsymbol{c}_{t,k,m}$ and $\underline{\boldsymbol{y}}_{t,k,m}$ 
conditioned on $b_{t,k,m}$ and 
$\underline{\mathcal{I}}_{\mathcal{T}_{\tau},k,m}$ 
are zero and $\underline{\hat{\boldsymbol{h}}}_{k,m}^{\mathcal{T}_{\tau}}
b_{t,k,m}$, respectively. 
Also, the covariance matrix of $(\boldsymbol{c}_{t,k,m}^{T},\  
\underline{\boldsymbol{y}}_{t,k,m}^{T})^{T}$ conditioned on $b_{t,k,m}$ and 
$\underline{\mathcal{I}}_{\mathcal{T}_{\tau},k,m}$ is evaluated as the 
diagonal matrix  
$\mathrm{diag}\{(P/M)\xi^{2}\sigma_{\mathrm{c}}^{2}/ 
\{(P/M)\xi^{2}+\sigma_{\mathrm{c}}^{2}\}\boldsymbol{I}_{N},\ 
\{(P/M)\xi^{2}+\sigma_{\mathrm{c}}^{2}\}\boldsymbol{I}_{N}\}$. 
Therefore, $\boldsymbol{c}_{t,k,m}$ and $\underline{\boldsymbol{y}}_{t,k,m}$ 
conditioned on $b_{t,k,m}$ and 
$\underline{\mathcal{I}}_{\mathcal{T}_{\tau},k,m}$ 
are mutually independent. This fact indicates that the two terms on the 
right-hand side of (\ref{eq1}) conditioned on $b_{t,k,m}$ and 
$\underline{\mathcal{I}}_{\mathcal{T}_{\tau},k,m}$ 
are independent of each other, since the second term on the right-hand side of 
(\ref{eq1}) conditioned on $b_{t,k,m}$ and 
$\underline{\mathcal{I}}_{\mathcal{T}_{\tau},k,m}$ 
is a function of $\underline{\boldsymbol{y}}_{t,k,m}$. 
Hence, from (\ref{eq1}), we have 
\begin{equation}
\mathrm{MMSE}_{t,k,m}
= \frac{N(P/M)\xi^{2}\sigma_{\mathrm{c}}^{2}}
{(P/M)\xi^{2}+\sigma_{\mathrm{c}}^{2}}
+ \left(
 \frac{\sigma_{\mathrm{c}}^{2}}{(P/M)\xi^{2} +
 \sigma_{\mathrm{c}}^{2}}
\right)^{2}\mathbb{E}\left[
 \|\underline{\hat{\boldsymbol{h}}}_{k,m}^{\mathcal{T}_{\tau}}\|^{2}
 |b_{t,k,m} - \langle b_{t,k,m} \rangle|^{2}
\right]. \label{eq2} 
\end{equation} 
We substitute (\ref{eq2}) into the fixed-point 
equation~(\ref{appen_fixed_point}) to obtain (\ref{fixed_point_data}).

\section{Proof of Proposition~\ref{proposition2}}
\label{derivation_proposition2} 
%The equivalent channel between user~$k$ and the associated decoder depends on 
%realizations of the channel vectors~$\mathcal{H}_{k}$ and 
%the information $\mathcal{I}_{\mathcal{T}_{\tau}}$ in the training phase even 
%in the large-system limit. 
%The equivalent channel can be regarded as a random variable, 
%depending on $\mathcal{H}_{k}$ and $\mathcal{I}_{\mathcal{T}_{\tau}}$, on 
%the space of distributions. We show that the random variable is statistically 
%equivalent to  in the large-system limit. 

We take the expectation of (\ref{decoupling}) for $\mathcal{K}=\{k\}$ with 
respect to $p(\mathcal{H}_{k}|\mathcal{I}_{\mathcal{T}_{\tau}})$ to obtain 
\begin{equation} \label{decoupling_tmp} 
\lim_{K,L\rightarrow\infty}p(\tilde{\mathcal{B}}_{t,k}|
\mathcal{B}_{t,k},\mathcal{I}_{\mathcal{T}_{\tau}},\mathcal{S}_{t}) = 
p(\underline{\tilde{\mathcal{B}}}_{t,k}=\tilde{\mathcal{B}}_{t,k} | 
\mathcal{B}_{t,k}, \mathcal{I}_{\mathcal{T}_{\tau}})
\quad \hbox{in law.} 
\end{equation}
Applying this expression to (\ref{spectral_efficiency_MMSE}), 
we have 
\begin{equation} \label{spectral_efficiency_MMSE_tmp} 
\lim_{K,L\rightarrow\infty}C_{\mathrm{sep}} =
\beta\left(
 1 - \frac{\tau}{T_{\mathrm{c}}} 
\right)\lim_{K,L\rightarrow\infty}
I(\mathcal{B}_{\tau+1,1},\underline{\mathcal{Y}}_{\tau+1,1} | 
\mathcal{I}_{\mathcal{T}_{\tau}}), 
\end{equation} 
where we have used the fact that 
$\underline{\tilde{\mathcal{B}}}_{t,k}$ contains all information about 
the received vectors $\underline{\mathcal{Y}}_{t,k}$ in the single-user SIMO 
channel (\ref{SIMO}) for the estimation of $\mathcal{B}_{t,k}$. 

In order to show that the right-hand side of 
(\ref{spectral_efficiency_MMSE_tmp}) is equal to 
(\ref{spectral_efficiency_MMSE_dec}), we regard the conditional pdf 
$p(\underline{\mathcal{Y}}_{t,k}|\mathcal{B}_{t,k}, 
\mathcal{I}_{\mathcal{T}_{\tau}})$ as a 
random variable and write it as $X_{k}(\mathcal{I}_{\mathcal{T}_{\tau}};\Theta)
\geq0$, with $\Theta = \{\mathcal{B}_{t,k},\  
\{\underline{\boldsymbol{n}}_{t,k,m}:\hbox{for all $m$}\}\}$, given by 
\begin{equation} \label{X_k_appen} 
X_{k}(\mathcal{I}_{\mathcal{T}_{\tau}};\Theta) = 
\int f_{k}(\mathcal{H}_{k};\Theta)
p(\mathcal{H}_{k} |\mathcal{I}_{\mathcal{T}_{\tau}})d\mathcal{H}_{k}, 
\end{equation}
with 
\begin{equation}
f_{k}(\mathcal{H}_{k};\Theta) = 
\prod_{m=1}^{M}p(\underline{\boldsymbol{y}}_{t,k,m}=
\boldsymbol{h}_{k,m}b_{t,k,m}+
\underline{\boldsymbol{n}}_{t,k,m} | \boldsymbol{h}_{k,m}, b_{t,k,m}), 
\end{equation} 
where $p(\underline{\boldsymbol{y}}_{t,k,m} | \boldsymbol{h}_{k,m},b_{t,k,m})$ 
represents the single-user SIMO channel~(\ref{SIMO}). 
There exists the moment generating function of (\ref{X_k_appen}) 
since (\ref{X_k_appen}) is bounded. Then, 
Lemma~\ref{lemma_channel_estimation} implies that 
$X_{k}\sim \prod_{m=1}^{M}\underline{X}_{k,m}
(\underline{\mathcal{I}}_{\mathcal{T}_{\tau},k,m}; \Theta)$ given $\Theta$, 
defined as 
\begin{equation} \label{decoupled_X_k_appen} 
\underline{X}_{k,m}(\underline{\mathcal{I}}_{\mathcal{T}_{\tau},k,m};\Theta) = 
\int p(\underline{\boldsymbol{y}}_{t,k,m}=\boldsymbol{h}_{k,m}b_{t,k,m}
+\underline{\boldsymbol{n}}_{t,k,m}
| \boldsymbol{h}_{k,m}, b_{t,k,m})
p(\boldsymbol{h}_{k,m}| \underline{\mathcal{I}}_{\mathcal{T}_{\tau},k,m})
d\boldsymbol{h}_{k,m}. 
\end{equation}
In evaluating (\ref{decoupled_X_k_appen}), 
$\sigma_{t}^{2}=\sigma_{\mathrm{tr}}^{2}(\tau)$ for $t\in\mathcal{T}_{\tau}$ is 
given by the solution to the fixed-point equation~(\ref{fixed_point_channel}). 
Applying this result to (\ref{spectral_efficiency_MMSE_tmp}), we find that 
the right-hand side of (\ref{spectral_efficiency_MMSE_tmp}) is equal to 
(\ref{spectral_efficiency_MMSE_dec}).

\section{List of Several Sets} \label{appen_set}
Table~\ref{table_set} lists several sets used in this paper. 
The other sets, such as $\mathcal{Y}$, $\mathcal{I}_{\mathcal{T}_{t}}$, and 
so on, are defined according to the rule described in 
Section~\ref{section_notation}. 
The indices of chips, symbol periods, users, transmit antennas, and replicas 
are denoted by $l$, $t$, $k$, $m$, and $a$, respectively. The indices 
$l$, $t$, $k$, and $m$ move from $1$ to the spreading factor~$L$, 
the coherence time~$T_{\mathrm{c}}$, the number of users~$K$, and 
the number of transmit antennas~$M$, respectively. 
The index $a$ runs from $0$ to $\tilde{n}$ ($n$) 
for Appendix~\ref{derivation_lemma_channel_estimation} 
(Appendix~\ref{appendix_derivation_proposition1}), which denotes the number 
of replicas. In this paper, 
indices themselves have meanings, as noted in Section~\ref{section_notation}. 
For example, $\mathcal{X}_{t}$ should not be confused with $\mathcal{X}_{k}$.  
The former denotes the pilot symbols in symbol period~$t$, while the latter 
represents all pilot symbols for user~$k$. 

\begin{table}[t]
\caption{ 
List of several sets. 
}
\label{table_set} 
\begin{center}
\begin{tabular}{|c|c|c|}
\hline Sets & Definitions & Eqs \\ 
\hline\hline 
$\mathcal{K}$ & finite subset of $\{1,\ldots,K\}$ & -- \\ 
\hline 
$\mathcal{T}_{t}$ & $\{1,\ldots,t\}$ & -- \\ 
\hline 
$\mathcal{C}_{t}$ & $\{t,\ldots,T_{\mathrm{c}}\}$ & -- \\ 
\hline 
$\{\mathcal{A}_{k}\}$ & disjoint subsets of $\{2,\ldots,n\}$ & (\ref{X_k}) \\ 
\hline\hline 
$\mathcal{Y}_{t}$ & 
$\{\boldsymbol{y}_{l,t}\in\mathbb{C}^{N}:\hbox{for all $l$}\}$ & 
(\ref{MIMO_DS_CDMA}) \\ 
\hline
$\tilde{\mathcal{Y}}_{t}$ & 
$\{\tilde{\boldsymbol{y}}_{l,t}\in\mathbb{C}^{N}:\hbox{for all $l$}\}$ 
& (\ref{MIMO_DS_CDMA_MMSE}) \\ 
\hline 
$\mathcal{S}_{t}$ & 
$\{s_{l,t,k,m}\in\mathbb{C}:\hbox{for all $t$, $k$, $m$}\}$ & 
(\ref{MIMO_DS_CDMA}) \\ 
\hline 
$\mathcal{H}_{k}$ & 
$\{\boldsymbol{h}_{k,m}\in\mathbb{C}^{N}:\hbox{for all $m$}\}$ & 
(\ref{MIMO_DS_CDMA}) \\
\hline 
$\mathcal{H}_{k}^{\{a\}}$ & replica of $\mathcal{H}_{k}$ & -- \\
\hline 
$\mathcal{H}_{k}^{\{0\}}$ & $\mathcal{H}_{k}$ & -- \\
\hline 
$\tilde{\mathcal{H}}_{k}$ & 
$\{\tilde{\boldsymbol{h}}_{k,m}\in\mathbb{C}^{N}:\hbox{for all $m$}\}$ & 
(\ref{postulated_posterior_H}) \\ 
\hline 
$\tilde{\mathcal{H}}_{k}^{\{a\}}$ & replica of $\tilde{\mathcal{H}}_{k}$ & --\\ 
\hline 
$\tilde{\mathcal{H}}_{k}^{\{0\}}$ & $\mathcal{H}_{k}$ & -- \\ 
\hline 
$\mathcal{X}_{t,k}$ & 
$\{x_{t,k,m}\in\mathbb{C}:\hbox{for all $m$}\}$ & (\ref{input_symbol}) \\
\hline 
$\mathcal{B}_{t,k}$ & 
$\{b_{t,k,m}\in\mathbb{C}:\hbox{for all $m$}\}$ & (\ref{input_symbol}) \\
\hline 
$\tilde{\mathcal{B}}_{t,k}$ & 
$\{\tilde{b}_{t,k,m}\in\mathbb{C}:\hbox{for all $m$}\}$ & 
(\ref{MIMO_DS_CDMA_MMSE}) \\ 
\hline 
$\mathcal{B}_{t,k}^{\{a\}}$ & replica of $\tilde{\mathcal{B}}_{t,k}$ & -- \\
\hline 
$\mathcal{B}_{t,k}^{\{0\}}$ & $\mathcal{B}_{t,k}$ & -- \\ 
\hline 
$\mathcal{U}_{t}$ & $\{u_{t,k,m}\in\mathbb{C}:\hbox{for all $k$, $m$}\}$ & 
(\ref{MIMO_DS_CDMA}) \\ 
\hline
$\mathcal{I}_{t}$ & $\{\mathcal{Y}_{t},\ \mathcal{S}_{t},\  
\mathcal{U}_{t}\}$ & -- \\ 
\hline 
$\overline{\mathcal{I}}_{t}$ & $\{\mathcal{Y}_{t},\ \mathcal{S}_{t}\}$ & -- \\ 
\hline\hline  
$\underline{\mathcal{Y}}_{t,k}$ & 
$\{\underline{\boldsymbol{y}}_{t,k,m}\in\mathbb{C}^{N}:
\hbox{for all $m$}\}$ & (\ref{SIMO}) \\ 
\hline 
$\underline{\mathcal{I}}_{t,k,m}$ & 
$\{u_{t,k,m},\ \underline{\boldsymbol{y}}_{t,k,m}\}$ & (\ref{SIMO}) \\ 
\hline 
\end{tabular}
\end{center}
\end{table}

% Can use something like this to put references on a page
% by themselves when using endfloat and the captionsoff option.
\ifCLASSOPTIONcaptionsoff
  \newpage
\fi

% trigger a \newpage just before the given reference
% number - used to balance the columns on the last page
% adjust value as needed - may need to be readjusted if
% the document is modified later
%\IEEEtriggeratref{8}
% The "triggered" command can be changed if desired:
%\IEEEtriggercmd{\enlargethispage{-5in}}

% references section

% can use a bibliography generated by BibTeX as a .bbl file
% BibTeX documentation can be easily obtained at:
% http://www.ctan.org/tex-archive/biblio/bibtex/contrib/doc/
% The IEEEtran BibTeX style support page is at:
% http://www.michaelshell.org/tex/ieeetran/bibtex/
\bibliographystyle{IEEEtran}
% argument is your BibTeX string definitions and bibliography database(s)
\bibliography{IEEEabrv,kt-it20101}
\end{document}